\documentclass[10pt]{article}

\usepackage{geometry}
\usepackage{lmodern}
\usepackage[T1]{fontenc}
\usepackage[mathletters]{ucs} 
\usepackage[utf8]{inputenc}

\usepackage{graphicx}
\graphicspath{{images/}}

\usepackage{amsmath, amsthm}
\usepackage{amssymb, amsbsy}
\usepackage[dvipsnames]{xcolor}
\usepackage{hyperref}
\hypersetup{
colorlinks=true,
linkcolor=Brown,
citecolor=OliveGreen,
urlcolor=RoyalBlue,
}

\usepackage{floatflt, wrapfig}
\usepackage{xcolor}

\usepackage{graphicx}
\graphicspath{ {images/} }
\usepackage{epstopdf}
\usepackage{color, import}
\usepackage{sectsty}
\usepackage{enumitem}
\usepackage{lineno}

\usepackage{cite, url}
\usepackage[edges]{forest}

\usepackage[linesnumbered,vlined,ruled]{algorithm2e}
\SetAlFnt{\footnotesize}
\SetAlCapFnt{\small}
\SetAlCapNameFnt{\small}

\usepackage{setspace}
\usepackage[list=true]{subcaption}
\usepackage[title]{appendix}
\usepackage{multirow}
\usepackage{array}

\usepackage{etoolbox}
\makeatletter
\patchcmd{\@maketitle}{\begin{center}}{\begin{flushleft}}{}{}
\patchcmd{\@maketitle}{\begin{tabular}[t]{c}}{\begin{tabular}[t]{@{}l}}{}{}
\patchcmd{\@maketitle}{\end{center}}{\end{flushleft}}{}{}
\makeatother

\let\oldnl\nl
\newcommand{\nonl}{\renewcommand{\nl}{\let\nl\oldnl}}

\renewenvironment{abstract}
{\small\section*
{\bfseries\noindent{\raisebox{-.15\baselineskip}{\normalsize\abstractname}}\hrulefill} 
}

\newtheorem{theorem}{Theorem}
\newtheorem{lemma}{Lemma}

\newtheorem{property}{Property}

\author{hkhj\\hjk}
\author{hjjkhj\\hjk}

\begin{document}
 \pagenumbering{gobble}

\title{Pattern Formation by Robots with Inaccurate Movements}
    \author{    
    Kaustav Bose\\
    \small \emph{Department of Mathematics, Jadavpur University, India}\\
    \small \emph{kaustavbose.rs@jadavpuruniversity.in}\\\\
    Archak Das\\
    \small \emph{Department of Mathematics, Jadavpur University, India}\\
    \small \emph{archakdas.math.rs@jadavpuruniversity.in}\\\\
    Buddhadeb Sau\\
    \small \emph{Department of Mathematics, Jadavpur University, India}\\
    \small \emph{buddhadeb.sau@jadavpuruniversity.in}\\\\
    }
    \date{}
    
  \maketitle

\begin{abstract}

\textsc{Arbitrary Pattern Formation} is a fundamental problem in autonomous mobile robot systems. The problem asks to design a distributed algorithm that moves a team of autonomous, anonymous and identical mobile robots to form any arbitrary pattern $F$ given as input. In this paper, we study the problem for robots whose movements can be inaccurate. Our movement model assumes errors in both direction and extent of the intended movement. Forming the given pattern exactly is not possible in this setting. So we require that the robots must form a configuration  which is close to the given pattern $F$. We call this the \textsc{Approximate Arbitrary Pattern Formation} problem. We show that with no agreement in coordinate system, the problem is unsolvable, even by fully synchronous robots, if the initial configuration 1) has rotational symmetry and there is no robot at the center of rotation or 2) has reflectional symmetry and there is no robot on the reflection axis. From all other initial configurations,  we solve the problem by 1) oblivious, silent and semi-synchronous robots and 2) oblivious,  asynchronous robots that can communicate using externally visible lights.  

\end{abstract}
%


%

\section{Introduction}

\subsection{Background and Motivation}
A robot swarm is a distributed system of autonomous mobile robots that collaboratively execute some complex tasks. Swarms of cheap, weak and generic robots are emerging as a viable alternative to using a single sophisticated and expensive robot. Robot swarms have potential use in a wide range of practical problems, e.g., structural health monitoring, search and rescue missions, environmental remediation, military application etc. As a result, distributed coordination of robot swarms has attracted considerable research interest. Early investigations of these problems were experimental in nature with the main emphasis being on designing good heuristics. However, the last two decades have seen a flurry of theoretical studies on the computability and complexity issues related to distributed computing by such system of robots. These studies are aimed at providing provably correct algorithmic solutions to fundamental coordination problems. The recent book \cite{thebook} provides a comprehensive survey of the large number of works that have been done in this direction.

The traditional framework of theoretical studies considers a very restrictive model of robots: 
\begin{itemize}
    \item the robots are \emph{anonymous} (they have no unique identifiers that they can use in computations), \emph{homogeneous} (they execute the same distributed algorithm) and \emph{identical} (they are indistinguishable by their appearance),
    
    \item the robots do not have access to any global coordinate system, each robot perceives its surroundings based on its  local coordinate system
    
    \item the robots are have either no memory or very little memory available to remember past observations and calculations
    
    \item the robots  either have no means of direct communication or  have some very weak communication mechanism (e.g., an externally visible light that can assume a small number of predefined colors)
    
\end{itemize}


The model assumes robots with such weak features because the theoretical studies usually intend to find the minimum capabilities necessary for the robots to solve a given problem. The objective of this approach is to obtain a clear picture of the relationship between different features and capabilities of the robots (such as memory, communication, sensing, synchronization, agreement among local coordinate systems etc.) and their exact role in solvability of fundamental problems.  Adopting such restrictive model also makes sense from a practical perspective since the individual units of robot swarms are low-cost generic robots with limited capabilities. Although certain assumptions, such as obliviousness (having no memory of past observations and calculations), may seem to be overly restrictive even for such weak robots, there are specific motivations for these assumptions. For example, the assumption of oblivious robots ensures  self-stabilization. This is because any algorithm that works correctly for oblivious robots is inherently self-stabilizing as it tolerates errors that alter the  local states of the robots. 


While the robots are assumed to be very weak with respect to memory, communication etc., certain aspects of the model are overly strong. In particular, the assumed mobility features of the robots are very strong. Two standard models regarding the movement of the robots are \textsc{Rigid} and \textsc{Non-Rigid}. In \textsc{Rigid}, if a robot $x$ wants to go to any point $y$, then it can move to exactly that point in one step. This means that the robots are assumed to be able to execute error-free movements in any direction and by any amount. Certain studies also permit the robots to move along curved trajectories. The algorithms in this model rely on the accurate execution of the movements and are not robust to movement errors that real life robots are susceptible to. Furthermore, the error-free movements of the robots have surprising theoretical consequences as shown in the remarkable results obtained in \cite{LunaFSVY20a}. A `positional encoding' technique was developed in \cite{LunaFSVY20a} that allows a robot, that have very limited or no memory to store data, to implicitly store unbounded amount of information by encoding the data in the binary representation of its distance from another robot or some other object, e.g., the walls of the room inside which it is deployed. Exact movements allow the robots to preserve and update the data. This gives the robots remarkable computational power that allows them to solve complex problems which appear to be unsolvable by robots with limited or no memory, e.g., constructing a map of a complex art gallery by an oblivious robot. Obviously these techniques are impossible to implement in practice. Also, for problems that we expect to be unsolvable by real life robots with certain restrictions in memory, communication etc., it may become difficult or impossible to theoretically establish a hardness or impossibility result due to the strong model. The \textsc{Non-Rigid} model assumes that a robot may stop before reaching its intended destination. However, there exists a constant $\delta > 0$ such that if the destination is at most $\delta$ apart, the robot will reach it; otherwise, it will move towards the destination  by at least $\delta$.  The assumption of existence of such a $\delta$ is necessary, because otherwise, the model will not allow any robot to traverse a distance $\geq d$, for any for any distance $d > 0$.  This follows from a classical Zenonian argument as the model allows interruptions after moving distances $\frac{d}{2}, \frac{d}{4}, \frac{d}{8}, \ldots$. Notice that in the \textsc{Non-Rigid} model, 1) the movement is still error-free if the destination is close enough, i.e., within $\delta$, and 2) there is no error whatsoever in the direction of the movement even if the destination is far away. In \cite{siroccoBoseAKS19}, it was shown that these two properties allow robots to implement positional encoding even in the \textsc{Non-Rigid} model. This motivates us to consider a new movement model allowing inaccurate movements.

\subsection{Our Contribution}

We consider a movement model that assumes errors in both direction and extent of the intended movement. Also, the errors can occur no matter what the extent of the attempted movement is. The details of the model are presented in Section \ref{Section mod}. In this model, we study the \textsc{Arbitrary Pattern Formation} problem. \textsc{Arbitrary Pattern Formation} is a fundamental robot coordination problem and has been extensively studied in the literature. The goal is to design a distributed algorithm that  allows the robots to form any pattern $F$ given as input. This problem has been extensively studied in the literature in the \textsc{Rigid} and \textsc{Non-Rigid} model. However, the techniques used in these algorithms are not readily portable in our setting. For example, in most of these algorithms, the minimum enclosing circle of the configuration plays an important role. The center of the minimum enclosing circle is set as the origin of the coordinate system with respect to which the pattern is to be formed. So the minimum enclosing circle is kept invariant throughout the algorithm. The robots inside the minimum enclosing circle move to form the part of the pattern inside the circle, without disturbing it. For the pattern points on the minimum enclosing circle, robots from the inside may have to move on to the circle. Also, the robots on the minimum enclosing circle, in order to reposition themselves in accordance with the pattern to be formed, will move along the circumference so that the minimum enclosing circle does not change. Notice that while moving along the circle, an error prone robot might skid off the circle. Also, when a robot from the inside attempts to move exactly on to the circle, it may move out of the circle due to the error in movement. In both cases, the minimum enclosing circle will change and the progress made by the algorithm will be lost. In fact, we face difficulty at a more fundamental level: exactly forming an arbitrary pattern is impossible by robots with inaccurate movements. Therefore, we consider a relaxed version of the problem called  \textsc{Approximate Arbitrary Pattern Formation}  where the robots are required to form an approximation of the input pattern $F$. We show that with no agreement in coordinate system, the problem is unsolvable, even by fully synchronous robots, if the initial configuration 1) has rotational symmetry and there is no robot at the center of rotation, or 2) has reflectional symmetry and there is no robot on the reflection axis. From all other initial configurations,  we solve the problem in  $\mathcal{OBLOT + SSYNC}$ (the robots are oblivious, silent and semi-synchronous) and $\mathcal{FCOM + ASYNC}$ (the robots are oblivious, asynchronous and can communicate using externally visible lights). 

\subsection{Related Works}
      
The \textsc{Arbitrary Pattern Formation} problem has been extensively studied in the literature in the \textsc{Rigid} and \textsc{Non-Rigid} model \cite{Yamashita99,Yamashita10,Flocchini08,Dieudonne10,Cicerone17,VaidyanathanST18,Bramas16,Fujinaga15,Yamauchi14,Yamauchi13,BoseKAS21,caldam/BoseAKS20}. As mentioned earlier, the techniques used in these studies are not applicable in our setting. The issue of overly strong movement model was addressed in \cite{bose18} where the problem was studied with discrete movements. In particular, the \textsc{Arbitrary Pattern Formation} problem was considered on a grid based terrain where the movements of the robots are restricted only along grid lines and only to a neighbouring grid point in each step. There can be two criticisms towards the grid model. First, from a theoretical perspective, although we assume that the robots do not have access to any global coordinate system, the grid structure of the terrain provides a partial agreement among the local coordinate systems of the robots. In this sense, the grid model is stronger than the continuous plane model with no agreement on coordinate system. Secondly, from a practical perspective, there can be application scenarios where such grid type floor layouts are not available or cannot be set up. 

Movement error was previously considered in \cite{CohenP08}, but in the context of the \textsc{Convergence} problem which requires the robots to converge towards a single point. The error model in \cite{CohenP08} also considers errors in both direction and extent of the intended movement. However, there is some difference between the error model of \cite{CohenP08} and the one introduced in this paper. In particular, the maximum possible error in direction is independent of the extent of the intended movement in \cite{CohenP08}. In our model, the maximum possible error in both direction and extent, depend upon the extent of the intended movement. We believe that this is a reasonable assumption as the error is expected to be less if the  destination of the intended movement is not far away.

\section{Robot Model}\label{Section mod}

\paragraph{The Robots}

A set of $n$ mobile computational entities, called \emph{robots}, are initially positioned at distinct points in the plane. The robots are assumed to be anonymous (they have no unique identifiers that they can use in a computation), identical (they are indistinguishable by their physical appearance), autonomous (there is no centralized control) and homogeneous (they execute the same deterministic algorithm). The robots are modeled as points in the plane, i.e., they do not have any physical extent. The robots do not have access to any global coordinate system. Each robot has its own local coordinate system centered at its current position. There is no consistency among the local coordinate systems of the robots except for a common unit of distance. We call this the \emph{standard unit of distance}.

\paragraph{Memory and Communication}

Based on the memory and communication capabilities, there are four models: $\mathcal{OBLOT}$, $\mathcal{LUMI}, \mathcal{FSTA}$ and $\mathcal{FCOM}$. In $\mathcal{OBLOT}$, the robots are \emph{silent} (they have no explicit means
of communication) and \emph{oblivious} (they have no memory of past observations and computations). In $\mathcal{LUMI}$, the robots are equipped with visible lights which can assume a constant number of colors. The lights serve both as a weak explicit communication mechanism and a form of internal memory. In $\mathcal{FSTA}$, a robot can only see the color of its own light, i.e., the light is \emph{internal} and in $\mathcal{FCOM}$, a robot can only see  the light of other robots, i.e., the light is \emph{external}. Therefore in $\mathcal{FSTA}$, the robots are silent, but have finite memory, while in $\mathcal{FCOM}$, the robots are oblivious, but have finite communication capability.

\paragraph{Look-Compute-Move Cycles}

The robots, when active, operate according to the so-called \textsc{Look-Compute-Move} cycles. In each cycle, a previously idle or inactive robot wakes up and executes the following steps. In the \textsc{Look} phase, the robot takes the snapshot of the positions (and their lights in case of $\mathcal{LUMI}$ and $\mathcal{FCOM})$ of all the robots. Based on the perceived configuration (and its own light in case of $\mathcal{LUMI}$ and $\mathcal{FSTA})$, the robot performs computations according to a deterministic algorithm to decide a destination point and (a color in case of $\mathcal{LUMI}, \mathcal{FSTA}$ and $\mathcal{FCOM}$). Based on the outcome of the algorithm, the robot (sets its light to the computed color in case of  $\mathcal{LUMI}, \mathcal{FSTA}$ and $\mathcal{FCOM}$, and ) either remains stationary or attempts to move to the computed destination.

\paragraph{Scheduler}

Based on the activation and timing of the robots, there are three types of schedulers considered in the literature. In $\mathcal{FSYNC}$ or  fully synchronous, time can be logically divided into global rounds. In each round, all the robots are activated. They take the snapshots at the same time, and then perform their moves simultaneously. $\mathcal{SSYNC}$ or semi-synchronous coincides with  $\mathcal{FSYNC}$, with the only difference that not all robots are necessarily activated in each round. However, every robot is activated infinitely often.
In $\mathcal{ASYNC}$ or asynchronous, there are no assumptions except that every robot is activated infinitely often. In particular, the robots are activated independently and each robot executes its cycles independently. The amount of time spent in \textsc{Look}, \textsc{Compute}, \textsc{Move} and inactive states is finite but unbounded, unpredictable and not same for different robots.

\paragraph{Movement of the Robots}

\begin{figure}[h!]
\centering
\subcaptionbox[Short Subcaption]{
        \label{fig: error0}
}
[
    0.24\textwidth 
]
{
    \fontsize{8pt}{8pt}\selectfont
    \def\svgwidth{0.24\textwidth}
    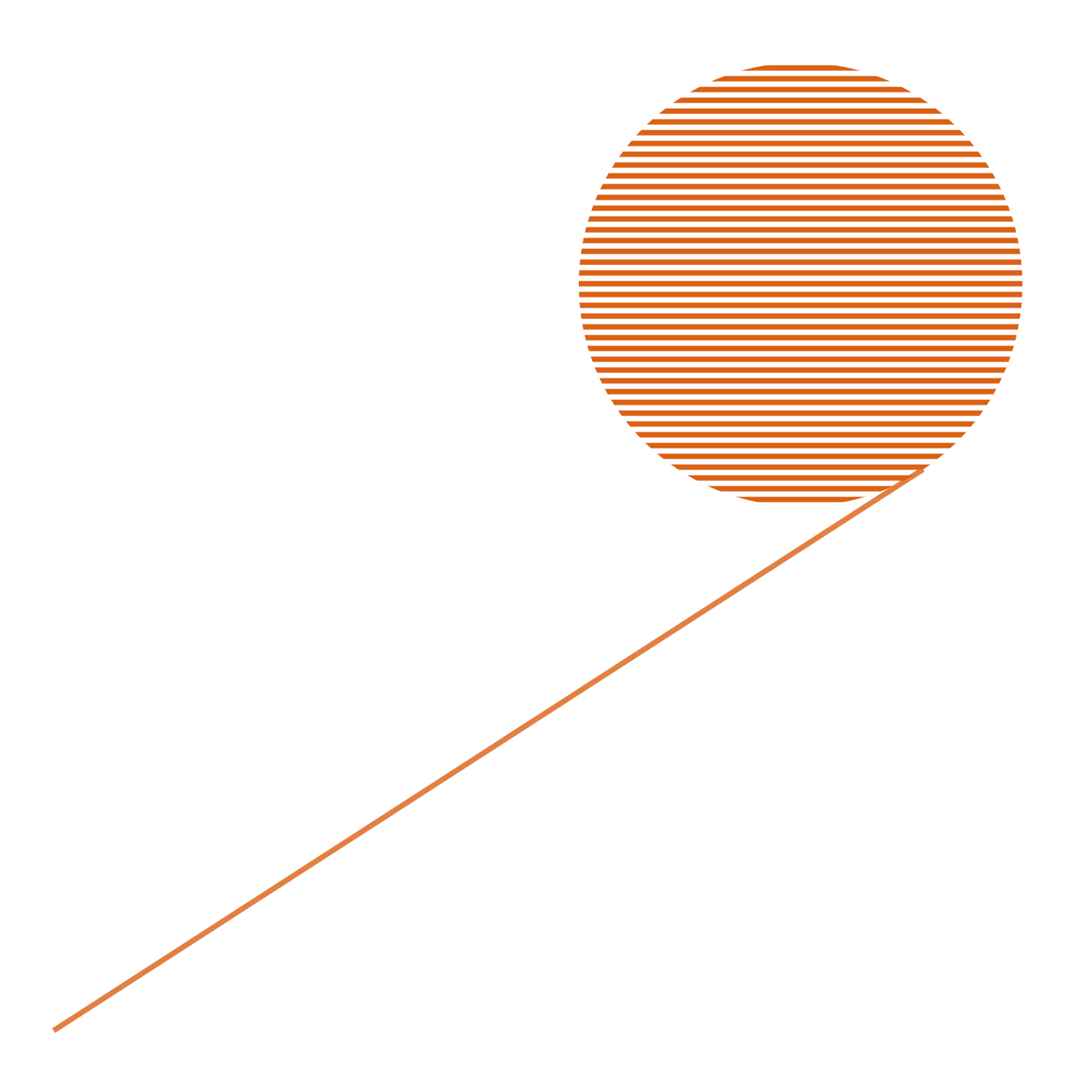
}
\hfill
\subcaptionbox[Short Subcaption]{
     \label{fig: error}
}
[
    0.75\textwidth 
]
{
    \fontsize{8pt}{8pt}\selectfont
    \def\svgwidth{0.75\textwidth}
    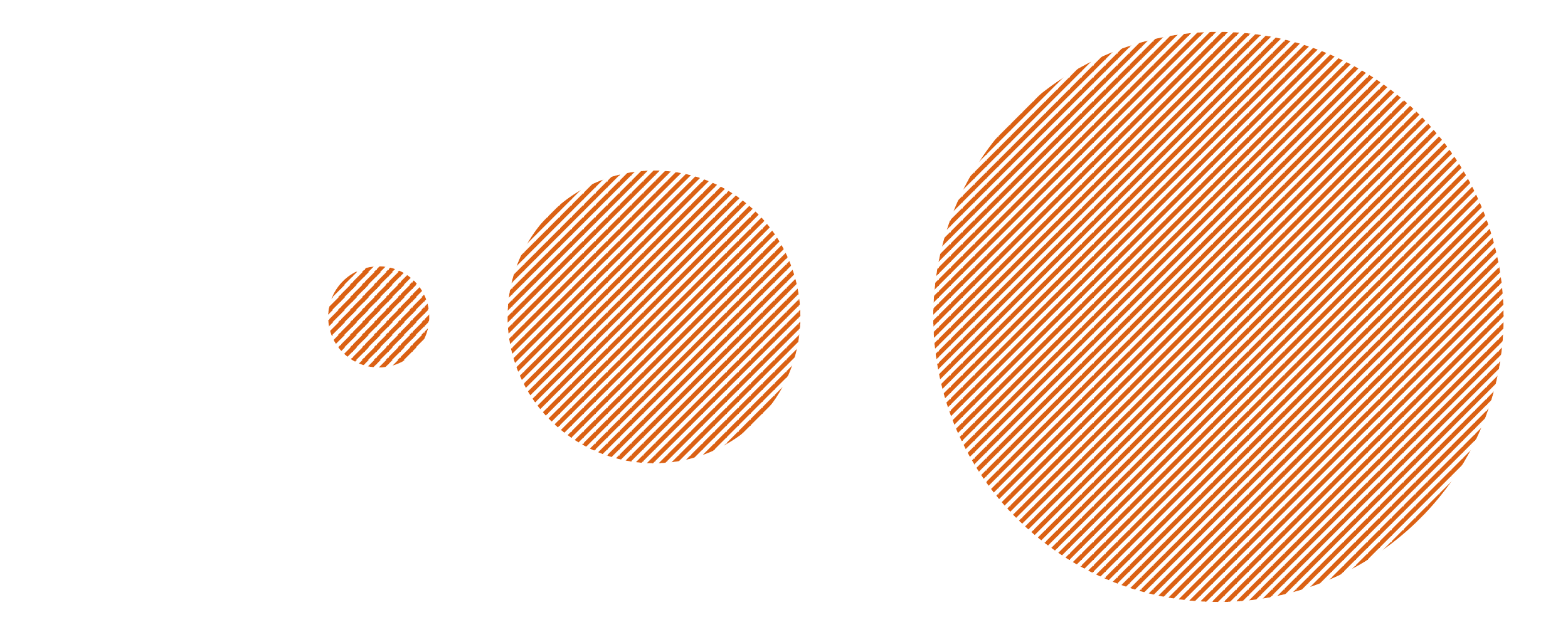
}
\caption[]{a) If a robot attempts to move from $x$ to $y$, then it will reach at some point $z$ in the shaded region $\mathcal{Z}(x,y)$.   b) If a robot attempts to move from $x$ to $y_i$, then it will reach at some point in $\mathcal{Z}(x,y_i)$ which is the shaded region around $y_i$. Observe that $error_d(x,y_1) < error_d(x,y_2) < error_d(x,y_3)$, but $error_a(x,y_1) < error_a(x,y_2) = error_a(x,y_3)$.}
\label{}
\end{figure}

There are known constants $0 < \lambda < 1$, $0 < \Delta < 1$, such that if a robot at $x$ attempts to move to $y$, then it will reach a point $z$ where $d(z,y) < \mu(x,y) d(x,y)$ where $\mu(x,y) =$ min$\{\Delta, \lambda d(x,y)\}$. Here $d(x,y)$ denotes (the numerical value of) the distance between the points $x$ and $y$ measured in the standard unit of distance. The movement of the robot will be along the straight line joining $x$ and $z$. We denote by $\mathcal{Z}(x,y)$ the set of all points where a robot may reach if it attempts to move from $x$ to $y$. So $\mathcal{Z}(x,y)$ is the open disk $\{z \in \mathbb{R}^2 \mid d(z,y) < \mu(x,y) d(x,y)\}$ (c.f. Fig. \ref{fig: error0}). We denote by $error_d(x,y)$ and $error_a(x,y)$ the supremums of the possible distance errors (i.e., the deviation from the intended amount of distance to be traveled) and angle  errors (i.e., the angular deviation from the intended trajectory) respectively when a robot intends to travel from $x$ to $y$. Notice that $error_d(x,y)$ is equal to the radius of $\mathcal{Z}(x,y)$ and $error_a(x,y)$ is equal to the angle between $line(x,y)$ and a tangent on $\mathcal{Z}(x,y)$ passing through $x$. Hence, $error_d(x,y) = \mu(x,y) d(x,y)$ and $error_a(x,y) = sin^{-1}(\frac{\mu(x,y) d(x,y)}{d(x,y)}) = sin^{-1}(\mu(x,y))$. Also notice that 1) $error_d(x,y)$ increases with $d(x,y)$, and 2) $error_a(x,y)$ increases with $d(x,y)$ only up to a certain value, i.e., $sin^{-1}(\Delta)$ and then remains constant (c.f. Fig. \ref{fig: error}). So, $error_a(x,y) \leq sin^{-1}(\Delta)$, for any $x, y$.

\section{Definitions and Notations}\label{sec: def}

We denote the configuration of robots by $R = \{r_1, r_2, \ldots , r_n\}$ where each $r_i$ denotes a robot as well as the point in the plane where it is situated. The input pattern given to the robots will be denoted by $F = \{f_1, f_2, \ldots f_n\}$ where each $f_i$ denotes an element from $\mathbb{R}^2$. 

\begin{figure}[h!]
\centering
\subcaptionbox[Short Subcaption]{
        \label{fig: cone}
}
[
    0.22\textwidth 
]
{
    \fontsize{8pt}{8pt}\selectfont
    \def\svgwidth{0.21\textwidth}
    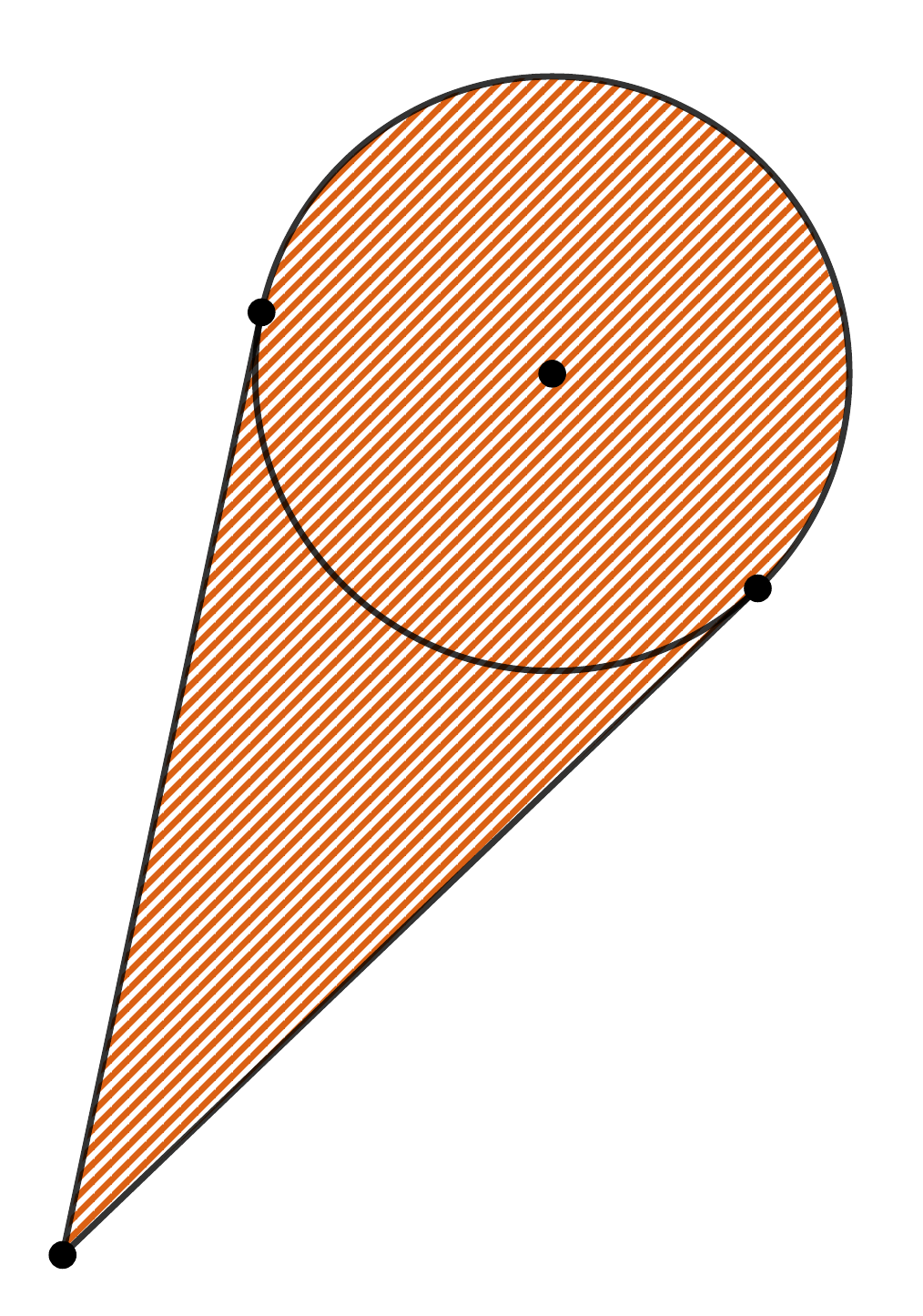
}
\hfill
\subcaptionbox[Short Subcaption]{
        \label{}
}
[
    0.38\textwidth 
]
{
    \fontsize{8pt}{8pt}\selectfont
    \def\svgwidth{0.38\textwidth}
    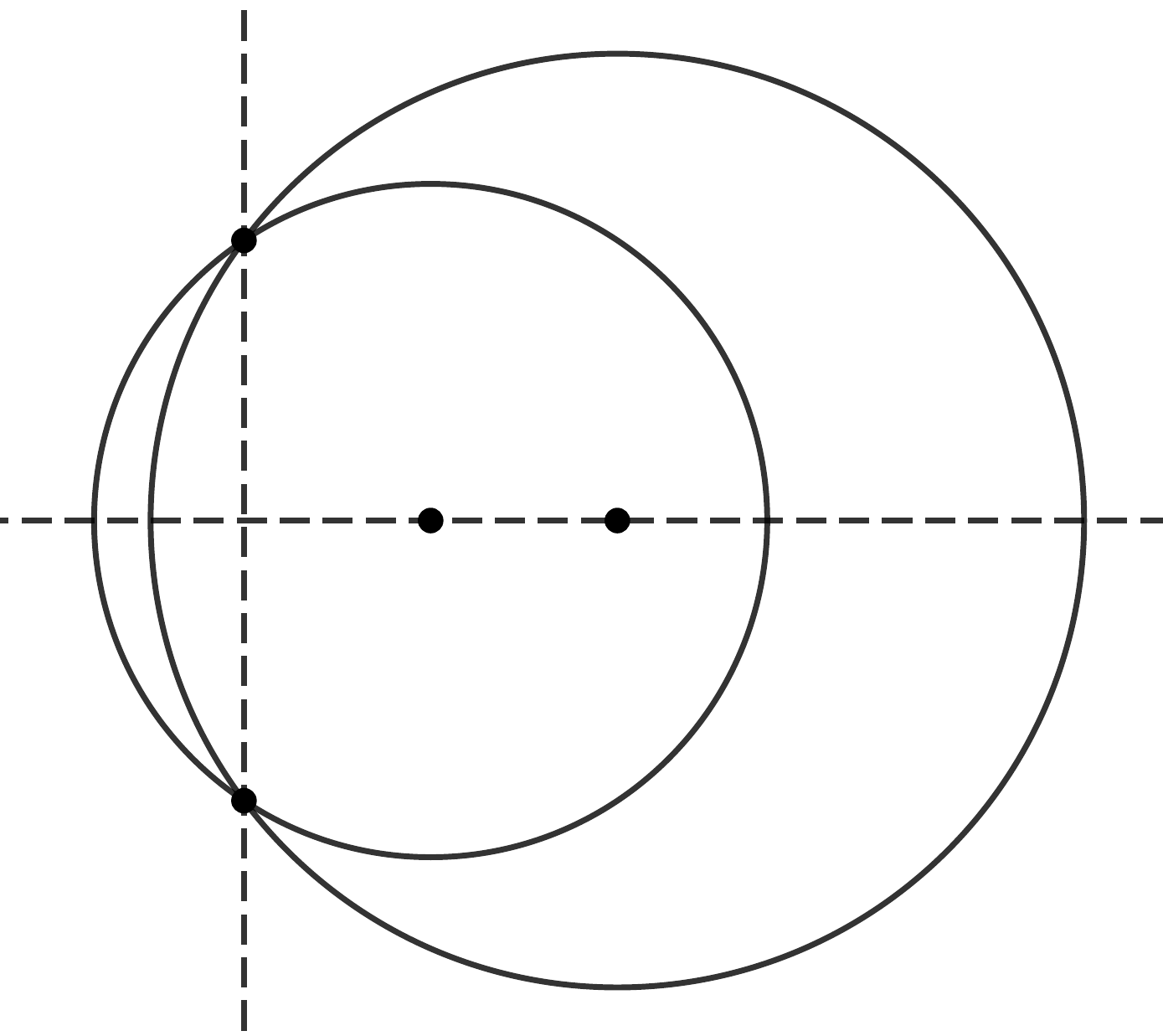
}
\hfill
\subcaptionbox[Short Subcaption]{
        \label{}
}
[
    0.38\textwidth 
]
{
    \fontsize{8pt}{8pt}\selectfont
    \def\svgwidth{0.38\textwidth}
    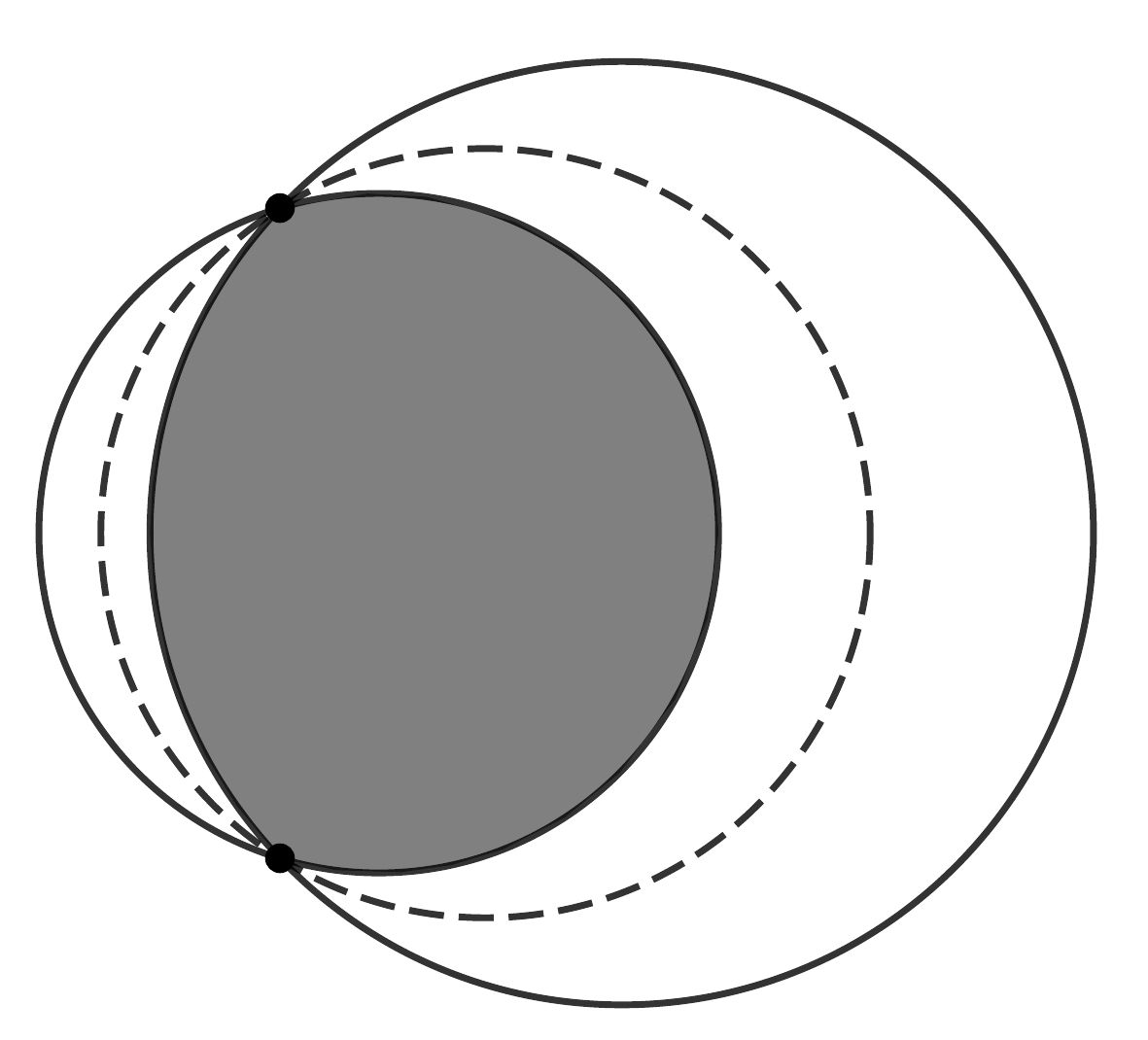
}

\caption[Short Caption]{a) $Cone(x, B(y,l))$ is the shaded open region. b)-c) Illustrations supporting Property \ref{fam1} and \ref{fam} respectively.}
\label{}
\end{figure}

Given two points $x$ and $y$ in the Euclidean plane, let $d(x, y)$ denote the distance between the points $x$ and $y$ measured in the standard unit of distance. For three points $x, y$ and $c$, $\angle_{\circlearrowright} xcy$ ($\angle_{\circlearrowleft} xcy$) is the angle centered at $c$ measured from $x$ to $y$ in the clockwise (resp. counterclockwise) direction. Also, $\angle xcy =$ min$\{\angle_{\circlearrowright} xcy$, $\angle_{\circlearrowleft} xcy\}$.   We denote by $line(x, y)$ the straight line passing through $x$ and $y$. By $seg(x, y)$ ($\overline{seg}(x, y)$) we denote the line segment joining $x$ and $y$  excluding (resp. including) the end points. If $\ell_1$ and $\ell_2$ be two parallel lines, then $\mathcal{S}(\ell_1,\ell_2)$ denotes the open region between these two lines. For any point $c$ in the Euclidean plane and a length $l$, $C(c, l) = \{z \in \mathbb{R}^2 \mid d(c,z) = l\}$, $B(c, l) = \{z \in \mathbb{R}^2 \mid d(c,z) < l\}$ and $\overline{B}(c, l) = \{z \in \mathbb{R}^2 \mid d(c,z) \leq l\}$ $= B(c, l) \cup C(c, l)$. If $C$ is a circle then $encl(C)$ and $\overline{encl}(C)$ respectively denote the open and closed region enclosed by $C$. Also, $ext(C) = \mathbb{R}^2 \setminus \overline{encl}(C)$ and $\overline{ext}(C) = \mathbb{R}^2 \setminus encl(C)$. Hence, $\overline{encl}(C) = encl(C) \cup C$ and $\overline{ext}(C) = ext(C) \cup C$. Let $x, y$ be two points in the plane and $d(x, y) > l$. Suppose that the tangents from $x$ to $C(y, l)$ touches $C(y, l)$ at $a$ and $b$. The $Cone(x, B(y,l))$ is the open region enclosed by $B(y,l), \overline{seg}(x, a)$ and $\overline{seg}(x, b)$, as shown in Fig. \ref{fig: cone}. Also, $\angle Cone(x, B(y,l)) = \angle axb$.

\begin{property}\label{prop: cone reduce}
If $x' \in Cone(x, B(y,l))$, then $d(x', y) < d(x,y)$.
\end{property}

\begin{property}\label{prop: cone}
If $x' \in Cone(x, B(y,l)) \setminus \overline{B}(y,l)$, then $Cone(x', B(y,l)) \subset Cone(x, B(y,l))$.
\end{property}

Given two points $x$ and $y$ in the Euclidean plane, we denote by $\mathcal{F}(x,y)$ the family of circles passing through $x$ and $y$. The center of all the circles lie on the perpendicular bisector of $\overline{seg}(x,y)$, say $\ell$. Each point $c \in \ell$ defines a unique circle of the family $\mathcal{F}(x,y)$, i.e., the circle from $\mathcal{F}(x,y)$ having center at $c$. If $C_1, C_2 \in \mathcal{F}(x,y)$ and $c_1, c_2$ be their centers respectively,  then $(C_1, C_2)_{\mathcal{F}(x,y)}$ and $[C_1, C_2]_{\mathcal{F}(x,y)}$ will denote respectively the family of circles $\{C \in \mathcal{F}(x,y) \mid c \in seg(c_1,c_2), \text{ where $c$ is the center of $C$} \}$ and $\{C \in \mathcal{F}(x,y) \mid c \in \overline{seg}(c_1,c_2), \text{ where $c$ is the center of $C$} \}$. 

\begin{property}\label{fam1}
Let $C_1, C_2, \in \mathcal{F}(x,y)$ be such that 1) their centers lie in the same closed half plane $\mathcal{H}$ delimited by $line(x,y)$ and 2) the distance of the center of $C_2$ from $line(x,y)$ is more than that of the center of $C_1$. Then $encl(C_1) \cap \mathcal{H} \subseteq encl(C_2) \cap \mathcal{H}$.    
\end{property}

\begin{property}\label{fam}
Let $C_1, C_2, C_3 \in \mathcal{F}(x,y)$ be such that $C_2 \in (C_1, C_3)_{\mathcal{F}(x,y)}$. Then $encl(C_1) \cap encl(C_3)  \subset encl(C_2)$ and $(\overline{encl(C_1)} \cap \overline{encl(C_3)}) \setminus \{x,y\}  \subset {encl(C_2)}$.    
\end{property}



For a set $P$ of points in the plane, $C(P)$ and $c(P)$ will respectively denote the minimum enclosing circle of $P$ (i.e., the smallest circle $C$ such that $P \subset \overline{encl}(C)$) and its center. The smallest enclosing circle $C(P)$ is unique and can be computed in linear time. For $P$, with $2 \leq |P| \leq 3$, $CC(P)$ denotes the circumcircle of $P$ defined as the following. If $P = \{p_1, p_2\}$, $CC(P)$ is the circle having $\overline{seg}(p_1,p_2)$ as the diameter and if $P = \{p_1, p_2, p_3\}$, $CC(P)$ is the unique circle passing through $p_1, p_2$ and $p_3$.

\begin{property} \label{prop: cc}
If $P' \subseteq P$ such that 1) $P'$ consists of two points or $P'$ consists of three points that form an acute angled triangle, and 2) $P \subset \overline{encl}(CC(P'))$, then $CC(P') = C(P)$. Conversely, for any $P$, $\exists P' \subseteq P$ so that 1) $P'$ consists of two points or $P'$ consists of three points that form an acute angled triangle and 2) $CC(P') = C(P)$.    
\end{property}

From Property \ref{prop: cc} it follows that $C(P)$ passes either through two of the points of $P$ that are on the same diameter (antipodal points), or through at least three points so that some three of them form an acute angled or right angled triangle. Also, $C(P)$ does not change by adding points inside $\overline{encl}(C(P))$ or deleting points in ${encl}(C(P))$. However, $C(P)$ may be changed by deleting points from $C(P)$. A point $p \in P$ is said to be \emph{critical} if $C(P) \neq C(P \setminus \{p\})$. Obviously, $p \in P$ is critical only if $p \in C(P)$. 


\begin{property} \label{prop: cr1}
  If $|P \cap C(P)| \geq 4$ then there exists at least one point from $P \cap C(P)$ which is not critical.
\end{property}

\begin{property} \label{prop: cr2}
Let $p_1, p_2, p_3 \in P$ be three consecutive points in clockwise order on $C(P)$. If $\angle_{\circlearrowright} p_1c(P)p_3 \leq \pi$, then $p_2$ is non-critical.
\end{property}

Consider all the concentric circles that are centered at $c(P)$ and have at least one point of $P$ on them. Let $C_i^{\downarrow}(P)$ ($C_i^{\uparrow}(P)$) denote the $i$th ($i \geq 1$) of these circles so that $C_{i+1}^{\downarrow}(P) \subset encl(C_i^{\downarrow}(P))$ ($C_{i}^{\uparrow}(P) \subset encl(C_{i+1}^{\uparrow}(P))$). We shall denote $c(P)$ by $C_0^{\uparrow}(P)$. So we have $C_1^{\downarrow}(P) = C(P)$ and if there is a point at $c(P)$, then $C_1^{\uparrow}(P) = c(P) = C_0^{\uparrow}(P)$.

We say that a configuration of robots $R$ is \emph{symmetry safe} if one of the following three is true.

\begin{enumerate}
 \item 
 \begin{itemize}
 \item there is some non-critical robot on $C(R)$, hence $|R \cap C(R)| \geq 3$ 
  \item there is no robot at $c(R)$
  \item $|R \cap C_{1}^{\uparrow}(R)| = 1$ and $|R \cap C_{2}^{\uparrow}(R)| = 1$
  \item if $R \cap C_{1}^{\uparrow}(R) = \{r_1\}$ and $R \cap C_{1}^{\uparrow}(R) = \{r_2\}$, then $r_1, r_2, c(R)$ are not collinear.
 \end{itemize}
\item
\begin{itemize}
 \item all robots on $C(R)$ are critical and $R \cap C(R) = \{r_1, r_2, r_3\}$
 \item $\Delta r_1r_2r_3$ is scalene, i.e., all three sides have different lengths
\end{itemize}

\item 
\begin{itemize}
  \item all robots on $C(R)$ are critical and $R \cap C(R) = \{r_1, r_2\}$
  \item $|R \cap C_{1}^{\uparrow}(R)| = 1$ 
  \item if $R \cap C_{1}^{\uparrow}(R) = \{r\}$, $r \notin line(r_1, r_2) \cup \ell$, where $\ell$ is the line passing through $c(R)$ and perpendicular to $line(r_1, r_2)$.
 \end{itemize}

\end{enumerate}

\begin{figure}[h!]
\centering
\subcaptionbox[Short Subcaption]{
        \label{}
}
[
    0.32\textwidth 
]
{
    \fontsize{8pt}{8pt}\selectfont
    \def\svgwidth{0.32\textwidth}
    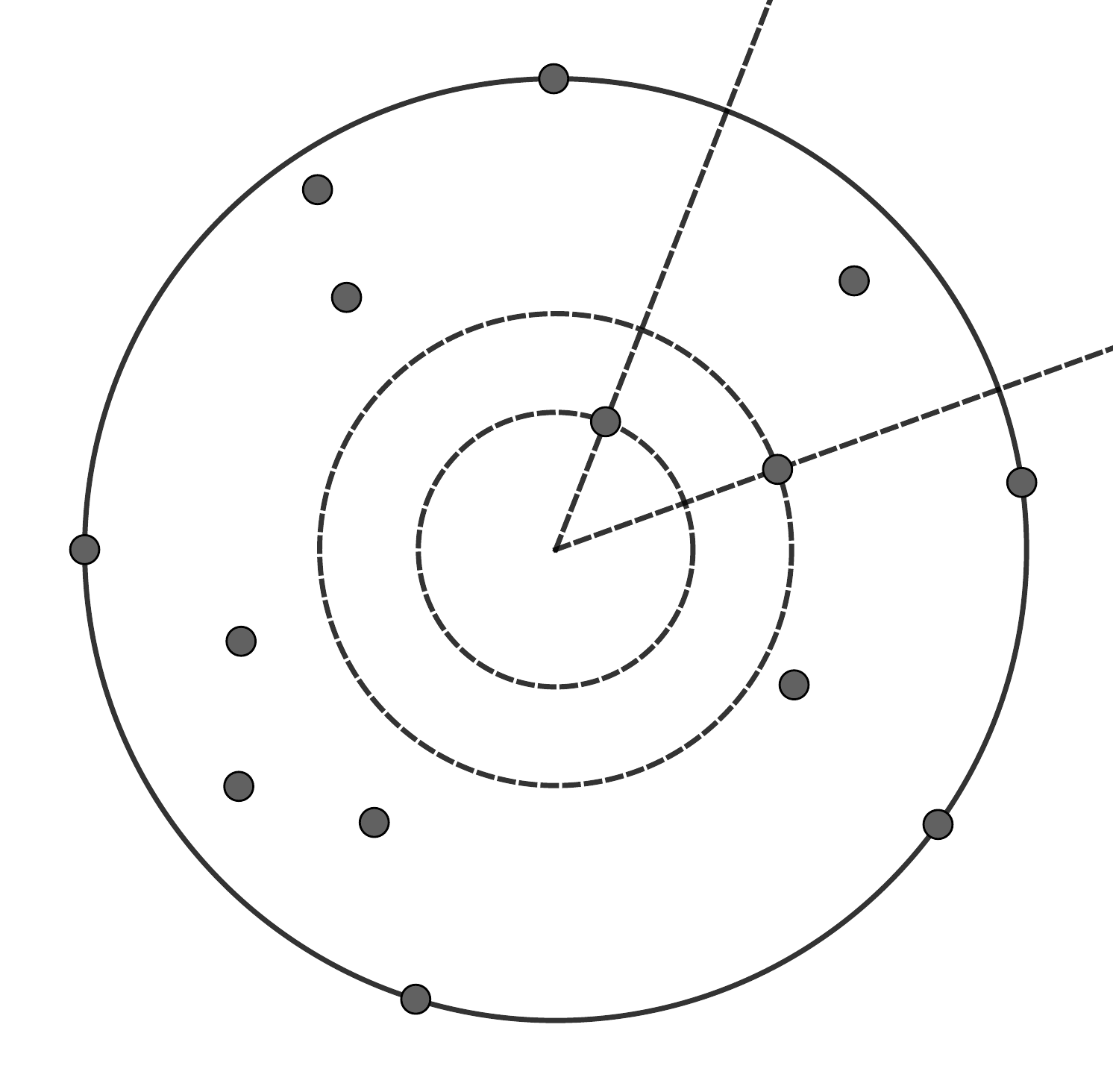
}
\hfill
\subcaptionbox[Short Subcaption]{
        \label{}
}
[
    0.32\textwidth 
]
{
    \fontsize{8pt}{8pt}\selectfont
    \def\svgwidth{0.32\textwidth}
    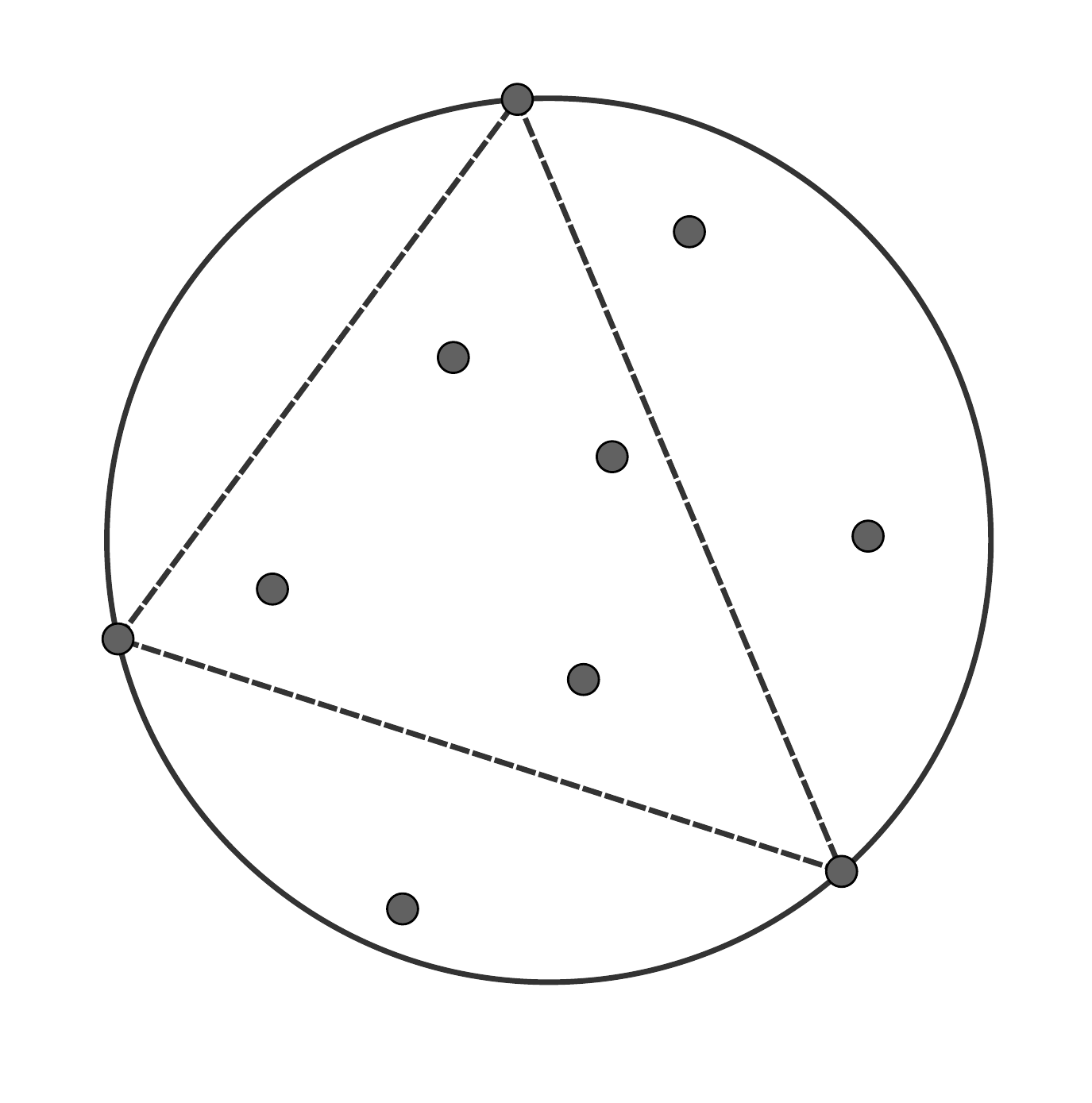
}
\hfill
\subcaptionbox[Short Subcaption]{
        \label{}
}
[
    0.32\textwidth 
]
{
    \fontsize{8pt}{8pt}\selectfont
    \def\svgwidth{0.32\textwidth}
    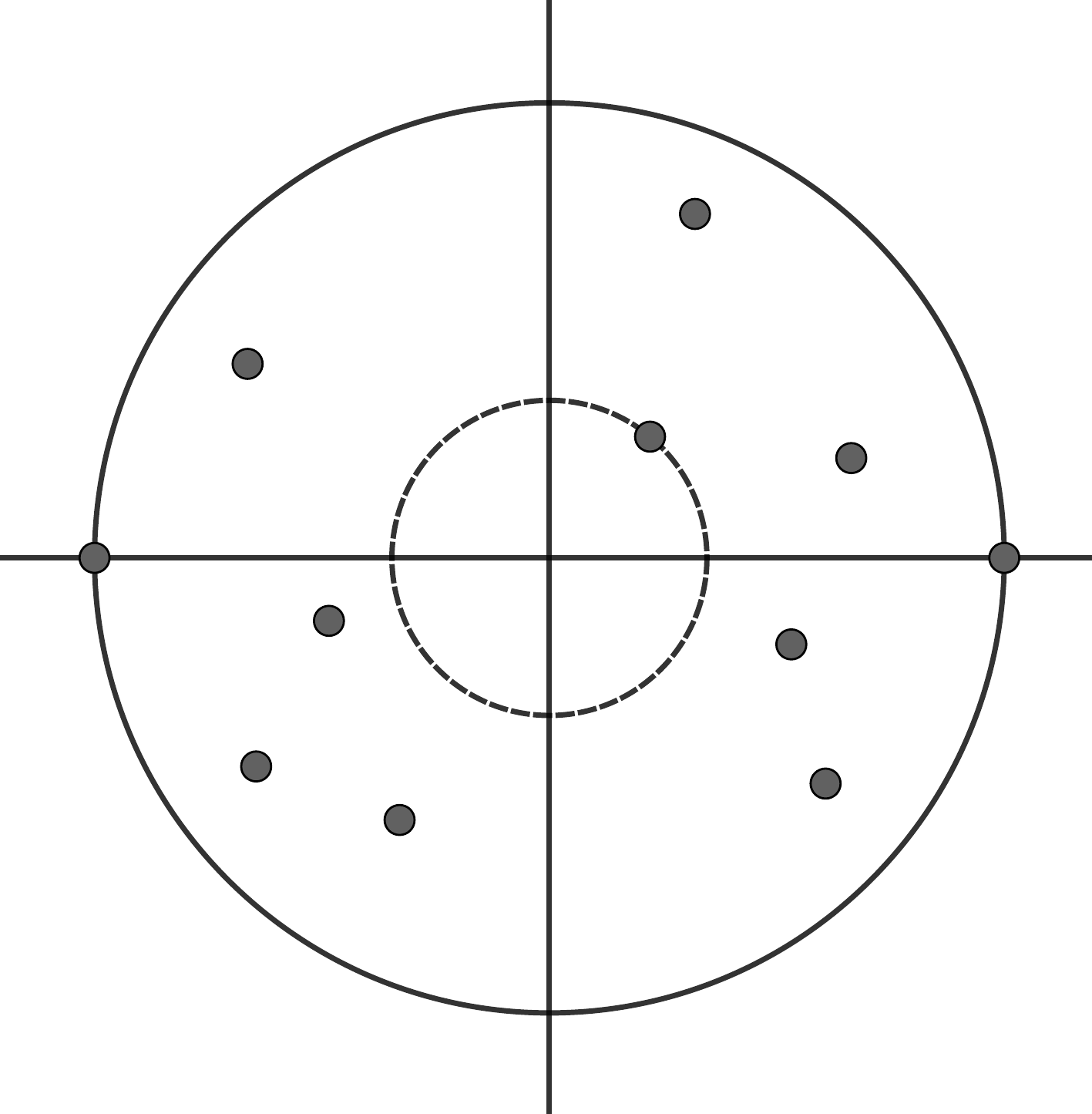
}

\caption[Short Caption]{a) The input pattern $F$. The bounding structure $b_F$ consists of the blue pattern points. b)-c) The bounding structure is formed by the robots. d) To obtain a final configuration, each shaded region must have a robot inside it.}
\label{fig: ss}
\end{figure}

\section{Basic Properties}\label{sec: pr}

\subsection{Approximate Arbitrary Pattern Formation}

The \textsc{Arbitrary Pattern Formation} problem in its standard form is the following. Each robot of a team of $n$ robots is given a pattern $F$ as input. The input pattern $F$ is a list of $n$ distinct elements from $\mathbb{R}^2$. The problem asks for a distributed algorithm that guides the robots to a configuration that is similar to $F$ with respect to translation, reflection, rotation and uniform scaling. We refer to this version of the problem as the \textsc{Exact Arbitrary Pattern Formation} problem, highlighting the fact that the configuration of the robots is required to be exactly similar to the input pattern. However, it is not difficult to see that \textsc{Exact Arbitrary Pattern Formation} is unsolvable in our model where the robot movements are inaccurate.

\begin{theorem}
 \textsc{Exact Arbitrary Pattern Formation} is unsolvable by robots with inaccurate movements.
\end{theorem}

\begin{proof}
 For simplicity, let $n = 3$.  Suppose that the input pattern $F$ is a equilateral triangle. Also, assume that the initial configuration of the robots is not a equilateral triangle, i.e., the required pattern is not already formed. Assume that exactly one robot is activated at each round. Suppose that the pattern is formed at round $i$. Let $r_1$ be the robot which is activated at this round, and $r_2, r_3$ are inactive. There are exactly two points on the plane, say $P_1, P_2$, where $r_1$ needs to go to form the given pattern. But it is not possible to go to exactly at one of these points from any point $P \notin \{P_1, P_2\}$ on the plane. So the pattern is not formed at round $i$. The same arguments hold for any round that follows.
\end{proof}

Therefore, we introduce a relaxed version of the problem called the \textsc{Approximate Arbitrary Pattern Formation}. Intuitively, we want the robots to form a pattern that is close to the given pattern, but may not be exactly similar to it. Formally, the robots are given as input a pattern $F$ and a number $0 < \epsilon < 1$. The number $\epsilon$ is small enough so that the distance between no two pattern points is less than $2\epsilon D$ where $D$ is the diameter of $C(F)$. Given the input $(F, \epsilon)$, the problem requires the robots to a form a configuration $R = \{r_1, \ldots, r_n\}$  such that there exists an embedding (subject to translation, reflection, rotation and uniform scaling) of the pattern $F$ on the plane, say $P = \{p_1, \ldots, p_n\}$, such that $d(p_i,r_i) \leq \epsilon \overline{D}$ for all $i = 1, \ldots, n$, where $\overline{D}$ is the diameter of $C(P)$. In this case, we say that \emph{the configuration $R$ is $\epsilon$-close to the pattern $F$}. Recall that the number $\epsilon$ is such that the disks $B(f_i, \epsilon D)$ are disjoint. Since $P$ is similar to $F$, disks $B(p_i, \epsilon \overline{D})$ are also disjoint. The problem requires that exactly one robot is placed inside each disk. Furthermore, the movements should be collisionless.

\subsection{Symmetries and Basic Impossibilities}
   
We first present the concept of \emph{view} (defined similarly as in \cite{Cicerone17}) of a point in a pattern or a robot in a configuration. The view if a point/robot can be used to determine whether the pattern/configuration is symmetric or asymmetric. Let $R = \{r_1, \ldots, r_n\}$ be a configuration of robots or a pattern of points. A map $\varphi : R \rightarrow R$ is called an \emph{isometry} or \emph{distance preserving} if $d(\varphi(r_i), \varphi(r_j)) = d(r_i, r_j)$ for any $r_i, r_j \in R$. $R$ is said to be \emph{asymmetric} if $R$ admits only the identity isometry, and otherwise it is called \emph{symmetric}. The possible symmetries that a symmetric pattern/configuration can admit are reflections and rotations.

For any $r \in R$, its \emph{clockwise view}, denoted by $\mathcal{V}^{\circlearrowright}(r)$, is a string of $n+1$ elements from $\mathbb{R}^2$ defined as the following. For $r \neq c(R)$, consider the polar coordinates of the points/robots in the coordinate system with origin at $c(R)$, $\overrightarrow{c(R)r}$ as the reference axis and the angles measured in clockwise direction. The first element of the string $\mathcal{V}^{\circlearrowright}(r)$ is the coordinates of $r$ and next $n$ elements are the coordinates of the $n$ points/robots ordered lexicographically. For $r = c(R)$, all $n+1$ elements are taken $(0,0)$. The \emph{counterclockwise view}  $\mathcal{V}^{\circlearrowleft}(r)$ is defined analogously. Among $\mathcal{V}^{\circlearrowright}(r)$ and $\mathcal{V}^{\circlearrowleft}(r)$, the one that is lexicographically smaller is called the \emph{view} of $r$ and is denoted as $\mathcal{V}(r)$. In a configuration, each robot can compute its view as well as the views of all other robots. Hence, the following properties can be used by the robots to detect whether the configuration is symmetric or not.

\begin{property}\label{lemm: view}
\begin{enumerate}
 \item $R$ admits a reflectional symmetry if and only if there exist two points $r_i, r_j \in R$, $r_i, r_j \neq c(R)$, not necessarily distinct, such that  $\mathcal{V}^{\circlearrowright}(r_i) = \mathcal{V}^{\circlearrowleft}(r_j)$. 
 
 \item $R$ admits a rotational symmetry if and only if there exist two points $r_i, r_j \in R$, $r_i \neq r_j$, $r_i, r_j \neq c(R)$, such that  $\mathcal{V}^{\circlearrowright}(r_i) = \mathcal{V}^{\circlearrowright}(r_j)$.
\end{enumerate}
\end{property}

A problem that is closely related is the \textsc{Leader Election} problem where a unique robot from the team is to be elected as the leader. The following theorem characterizes the configurations from where \textsc{Leader Election} can be deterministically solved. 

%

\begin{theorem}\cite{Cicerone17} \label{th: leader}
 \textsc{Leader Election} is deterministically solvable, if and only if the initial configuration $R$ does not have i) rotational symmetry with no robot at $c(R)$ or ii) reflectional symmetry with respect to a line $\ell$ with no robot on $\ell$.
\end{theorem}

We call the symmetries i) and ii)  \emph{unbreakable symmetries}. If a configuration does not have such symmetries, then the robots can use the views to elect a unique leader. It follows from Property \ref{lemm: view} that if $R$ is asymmetric then the views of the robots are all different and hence the unique robot with lexicographically minimum view can be elected as the leader. If the configuration has rotational symmetry and there is a robot at $c(R)$, then that robot has the unique minimum view and can be elected as the leader. Now assume that the configuration is not asymmetric but does not have rotational symmetry. So it has reflectional symmetry with respect to a single line $\ell$. Notice that if a configuration has reflectional symmetry with respect to multiple lines then it has rotational symmetry. The robots on $\ell$ have different views as otherwise $R$ would admit another axis of symmetry. So the one among them that has minimum view can be elected as the leader.

It is known that \textsc{Exact Arbitrary Pattern Formation} is deterministically unsolvable by robots, even with \textsc{Rigid} movements, if the initial configuration has unbreakable symmetries. The same result holds for \textsc{Approximate Arbitrary Pattern Formation}.

\begin{theorem}\label{th: imp}
 \textsc{Approximate Arbitrary Pattern Formation} is deterministically unsolvable, even in $\mathcal{LUMI} + \mathcal{FSYNC}$ and with \textsc{Rigid} movements, if the initial configuration has unbreakable symmetries. 
\end{theorem}

\begin{proof}
 For any configuration of robots $R$, define $\gamma(r)$ for any $r \in R$ as $\gamma(r) = \Sigma_{r' \in R \setminus \{r\}} d(r,r')$. Let $R_0$ be an initial configuration of $n$ robots that has an unbreakable symmetry. For the sake of contradiction, assume that there is a distributed algorithm $\mathcal{A}$ that solves \textsc{Approximate Arbitrary Pattern Formation} for any input $(F, \epsilon)$ from this configuration, i.e., it forms a configuration that is $\epsilon$-close to $F$. Consider the following input pattern $F = \{f_1, f_2, \ldots, f_n\}$, where $f_1, f_2, f_3$ form an isosceles triangle with $d(f_1, f_2) = d(f_1, f_3) > d(f_2,f_3)$ and $f_4, \ldots, f_n$ are arranged on the smaller side of the triangle. If $d(f_1, f_2) = d(f_1, f_3)$ is sufficiently large compared to  $d(f_2,f_3)$ and  $\epsilon$ is sufficiently small, then for any configuration $R'$ of robots that is $\epsilon$-close to $F$, we have $\gamma(r_1) > \gamma(r)$ for all $r \in R' \setminus \{r_1\}$, where $r_1$ is the robot approximating $f_1$. This property can be used to elect $r_1$ as the leader. Hence, \textsc{Approximate Arbitrary Pattern Formation} can be used to solve \textsc{Leader Election} from the initial configuration $R_0$. This is a contradiction to Theorem \ref{th: leader}. 
\end{proof}

So a necessary condition for solvability of \textsc{Approximate Arbitrary Pattern Formation} by robots with inaccurate movements is that the initial configuration must not have unbreakable symmetries. We shall show that this condition is also sufficient in $\mathcal{OBLOT} + \mathcal{SSYNC}$ and $\mathcal{FCOM} + \mathcal{ASYNC}$.

 \subsection{Moving Through Safe Zone}\label{sec: safe zone}
 
 In this section, we present some movement strategies that will be used several times in the main algorithm. Suppose that a robot needs to move to or close to some point in the plane. If the point is far away from the robot and it attempts to reach it in one step, the error would be very large and it will miss the target by a large distance. Furthermore, it may reach a point which  makes the configuration to loose some desired property or due to the large deviation from the intended trajectory, it may collide with other robots. So the robot needs to move towards its target in multiple steps and move through a `safe' region where it does not collide with any robot and the desired properties of the configuration are preserved.

  We first discuss the following problem. Let $x_0$ and $y$ be two points in the plane so that $d(x_0,y) > l$. Suppose that a robot $r$ is initially at $x_0$ and the objective is that it has to move to a point inside ${B}(y, l)$ via a trajectory which should lie inside $Cone(x_0, B(y, l))$. A pseudocode description of the algorithm that solves the problem is presented in Algorithm \ref{algo: safe}.

  \begin{algorithm}[H]
    \setstretch{.1}
    \SetKwInOut{Input}{Input}
    \SetKwInOut{Output}{Output}
    \SetKwProg{Fn}{Function}{}{}
    \SetKwProg{Pr}{Procedure}{}{}

    \Input{A point $y$ on the plane and a distance $l$}
    

    $r \leftarrow$ myself

    \If{$d(r,y) > l$}{
    
      \uIf{$d(r,y) = l$ or $\frac{l}{d(r,y)} \geq sin(error_a(r,y))$}{Move to $y$}
      \Else{
      
	  $p \leftarrow$ point on $seg(r,y)$ so that $\frac{l}{d(r,y)} = sin(error_a(r,p))$\\
	  Move to $p$
      
      }
      
    }
    


\caption{}
    \label{algo: safe} 
\end{algorithm}

 \begin{theorem}\label{movcort}
 Algorithm \ref{algo: safe} is correct.
 \end{theorem}
 
 \begin{proof}
  Initially we have $r$  at point $x_0$ and $d(x_0,y) > l$. First assume that $\frac{l}{d(x_0,y)} \geq sin(error_a(x_0,y))$. This implies that $\angle(Cone(x_0, B(y, l))) = 2 sin^{-1}(\frac{l}{d(x_0,y)}) \geq 2 error_a(x_0,y)$, i.e., if $r$ attempts to move to $y$, its angular deviation will not take it outside $Cone(x_0, B(y, l))$. In this case, it will decide to move to $y$. So it will move to a point $z$ where $d(z,y) <$ $ \mu(x_0,y) d(x_0,y)$ $= sin(error_a(x_0,y)) d(x_0,y)$ $\leq \frac{l}{d(x_0,y)} d(x_0,y) = l$. Hence, $r$ gets inside $B(y, l)$ in one step and clearly its trajectory is inside $Cone(x_0, B(y, l))$ as required. Now consider the case where $\frac{l}{d(x_0,y)} < sin(error_a(r,y))$. Let $sin^{-1}(\frac{l}{d(x_0,y)}) = \theta_0$. It will decide to move to a point $p_1 \in seg(x_0,y)$ such that $error_a(x_0,p_1) = \theta_0$. So it will move to a point $x_1 \in B(p_1, l_1)$ where $l_1 = d(x_0,p_1)sin(\theta_0)$. We have $Cone(x_0, B(p_1, l_1)) \subset Cone(x_0, B(y, l))$. Hence $x_1 \in Cone(x_0, B(y, l))$ and the trajectory is inside $Cone(x_0, B(y, l))$.

  If $d(x_1, y) <l$, then we are done. If $d(x_1, y) = l$, then $r$ will decide to move to $y$ and will get inside $B(y, l)$ in one step. This is because it will reach a point $z$ where $d(z,y) <$ $\mu(x_1,y) l$ $< l$. So now assume that $d(x_1, y) > l$. It will again check if $\frac{l}{d(x_1,y)} \geq sin(error_a(x_1,y))$ or $\frac{l}{d(x_1,y)} < sin(error_a(x_1,y))$. In the first case, it will get inside ${B}(y, l)$ in one step and its complete trajectory is clearly inside $Cone(x_0, B(y, l))$. In the second case, it will decide to move to a point $p_2 \in seg(x_1,y)$ such that $error_a(x_1,p_2) = \theta_1$, where $\theta_1 = sin^{-1}(\frac{l}{d(x_1,y)})$. It will reach a point $x_2 \in B(p_2, l_2)$ where $l_2 = d(x_1,p_2)sin(\theta_1)$. We have $Cone(x_1, B(p_2, l_2)) \subset Cone(x_1, B(y, l))$. By Property \ref{prop: cone}, we have $Cone(x_1, B(y, l)) \subset Cone(x_0, B(y, l))$. Hence $x_1 \in Cone(x_0, B(y, l))$ and the complete trajectory is inside $Cone(x_0, B(y, l))$. 
  
  Continuing like this, $r$ moves along a polygonal path $x_0x_1x_2x_3\ldots$ which lies inside $Cone(x_0, B(y, l))$. In each step, $r$ gets closer to $y$. If $\theta_i = sin^{-1} (\frac{l}{d(x_i,y)})$, we have $\theta_0 < \theta_1 < \theta_2 < \theta_3 <\ldots$. This implies that, if it cannot attempt to move to $y$ in the $i$th step ($i > 1$), the amount of distance it intends to move towards $y$ in that step is more than that in the $(i-1)$th step. This implies that the minimum possible reduction in distance between $r$ and $y$ in the $i$th step is more than that in the $(i-1)$th step. Therefore, after finitely many steps, it will get inside $B(y, l)$.
 \end{proof}

\begin{figure}[h!]
\centering
\subcaptionbox[Short Subcaption]{
        \label{}
}
[
    0.48\textwidth 
]
{
    \fontsize{8pt}{8pt}\selectfont
    \def\svgwidth{0.48\textwidth}
    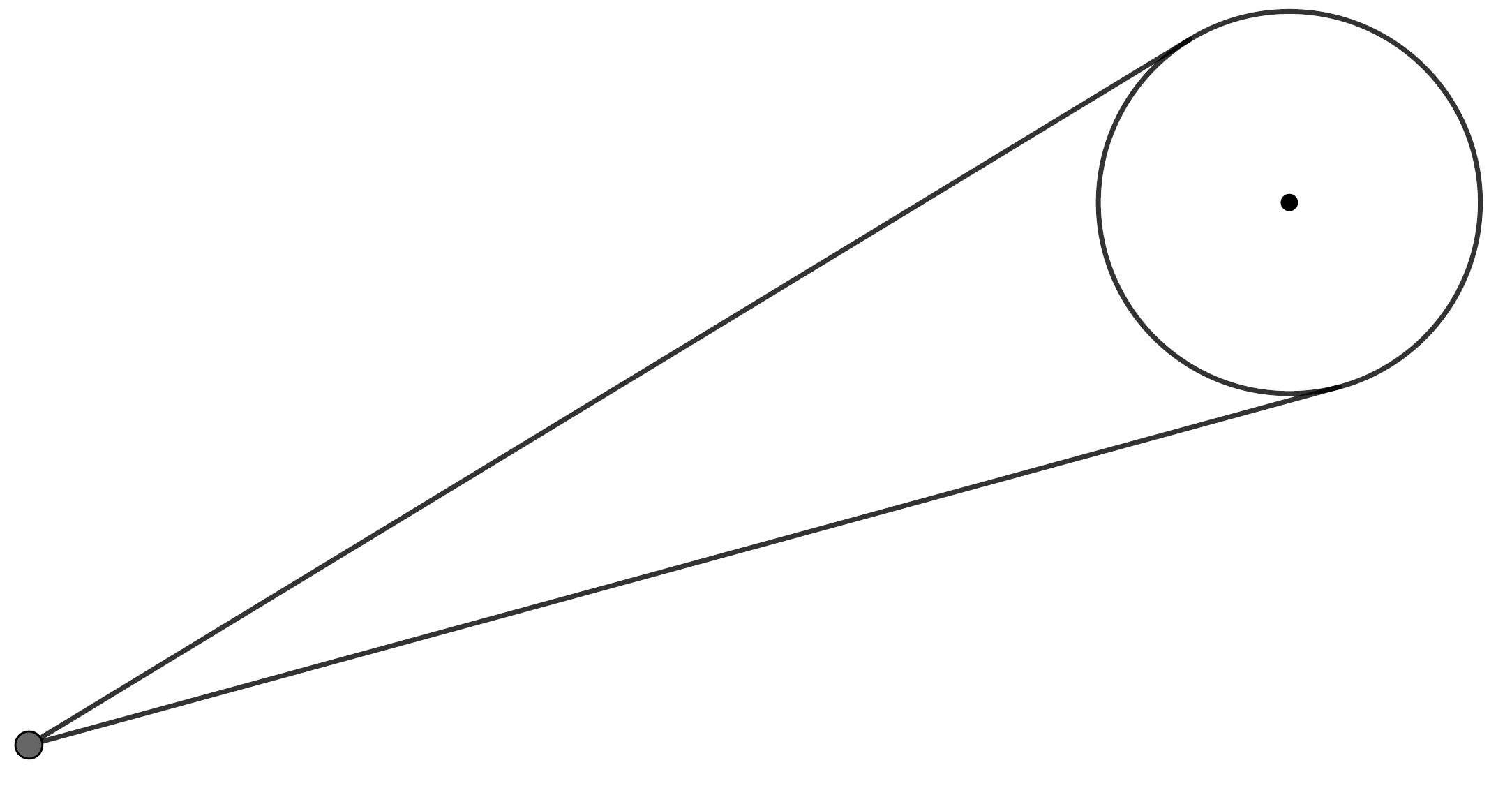
}
\hfill
\subcaptionbox[Short Subcaption]{
        \label{}
}
[
    0.48\textwidth 
]
{
    \fontsize{8pt}{8pt}\selectfont
    \def\svgwidth{0.46\textwidth}
    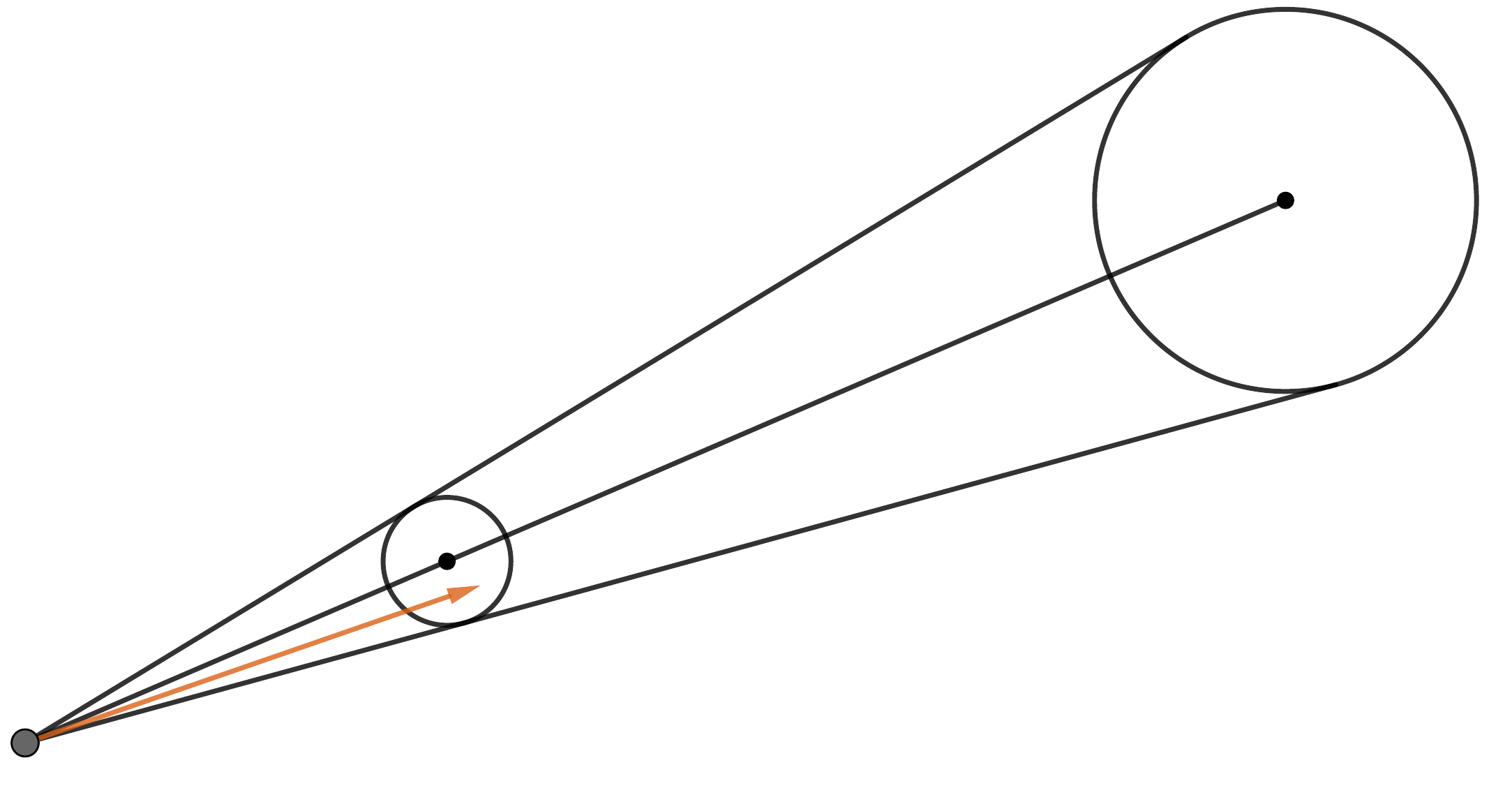
}
\\
\subcaptionbox[Short Subcaption]{
        \label{}
}
[
    0.48\textwidth 
]
{
    \fontsize{8pt}{8pt}\selectfont
    \def\svgwidth{0.48\textwidth}
    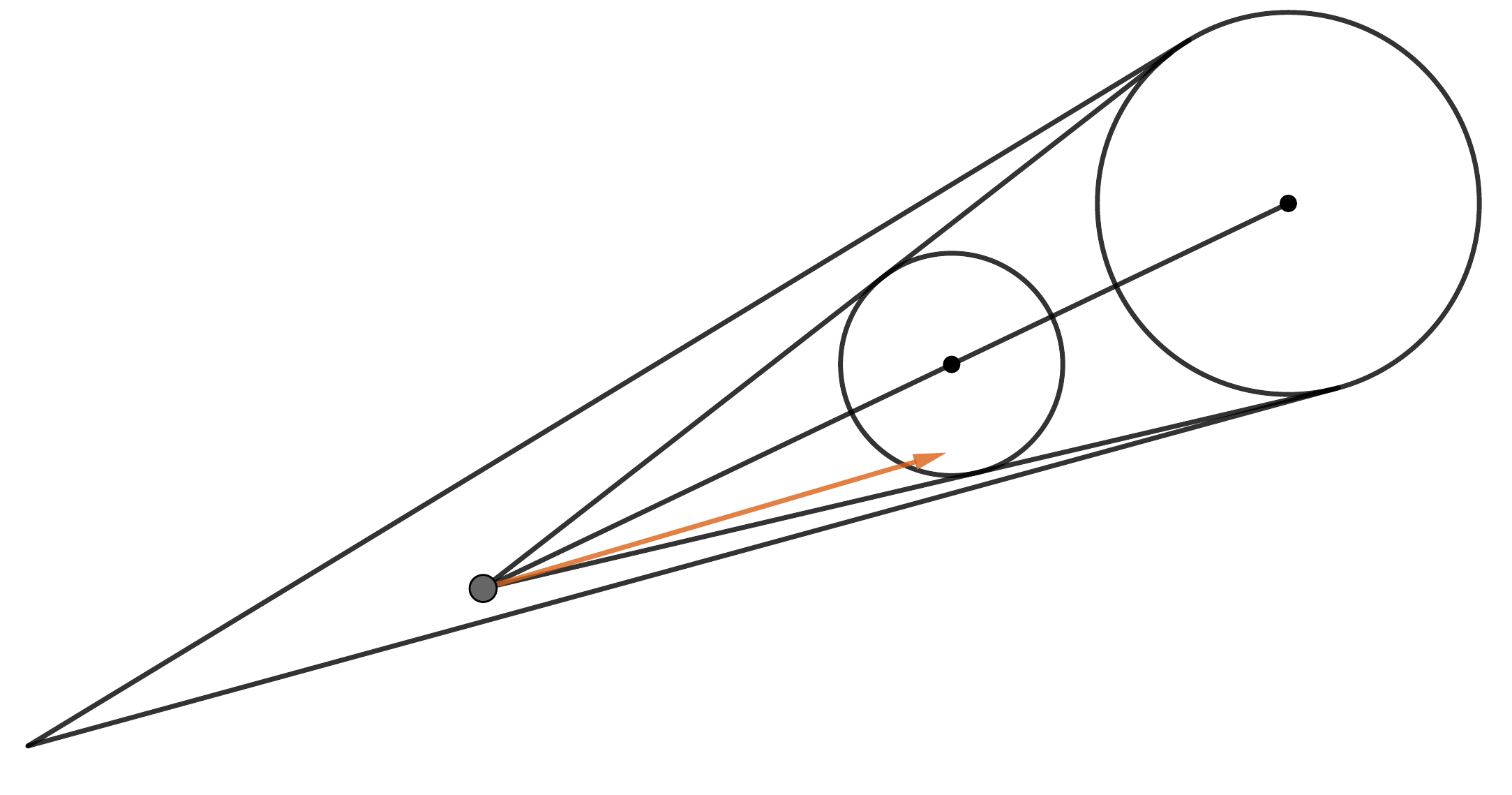
}
\hfill
\subcaptionbox[Short Subcaption]{
        \label{}
}
[
    0.48\textwidth 
]
{
    \fontsize{8pt}{8pt}\selectfont
    \def\svgwidth{0.48\textwidth}
    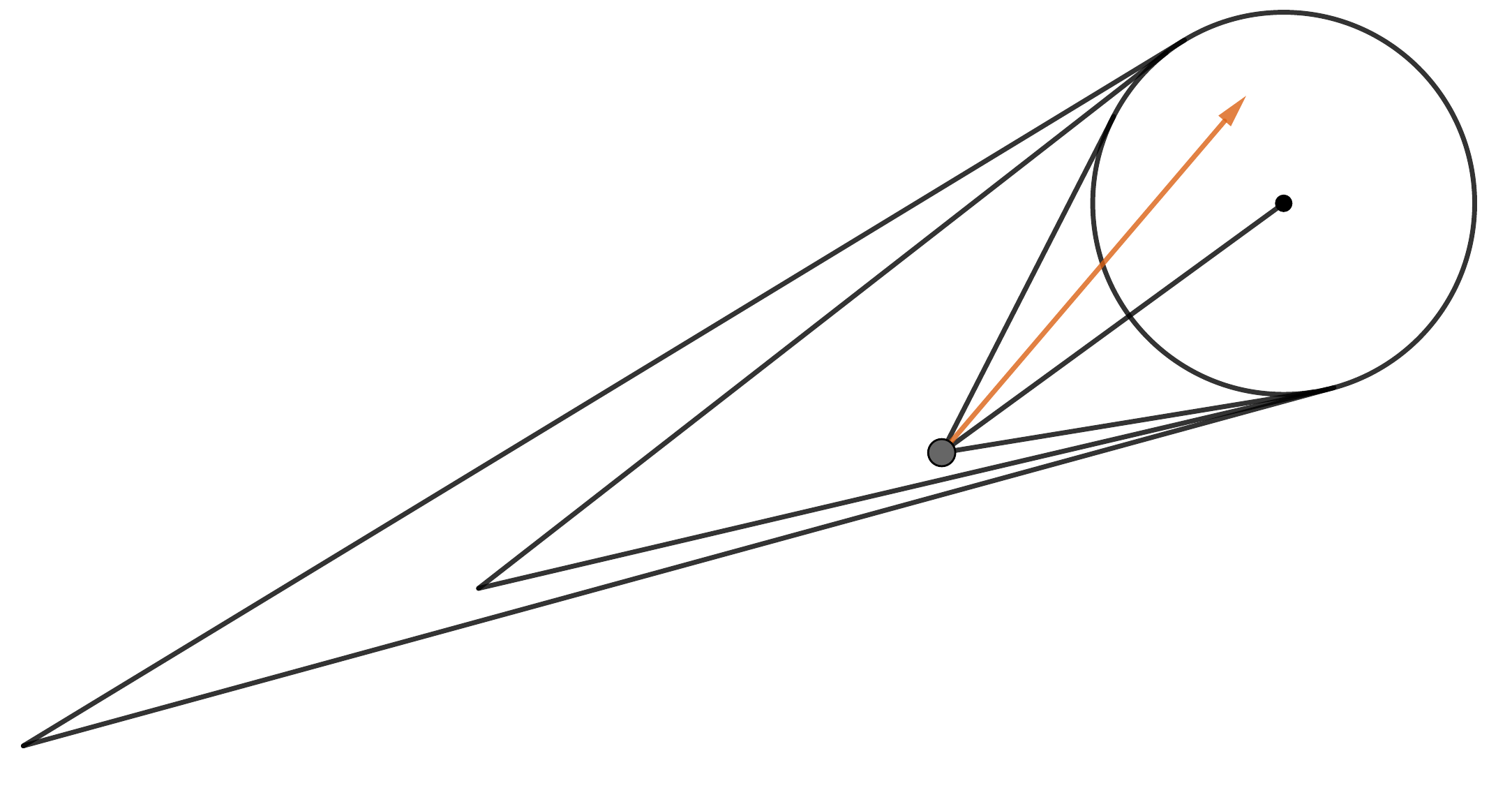
}
\caption[Short Caption]{An illustration of an execution of Algorithm \ref{algo: safe}}
\label{}
\end{figure}

 We now discuss some variants of the problem. They can be solved using the movement strategy of Algorithm \ref{algo: safe} subject to some modifications.
 
\begin{figure}[h!]
\centering
\subcaptionbox[Short Subcaption]{Starting from $x_0$, the robot has to move inside the shaded region.
        \label{fig: sz1}
}
[
    0.48\textwidth 
]
{
    \fontsize{8pt}{8pt}\selectfont
    \def\svgwidth{0.3\textwidth}
    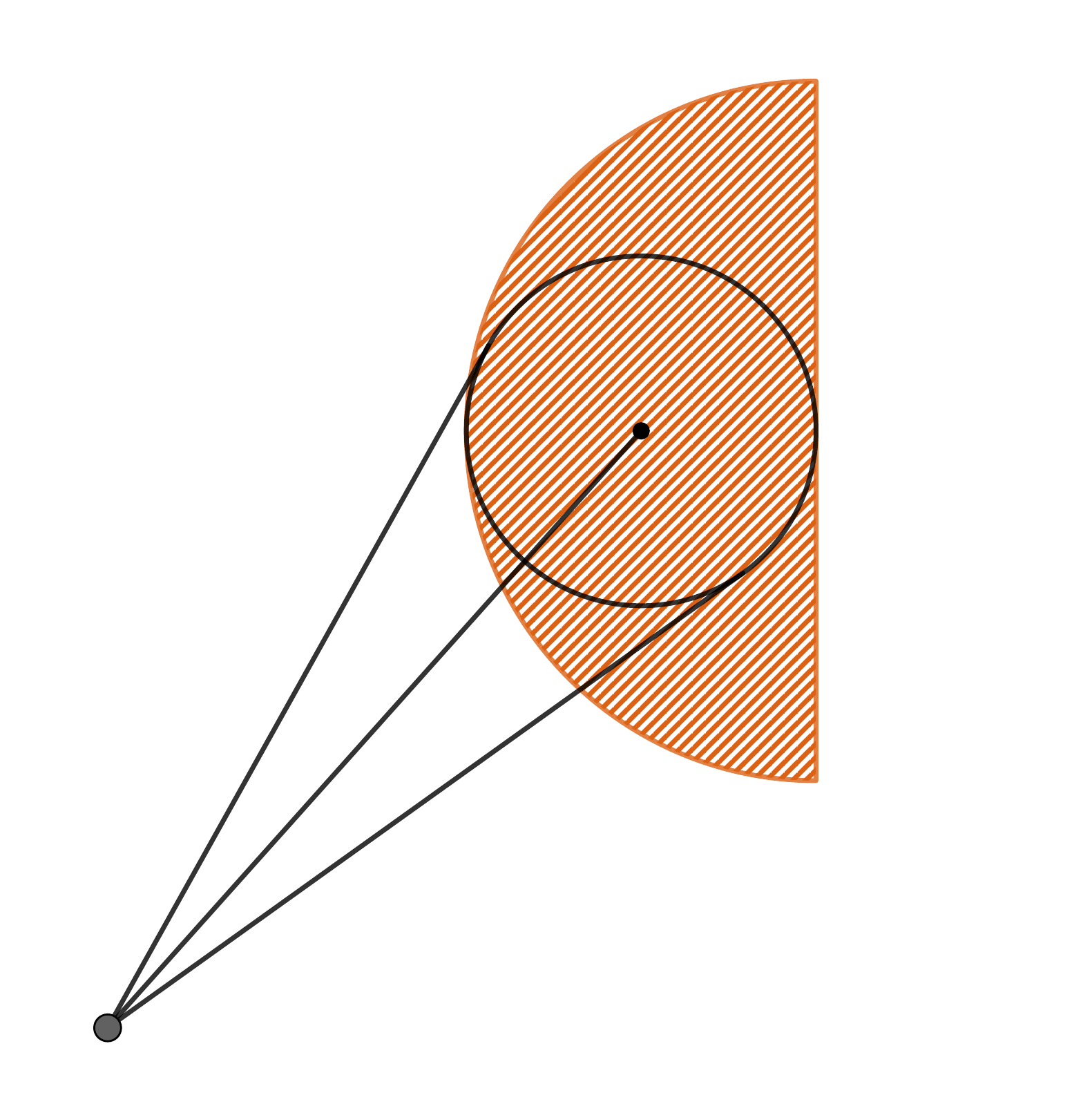
}
\hfill
\subcaptionbox[Short Subcaption]{Starting from $x_0$, the robot has to move inside $B(y,l)$ avoiding point obstacles. There are no obstacles on the line segment joining $x_0$ and $y$.
        \label{fig: sz2}
}
[
    0.48\textwidth 
]
{
    \fontsize{8pt}{8pt}\selectfont
    \def\svgwidth{0.3\textwidth}
    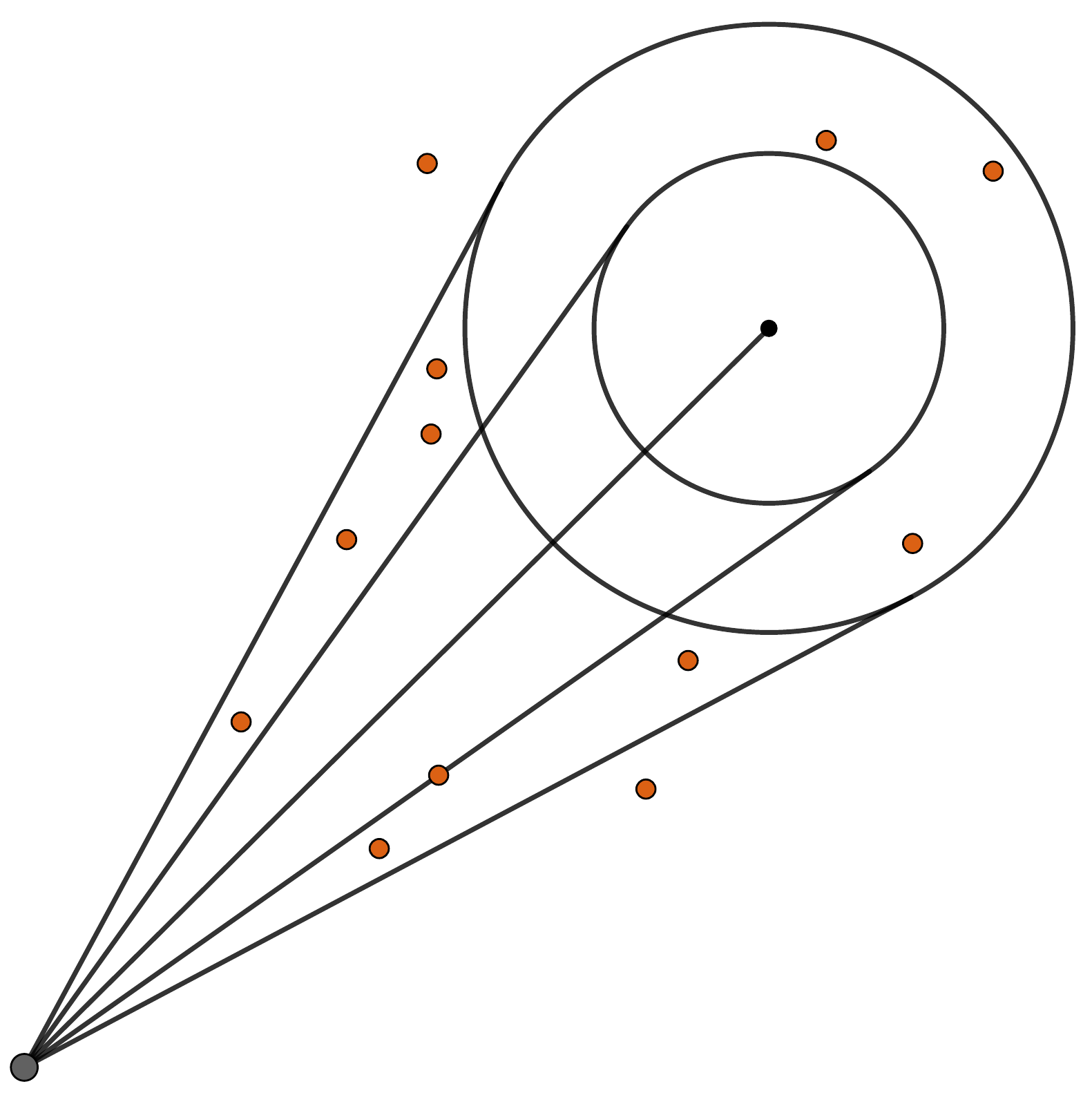
}
\\
\subcaptionbox[Short Subcaption]{Starting from $x_0$, the robot has to move inside $B(y,l)$ avoiding disk shaped obstacles. The  line segment joining $x_0$ and $y$ does not intersect any obstacle.
        \label{fig: sz3}
}
[
    0.48\textwidth 
]
{
    \fontsize{8pt}{8pt}\selectfont
    \def\svgwidth{0.3\textwidth}
    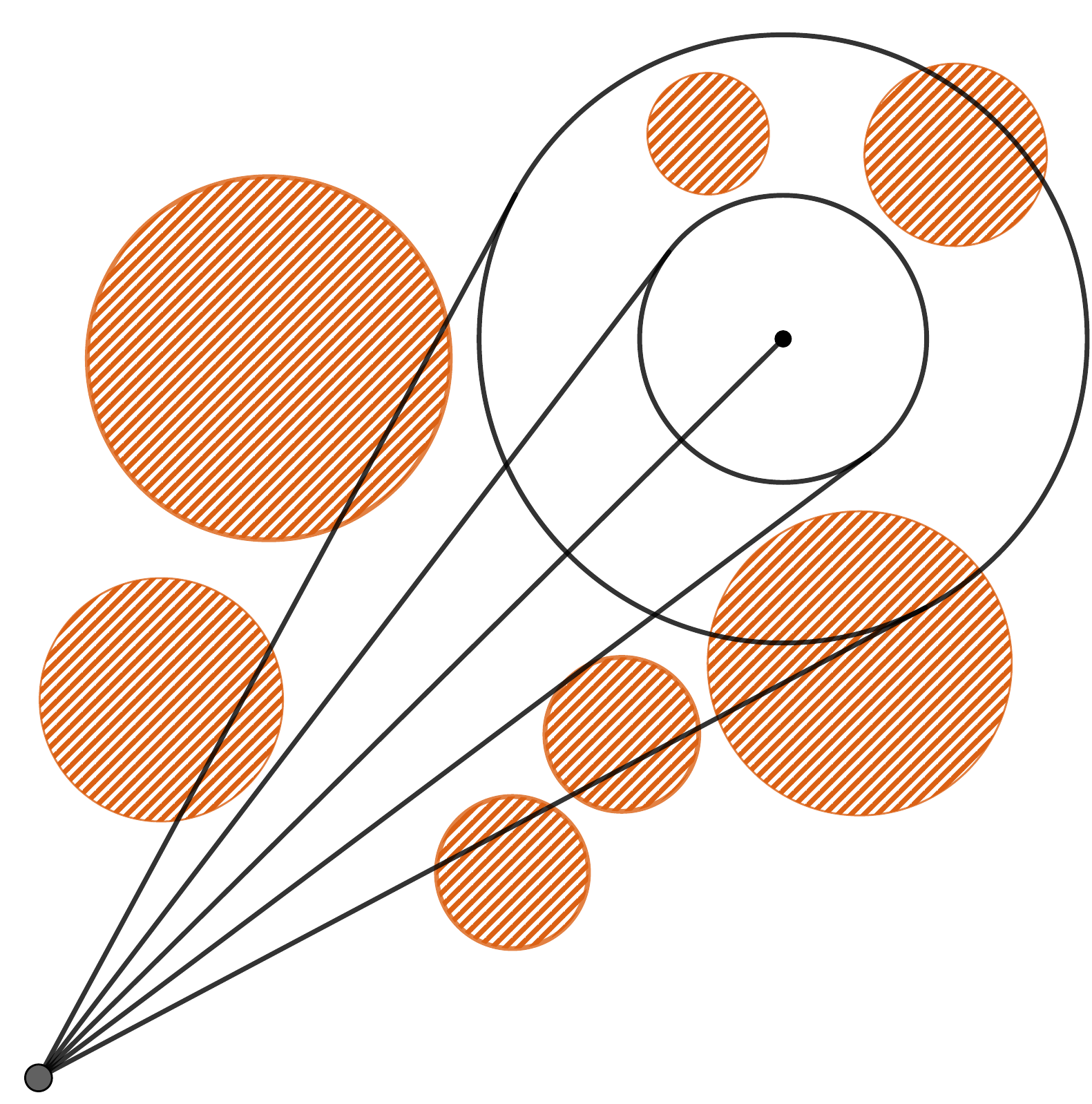
}
\hfill
\subcaptionbox[Short Subcaption]{Starting from $x_0$, the robot has to move inside $B(y,l)$ avoiding point obstacles. There are some obstacles on the line segment joining $x_0$ and $y$.
        \label{fig: sz4}
}
[
    0.48\textwidth 
]
{
    \fontsize{8pt}{8pt}\selectfont
    \def\svgwidth{0.3\textwidth}
    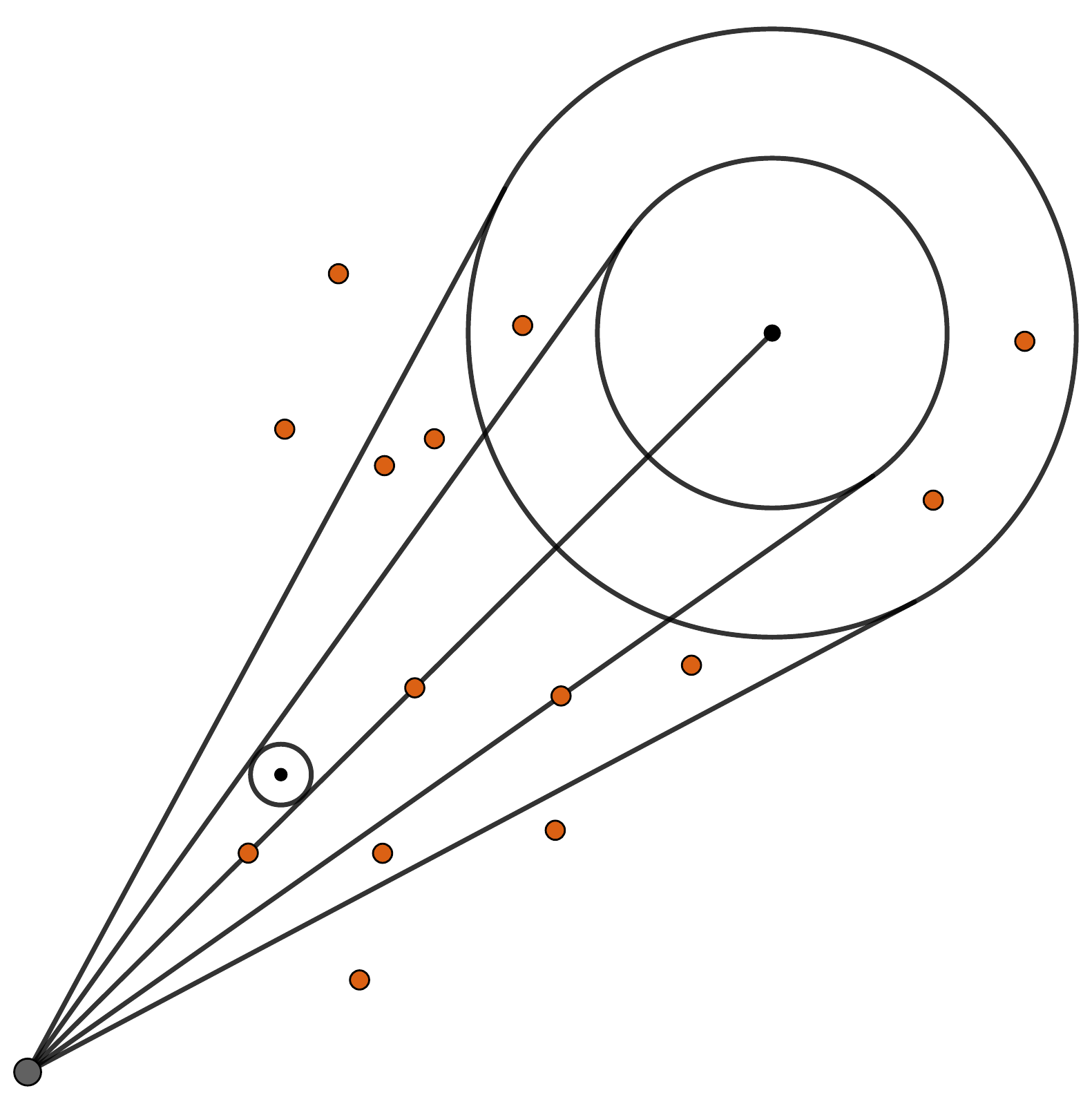
}
\caption[Short Caption]{Some variants of Algorithm \ref{algo: safe}}
\label{}
\end{figure}

 \begin{enumerate}
  \item Suppose that the robot $r$ is required to move inside some region other than a disk. Assume that the region is enclosed by some line segments and circular arcs. We can easily solve this problem using the same movement strategy, e.g., by fixing some disk $B(y, l)$ inside the region and following Algorithm \ref{algo: safe}. See Fig. \ref{fig: sz1}.
  
  \item Now consider the situation where  the robot $r$, starting from $x_0$, have to get inside a disk $B(y, l)$, but there are some point obstacles that it needs to avoid. Let $\mathcal{O} \subset \mathbb{R}^2$ be the set of obstacles. However, there are no obstacles on $\overline{seg}(x_0,y)$. Again a similar approach will work. Instead of $B(y, l)$, the robot $r$ just needs to consider $B(y, l')$ where $l' \in (0, l]$ is the largest possible length such that $Cone(r, B(y, l')) \cap \mathcal{O} = \emptyset$. See Fig. \ref{fig: sz2}.

  \item Instead of point obstacles, now consider disk shaped obstacles. Assume that none of the obstacles intersect $\overline{seg}(x_0,y)$. The same approach as in the previous problem would work here too. See Fig. \ref{fig: sz3}.
  
  \item Now again consider point obstacles, but  this time there might be some obstacles lying on $\overline{seg}(x_0,y)$. Let $\mathcal{O}' = \mathcal{O} \cap \overline{seg}(x_0,y)$. The robot will move to a point $x' \in Cone(x_0, B(y, l))$ so that there is no obstacle on $\overline{seg}(x',y)$. For this, it will move so that it reaches a point in $Cone(x_0, B(y, l')) \setminus \overline{seg}(x_0,y)$ where $l' \in (0, l]$ is the largest possible length such that $Cone(x_0, B(y, l')) \cap (\mathcal{O} \setminus \mathcal{O}') = \emptyset$. See Fig. \ref{fig: sz4}.
 \end{enumerate}

\section{The Algorithm for Semi-Synchronous Robots}\label{sec: main ssync}

In this section, we present an algorithm that solves the \textsc{Approximate Arbitrary Pattern Formation} problem in $\mathcal{OBLOT} + \mathcal{SSYNC}$ from any initial configuration that does not have any unbreakable symmetries. An important property of the algorithm is that it is \emph{sequential} in the sense that at any round at most one robot  moves. We shall show in Section \ref{sec: main async} that this property allows to design an algorithm for $\mathcal{FCOM} + \mathcal{ASYNC}$ using two colors. 



 We now present our algorithm for $\mathcal{OBLOT} + \mathcal{SSYNC}$. We assume that the initial configuration does not have any unbreakable symmetries. The algorithm works in three phases. These phases are described in detail in the next three subsections.  We introduce some Boolean predicates in Table \ref{table} that will be used to describe the structure of the phases.


\begin{table}
\begin{center}
    \begin{tabular}{ c m{1cm}  m{10cm} }
    \hline
    Variable & &  \hspace{3cm} Description \\ \hline \hline
    
    $\texttt{a}$ & & $R$ is an asymmetric configuration.   \\ \hline
    
    $\texttt{u}$ & & $R$ has an unbreakable symmetry.   \\ \hline
    
    $\texttt{c}$ & & All robots on $C(R)$ are critical.   \\ \hline
    
    $\texttt{s}$ & & $R$ is symmetry safe.   \\ \hline

    $\texttt{b}$ & & The bounding structure is formed.   \\ \hline

%
%
%
%
%
%
%
%
%
%
   
    \end{tabular}
\end{center}
\caption{Description of the variables used by the robots}\label{table}
\end{table}

\subsection{Phase 1}

\subsubsection{Motive and Overview}\label{}

The goal of Phase 1 is to create a configuration which is asymmetric and in which all robots on its minimum enclosing circle are critical. Phase 1 consists of three subphases, namely Subphase 1.1, Subphase 1.2 and Subphase 1.3. If the configuration is symmetric, our first step would be to get rid of the symmetry. Since the initial configuration cannot have any unbreakable symmetries, it is possible to choose some unique robot from the configuration. We can remove the symmetry by appropriately moving this robot. This is done in Subphase 1.1. Once we have an asymmetric configuration, the next objective is to bring inside some non-critical robots from the minimum enclosing circle so that all the remaining robots on the minimum enclosing circle are critical. However, we have to make sure that these moves do not create new symmetries in the configuration. For this, we first make the configuration symmetry safe, i.e., have unique robots $r_1$ and $r_2$ respectively closest and second closest  from the center of the minimum enclosing circle such that $r_1$ and $r_2$ are not on the same diameter. This is done Subphase 1.2. After this, in Subphase 1.3, we start bringing inside the robots from the circumference. The movements of the robots should be such that $r_1$ and $r_2$ remain the unique closest and second closest robot from the center. This ensures that these movements do not create any symmetries. The two properties that we achieved in Phase 1, namely, having an asymmetric configuration and not having any non-critical robot on the minimum enclosing circle, will play crucial role in our approach and hence, will be preserved during the rest of the algorithm. This will be the case even if the target pattern $F$ is symmetric or has non-critical robots on its minimum enclosing circle. This is not a problem as we are not required to exactly form the  pattern $F$. Any pattern $F$ can be approximated by a pattern that is asymmetric and has no non-critical points on its minimum enclosing circle. 

Now let us formally describe the structure of Phase 1. The algorithm is in Phase 1 if $\neg\texttt{u} \wedge (\neg\texttt{a} \vee \neg\texttt{c})$ holds. The objective is to create a configuration in which $\texttt{a} \wedge \texttt{c}$ holds, i.e., create an asymmetric configuration in which all robots on the minimum enclosing circle are critical. The algorithm is in Subphase 1.1 if $\neg\texttt{u} \wedge \neg\texttt{a}$ holds, in Subphase 1.2 if $\texttt{a} \wedge \neg \texttt{s} \wedge \neg \texttt{c}$ holds and in Subphase 1.3 if $\texttt{s} \wedge \neg \texttt{c}$ holds. These subphases are described in detail in Sections \ref{1.1}. \ref{1.2} and \ref{1.3} respectively.

\subsubsection{Subphase 1.1}\label{1.1}

 The algorithm is in Subphase 1.1 if $\neg\texttt{u} \wedge \neg\texttt{a}$ holds. Our objective is to create an asymmetric configuration, i.e., have $\texttt{a}$. As mentioned earlier, we will remove the symmetry by moving exactly one robot of the configuration, while all other robots will remain stationary. The fact that we have $\neg\texttt{u}$, allows us to select one such robot from the configuration. To describe the algorithm, we have to consider the following four cases. 
 
 Case 1 consists of the configurations in Subphase 1.1 where there is a robot at $c(R)$. Now consider the cases where there is no robot at $c(R)$. Notice that in this case, $R$ cannot have a rotational symmetry because $\neg \texttt{u}$ holds. So $R$  has a reflectional symmetry with respect to a unique line $\ell$. Since $\neg \texttt{u}$ holds, there are robots on $\ell$. If there is a non-critical robot on $\ell$ then we call it Case 2. For the remaining cases where there is no non-critical robot on $\ell$, we call it Case 3 if there are more than 2 robots on $C(R)$ and Case 4 if there are exactly 2 robots on $C(R)$.

\begin{figure}[h!]
\centering
\fontsize{8pt}{8pt}\selectfont
\def\svgwidth{0.35\textwidth}
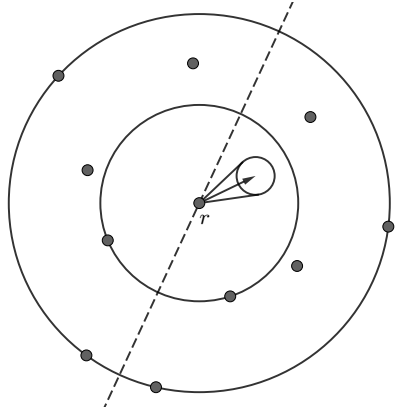
\caption{The movement in Subphase 1.1, Case 1.}
\label{fig: p1c1}
\end{figure}

\paragraph{Case 1.}

We have a robot $r$ at $x = c(R)$. In this case, $r$ will move away from the center and all other robots will remain static. The destination $y$ chosen by the robot $r$ should satisfy the following the conditions (See Fig. \ref{fig: p1c1}).

\begin{enumerate}[label=(\arabic*)]
 \item $\mathcal{Z}(x,y)  \subset encl(C^{2}_{\uparrow}(R)) \setminus \{c(R)\}$ 
 \item $\mathcal{Z}(x,y) \cap \ell = \emptyset$ for any reflection axis $\ell$ of $R \setminus \{r\}$.
\end{enumerate}

It is easy to see that such an $y$ exists. Furthermore, $r$ can easily compute such an $y$. 

%

\begin{lemma}\label{1.1.1}
 If the algorithm is in Subphase 1.1, Case 1 at some round, then after finitely many rounds we have $\texttt{a}$.
\end{lemma}

\begin{proof}
The robot $r$ is either active in that round or will be activated after some finitely many rounds. In the later case, all other robots will remain static till that round. When activated, $r$ will decide to move. The robot $r$ is initially at $x = c(R)$ and decides to move to $y$ as described. Then $r$ moves to a point $z \in \mathcal{Z}(x, y)$. We show that the resulting configuration, say $R^{new}$, is asymmetric. Notice that condition (1) implies that $r$ is not at $c(R^{new}) = c(R)$ and it is the unique robot closest to $c(R^{new})$. This implies that the configuration has no rotational symmetry. For the sake of contradiction, assume that $R^{new}$ has reflectional symmetry with respect to a line $\ell$. First assume that $r$ is on $\ell$. Then $R^{new} \setminus \{r\} = R \setminus \{r\}$ has reflectional symmetry with respect to $\ell$. This cannot happen because of condition (2). So now assume that $r$ is not on $\ell$. This implies that there is another robot $r'$ (its specular partner with respect to $\ell$) such that $d(r, c(R^{new})) = d(r', c(R^{new}))$. This contradicts the fact that it is the unique closest robot to $c(R^{new})$. Hence $R^{new}$ does not have reflectional symmetry. Hence we conclude that $R^{new}$ is asymmetric.
\end{proof}

\begin{figure}[h!]
\centering
\subcaptionbox[Short Subcaption]{
        \label{}
}
[
    0.45\textwidth 
]
{
    \fontsize{8pt}{8pt}\selectfont
    \def\svgwidth{0.35\textwidth}
    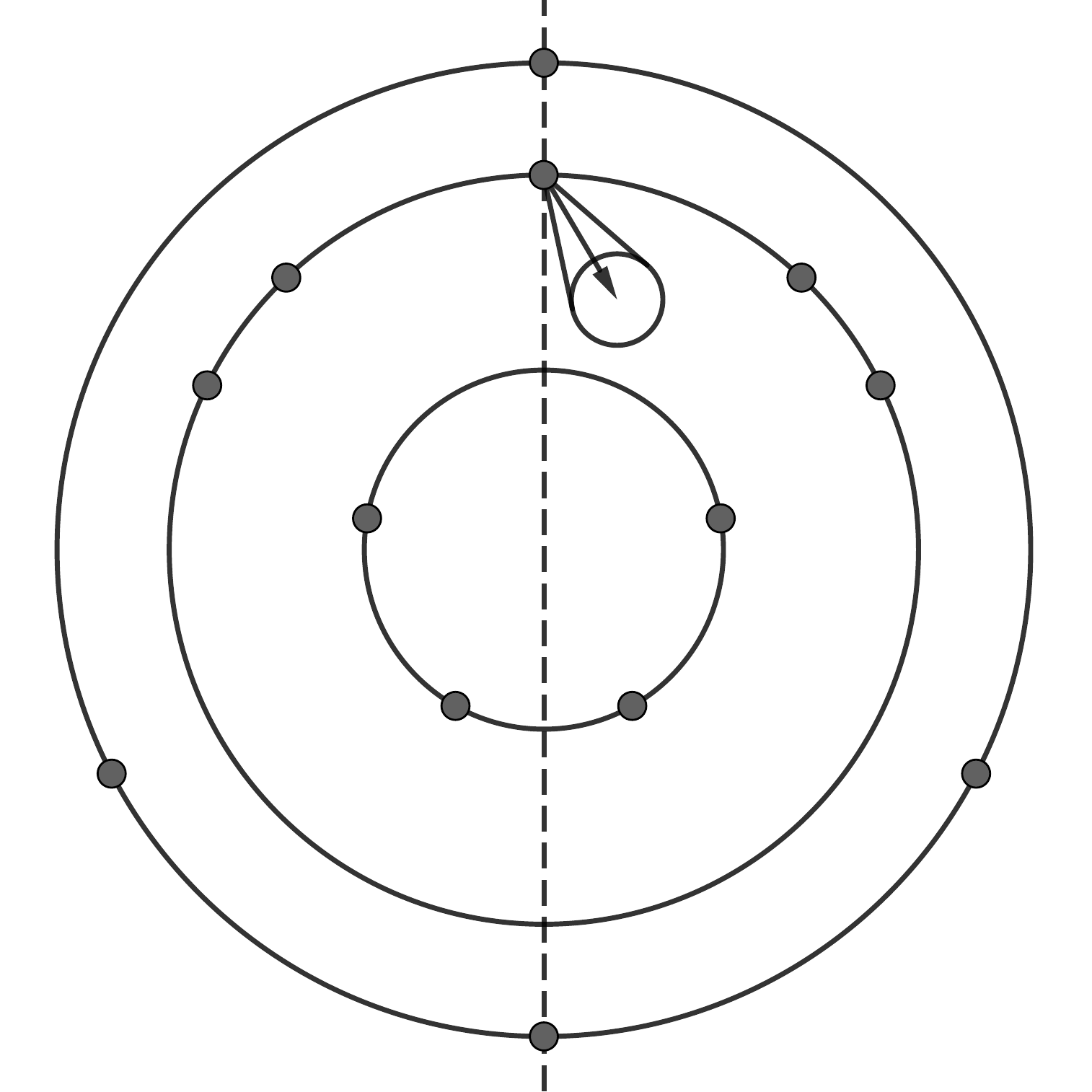
}
\hfill
\subcaptionbox[Short Subcaption]{
     \label{}
}
[
    0.45\textwidth 
]
{
    \fontsize{8pt}{8pt}\selectfont
    \def\svgwidth{0.35\textwidth}
    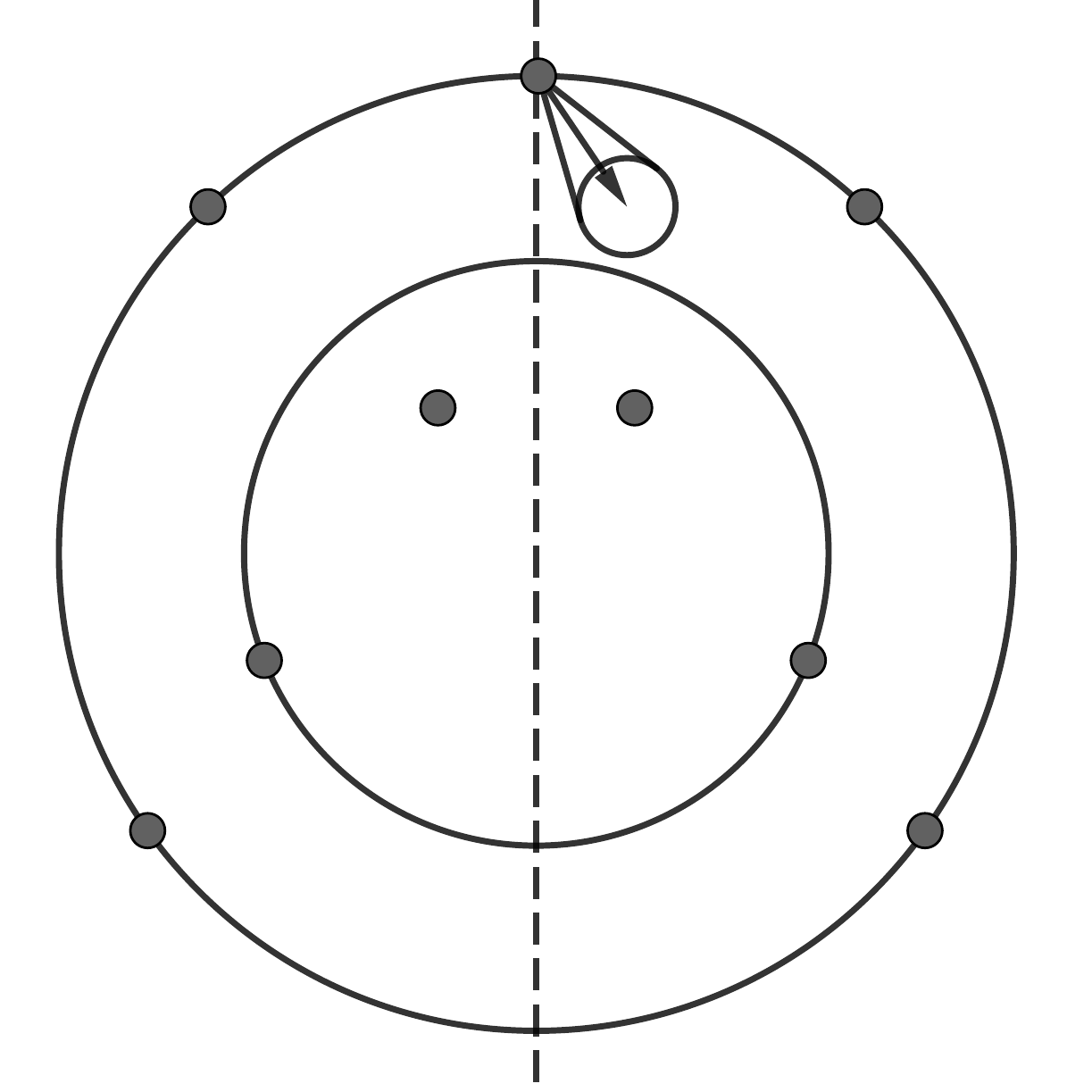
}
\caption[]{The movement in Subphase 1.1, Case 2.}
\label{fig: p1c2}
\end{figure}

\paragraph{Case 2.}
 In this case, there is no robot at $c(R)$, $R$ has reflectional symmetry with respect to a unique line $\ell$ and there is at least one non-critical robot on $\ell$. Since $\neg \texttt{u}$ holds, the views of the robots on $\ell$ are all distinct. So let $r$ be the non-critical robot on $\ell$ with minimum view. Only $r$ will move in this case. Suppose that $r$ is at point $x$. The destination $y$ chosen by $r$ should satisfy the following the conditions (See Fig. \ref{fig: p1c2}).

\begin{enumerate}[label=(\arabic*)]
 \item If $x \in C^{i}_{\uparrow}(R))$, then $Cone(x, \mathcal{Z}(x,y))  \subset encl(C^{i}_{\uparrow}(R)) \setminus \overline{encl}(C^{i-1}_{\uparrow}(R))$
 \item $\mathcal{Z}(x,y) \cap \ell = \emptyset$ for any reflection axis $\ell$ of $R \setminus \{r\}$.
\end{enumerate}

Such points clearly exist and $r$ can easily compute one.

\begin{lemma}
 If the algorithm is in Subphase 1.1, Case 2 at some round, then after finitely many rounds we have $\texttt{a}$.
\end{lemma}

\begin{proof}
As described, the robot $r$ will move to a point $z \in \mathcal{Z}(x,y)$. Let $R^{new}$ be the new configuration. Since $r$ was a non-critical robot $C(R) = C(R^{new})$. Notice that condition (1) implies that $r$ is a unique robot at a distance $d(z,c(R^{new})) > 0$ from $c(R^{new})$. This implies the configuration has no rotational symmetry. Similarly as in Lemma \ref{1.1.1} we can show that $R^{new}$ has no reflectional symmetry. Hence we conclude that $R^{new}$ is asymmetric.
\end{proof}

\paragraph{Case 3.}

We have no robot at $c(R)$, $R$ has reflectional symmetry with respect to a unique line $\ell$, there is no non-critical robot on $\ell$ and $C(R)$ has at least 3 robots on it. First we prove the following result.

\begin{lemma}\label{lemm p1c3}
In Subphase 1.1, Case 3, the following are true.

 \begin{enumerate}[label=(\arabic*)]
  \item There is no robot on $\ell \cap encl(C(R))$.
  \item There is exactly one robot on $\ell$.
  \item If $r$ is the unique robot on $\ell$, then $\frac{\pi}{2} < \text{max}\{\angle rc(R)r'' \mid r'' \in R \cap C(R)\} < \pi$.
 \end{enumerate}

\end{lemma}

\begin{proof}

1) This follows from the fact that any robot in $encl(C(R))$ is non-critical.

2) For the sake of contradiction assume that there are two robots on $\ell$, i.e., two antipodal robots on $\ell \cap C(R)$. Let these two robots be $r$ and $r'$. Let $\ell'$ be the line perpendicular to $\ell$ and passing through $c(R)$. Let $\mathcal{H}$ and $\mathcal{H}'$ be the closed half-planes delimited by $\ell'$ that contain $r$ and $r'$ respectively. By Property \ref{prop: cr2}  and the fact that $R$ has a reflectional symmetry with respect to $\ell$, $\mathcal{H} \cap C(R)$ has no robot other than $r$ because otherwise $r$ is non-critical. Similarly, $\mathcal{H}' \cap C(R)$ has no robot other than $r'$. This contradicts the fact that $C(R)$ has at least 3 robots on $C(R)$.

3) The robot $r$ is on $\ell \cap C(R)$. Let max$\{\angle rc(R)r'' \mid r'' \in R \cap C(R)\} = \theta$. We have $\theta < \pi$ because of 2). We have $\theta \neq \frac{\pi}{2}$ because then $r$ is not critical. Finally, we cannot have $\theta < \frac{\pi}{2}$ because of Property \ref{prop: cc}. 
\end{proof}

\begin{figure}[htb!]
\centering
\fontsize{8pt}{8pt}\selectfont
\def\svgwidth{0.4\textwidth}
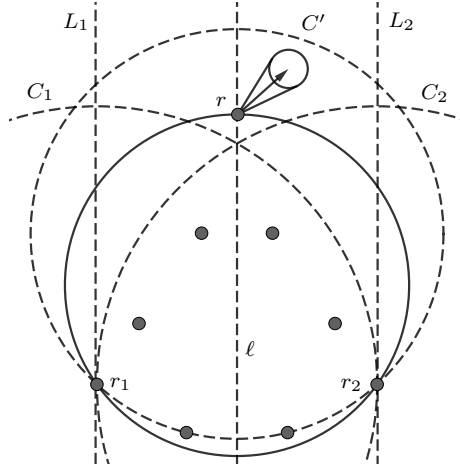
\caption[]{The movement in Subphase 1.1, Case 3.}
\label{fig: p1c3}
\end{figure}


Let $r$ be the unique robot on $\ell$. Recall that it is on $\ell \cap C(R)$. Let $x$ denote its position. Let $r_1, r_2$ be the two robots (specular with respect to $\ell$) on $C(R)$ such that  $\angle rc(R)r_1 = \angle rc(R)r_2  = max\{\angle rc(R)r'' \mid r'' \in R \cap C(R)\}$. Only $r$ will move in this case and the rest will remain static. Here the robot will move outside of the current minimum enclosing circle. The chosen destination $y$ should satisfy the following the conditions (See Fig. \ref{fig: p1c3}).

 \begin{enumerate}[label=(\arabic*)]
  \item $\mathcal{Z}(x,y)  \cap \ell = \emptyset$ 
  \item $Cone(x, \mathcal{Z}(x,y))  \subset ext(C(R)) \cap encl(C') \cap \mathcal{H}$ where $C'$ is the largest circle from $\{C \in \mathcal{F}(r_1,r_2) \mid R \subset \overline{encl}(C)\}$ and $\mathcal{H}$ is the open half-plane delimited by $line(r_1, r_2)$ that contains $x$.
  \item $\mathcal{Z}(x,y) \cap C_i = \emptyset$, where $C_i = C(r_i, d(r_1, r_2)), i=1,2$.
  \item $\mathcal{Z}(x,y) \subset \mathcal{S}(L_1,L_2)$, where $L_i$ is the line parallel to $\ell$ and passing through $r_i$, $i=1,2$.
  \end{enumerate}

 Again, it is staightforward to see that such an $y$ should exist and $r$ can easily compute one.

 \begin{lemma}\label{lemm case3}
 If the algorithm is in Subphase 1.1, Case 3 at some round, then after finitely many rounds we have $\texttt{a}$.
\end{lemma}

\begin{figure}[htb!]
\centering
\fontsize{8pt}{8pt}\selectfont
\def\svgwidth{0.3\textwidth}
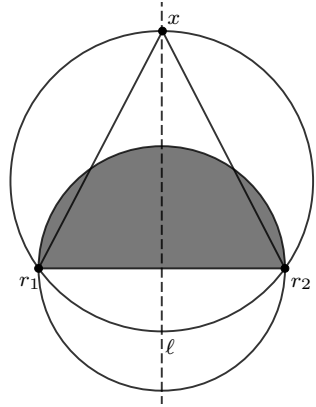
\caption[]{Illustrations supporting the proof of Lemma \ref{lemm case3}}
\label{fig: p1c3 new}
\end{figure}

\begin{proof}
 Suppose $r$ attempts to move to a point $y$ as described and reaches a point $z$.  Let $R$ be the initial configuration and $R^{new}$ be the new configuration. Let $C_1$ $=$ $CC(\{r_1, r_2,z\})$. We will first prove that  $C_1$ $=$ $C(R^{new})$. For this, we will show that (cf. Property \ref{prop: cc})  i) $R' \subset encl(C_1)$ where  $R'$ $=$  $R\setminus\{r,r_1,r_2\}$, and ii) $\Delta r_1 r_2 z$ is an acute angled triangle.
 
 Let us first prove i). Let $C'$ be as defined in the description. It follows from the definitions of $C(R)$ and $C'$ that $R' \subset \overline{encl}(C(R))$ and $R' \subset \overline{encl}(C')$. So we have $R' \subset \overline{encl}(C(R)) \cap \overline{encl}(C')$.  By condition (2), we have $C_1 \in (C(R), C')_{\mathcal{F}(r_1,r_2)}$. Therefore by  Property \ref{fam}, we have $R' \subset \overline{encl}(C(R)) \cap \overline{encl}(C') \subset encl(C_1)$. 
 
 Now for ii), observe that both $\angle zr_1r_2$ and $\angle zr_2r_1$ are $< \frac{\pi}{2}$ since $z \in \mathcal{S}(L_1,L_2)$ by condition (4). So it remains to show that $\angle r_1zr_2 < \frac{\pi}{2}$. Let $\mathcal{C}'' = CC(\{r_1,r_2\})$. By (3) of Lemma \ref{lemm p1c3}, $\angle xc(R)r_1 = \angle xc(R)r_2 > \pi/2$ $\implies \angle r_1xr_2 < \pi/2$. This implies that $c(R) \in \mathcal{H}$, where $\mathcal{H}$ be the open half-plane delimited by $line(r_1, r_2)$ that contains $x$. Then it follows from Property \ref{fam1} that $\mathcal{H} \cap \overline{encl(C'')} \subset \mathcal{H} \cap encl(C(R))$ (See Fig. \ref{fig: p1c3 new}). Notice that  $\mathcal{H} \cap \overline{encl(C'')}$ is precisely the set of all points $u \in \mathcal{H}$ such that $\angle r_1ur_2 \geq \frac{\pi}{2}$. Therefore, $\angle r_1zr_2 < \frac{\pi}{2}$ as $z \notin \mathcal{H} \cap encl(C(R))$ by condition (2).



 So we have $C_1$ $=$ $C(R^{new})$. Notice that $C_1$ has exactly three robots on it, namely $r$ (at $z$), $r_1$ and $r_2$. By condition (1), $d(r_1,z) \neq d(r_2,z)$. By condition $(3)$, $d(r_1,z) \neq d(r_1, r_2)$ and $d(r_2,z) \neq d(r_1,r_2)$. Hence  $\Delta r_1 r_2 P$ is a scalene triangle. This implies that $R^{new}$ is asymmetric.
\end{proof}

\paragraph{Case 4.}

In this case, we have no robot at $c(R)$, $R$ has reflectional symmetry with respect to a unique line $\ell$, there is no non-critical robot on $\ell$ and $C(R)$ has exactly 2 robots on it. 

\begin{lemma}
 In Subphase 1.1, Case 4 ,the following are true.
 \begin{enumerate}
  \item There is no robot on $\ell \cap encl(C(R))$.
  \item There are two antipodal robots on $\ell$.
 \end{enumerate}
\end{lemma}

\begin{proof}
1) The same as in Lemma \ref{lemm p1c3}.

2) Since $\neg \texttt{u}$ holds, $\ell$ has at least one robot $r$. By 1), it is on $C(R)$. There is exactly one robot $r'$ on $C(R)$ other than $r$ by our assumption. By Property \ref{prop: cc}, it must be diametrically opposite to $r$, i.e., on $\ell$. Hence $r$ and $r'$ are the antipodal robots on $\ell$.  
\end{proof}

Let $r$ and $r'$ are the antipodal robots on $\ell$. Since  $\neg \texttt{u}$ holds, the views of $r$ and $r'$ are different. So let $r$ be the robot with minimum view. Only $r$ will move in this case. Let $\ell'$ be the line perpendicular to $\ell$ and passing through $r$. For each $r'' \in R \setminus \{r, r'\}$, consider the line passing through $r''$ and perpendicular to $seg(r''r')$. Consider the points of intersection of these lines with $\ell'$. Let $P_1, P_2$ (specular with respect to $\ell$) be the two of these points that are closest to $\ell$. Let $L_1, L_2$ be the lines parallel to $\ell$ and passing through $P_1, P_2$  respectively. Assuming that $r$ is at point $x$, the destination $y$ chosen by $r$ should satisfy the following the conditions.


 \begin{enumerate}[label=(\arabic*)]
  \item $Cone(x,  \mathcal{Z}(x,y))  \subset ext(C(R))$ 
  \item $ \mathcal{Z}(x,y)  \cap \ell = \emptyset$ 
  \item $ \mathcal{Z}(x,y)  \subset \mathcal{S}(L_1,L_2)$ 
   \item $ \mathcal{Z}(x,y) \cap C(c, d(c,r')) = \emptyset$, where $c = c(R \setminus \{r,r'\})$.
 \end{enumerate}

\begin{figure}[h!]
\centering
\subcaptionbox[Short Subcaption]{
        \label{}
}
[
    0.8\textwidth 
]
{
    \fontsize{8pt}{8pt}\selectfont
    \def\svgwidth{0.43\textwidth}
    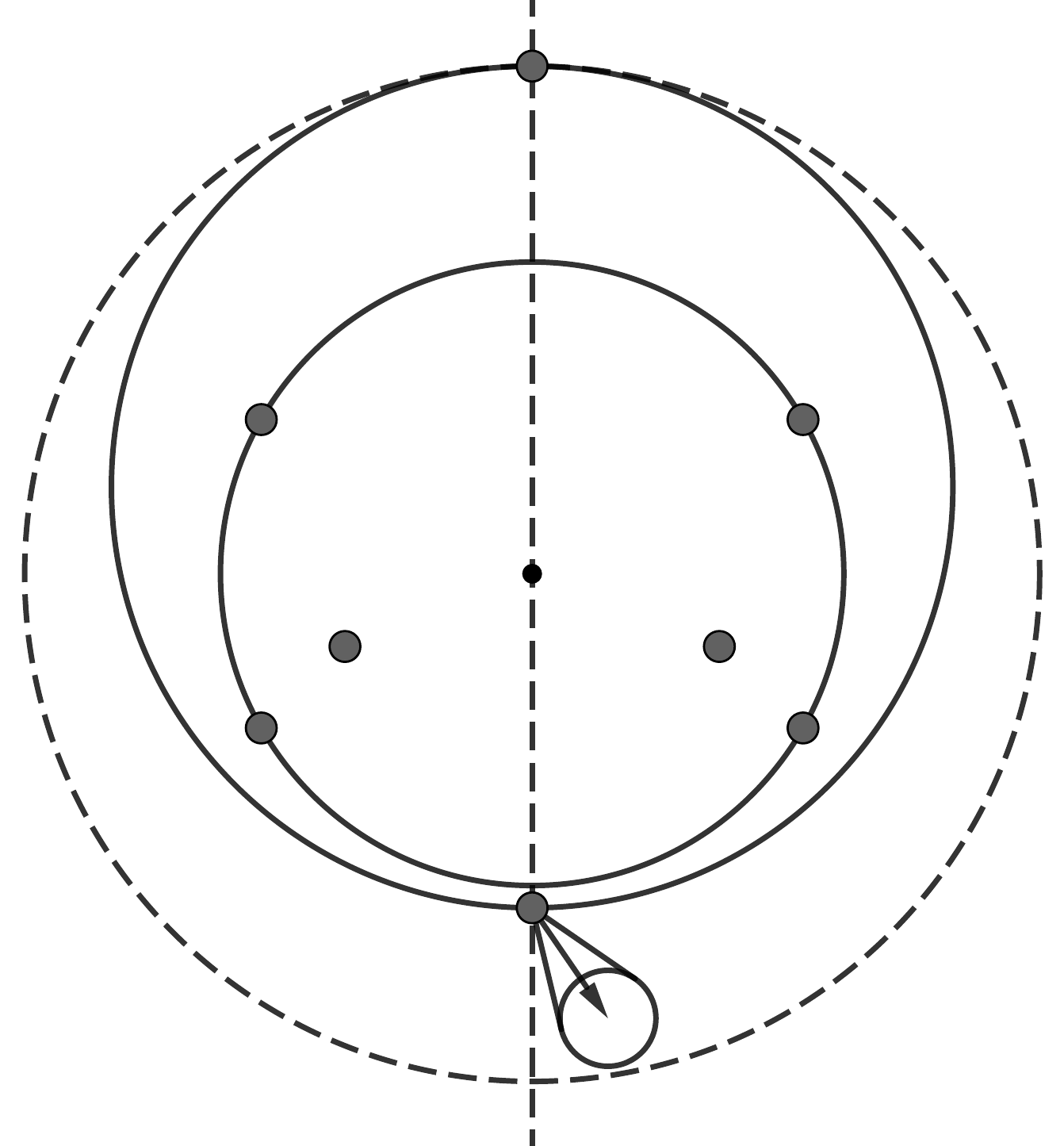
}
\\
\subcaptionbox[Short Subcaption]{
     \label{}
}
[
    0.8\textwidth 
]
{
    \fontsize{8pt}{8pt}\selectfont
    \def\svgwidth{0.8\textwidth}
    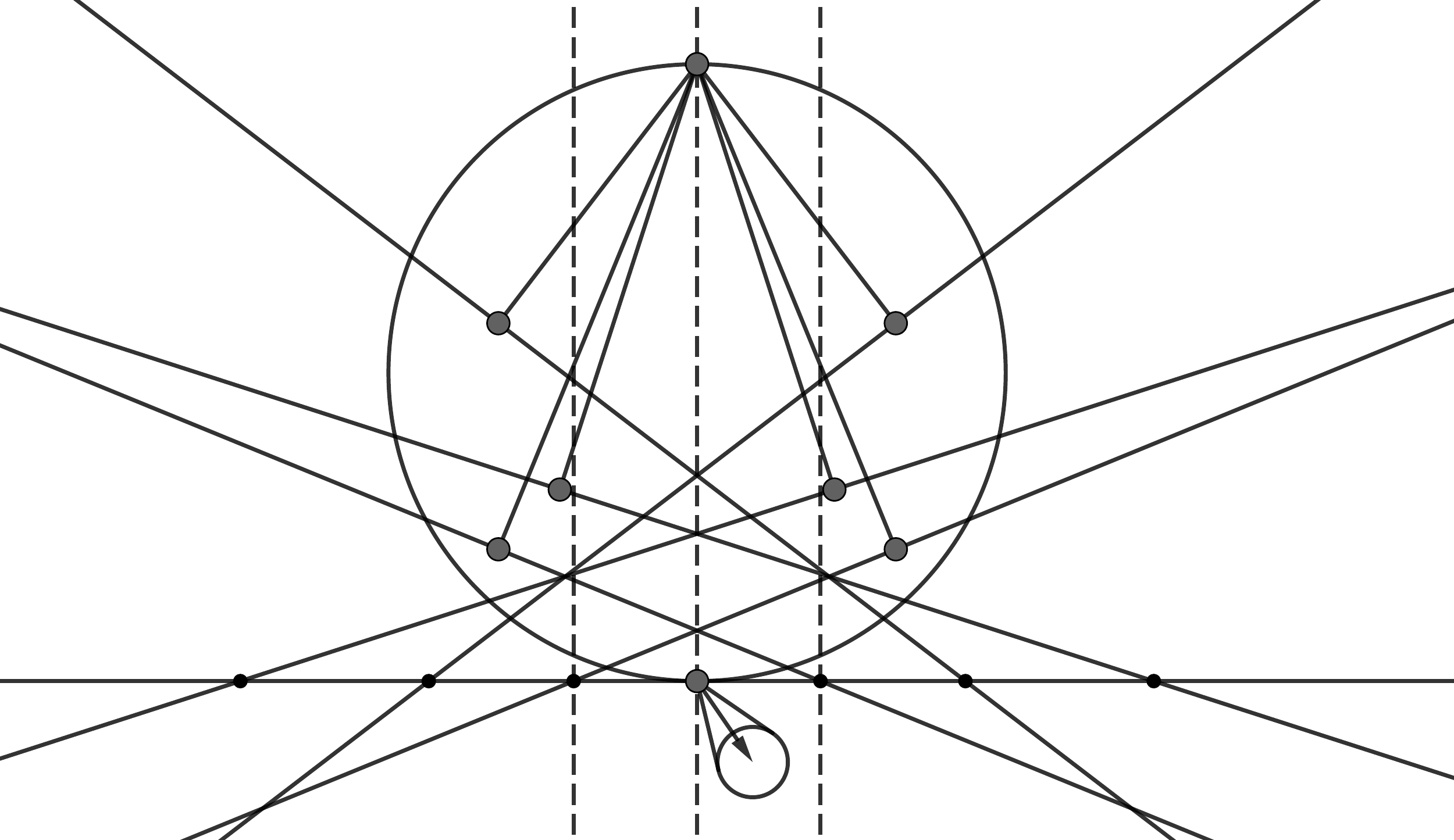
}
\caption[Short Caption]{The movement in Subphase 1.1, Case 4.}
\label{}
\end{figure}

%


\begin{lemma}\label{proof: l4}
 
 If the algorithm is in Subphase 1.1, Case 4 at some round, then after finitely many rounds we have $\texttt{a}$.

\end{lemma}

\begin{proof}

Let $r, r'$ be the antipodal robots on $\ell$ and $r$ be the robot with minimum view among them. Let $R' = R \setminus \{r,r'\}$. Let $A$ and $B$ denote the positions of $r$ and $r'$ respectively. Let $\ell'$ and $\ell''$ be the lines perpendicular to $\ell$ and passing through $A$ and $B$ respectively. Let $P_1, P_2$ (specular with respect to $\ell$) be the two points on $\ell'$ as defined in the description of Case 4. Suppose that $r$ decides to move to a point  satisfying (1)-(4) and reaches the point $P$. Let $\mathcal{H}_1$ and $\mathcal{H}_2$ be the open half planes delimited by $\ell$ that contain $P_1$ and $P_2$ respectively. Without loss of generality, assume that $P \in \mathcal{H}_1$. The line parallel to $\ell$ and passing through $P$ intersects $\ell'$ and $\ell''$  at $C$ and $D$ respectively. Let $C_1 = C(R) = CC(\{A, B\})$ (the red circle in Fig. \ref{fig: case 4 proof}), $C_2 = CC(\{A, B, C\}) = CC(\{B, C, D\}) = CC(\{B, C\})$ (the green circle in Fig. \ref{fig: case 4 proof}), $C_3 = CC(\{B, P\}) = CC(\{B, P, D\})$ (the orange circle in Fig. \ref{fig: case 4 proof}) and $C_4 = CC(\{A, B, P_1\})$ (the gray circle in Fig. \ref{fig: case 4 proof}). So, $C_1 \in \mathcal{F}(A, B)$, $C_2 \in \mathcal{F}(A, B)$, $C_4 \in \mathcal{F}(A, B)$, $C_2 \in \mathcal{F}(B, D)$ and $C_3 \in \mathcal{F}(B, D)$. We have $C_1 = C(R)$ before the move by $r$.  Let us denote the configuration after the move by $R^{new}$. We shall prove that  $C_3 = C(R^{new})$. To show this, we only need to prove that $R' \subset encl(C_3)$ (cf. Property \ref{prop: cc}).

  \begin{figure}[h!]
\centering
\fontsize{8pt}{8pt}\selectfont
\def\svgwidth{0.5\textwidth}
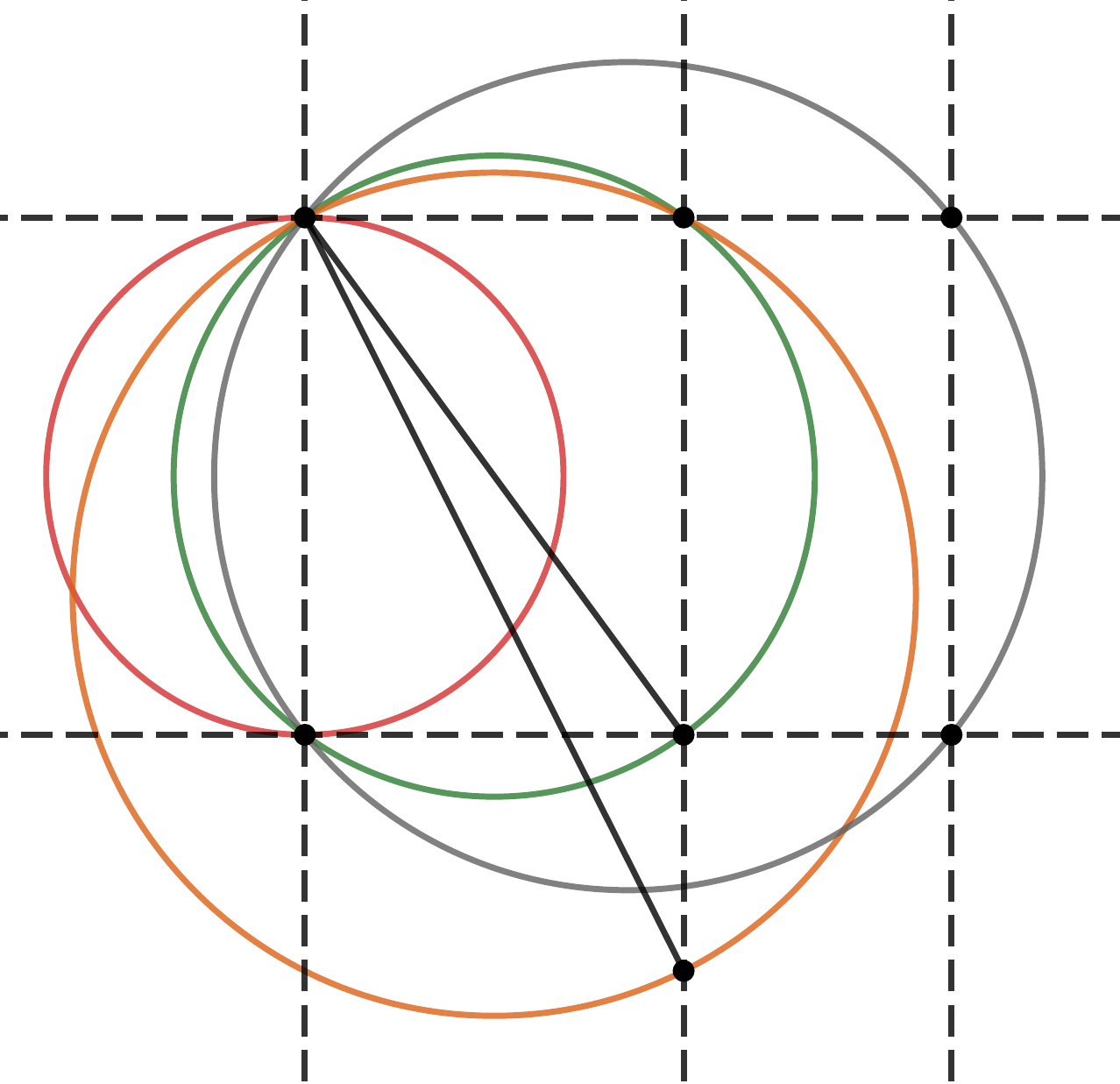
\caption{Illustrations supporting the proof of Lemma \ref{proof: l4}}
\label{fig: case 4 proof}
\end{figure}

 We claim that $R' \subset encl(C')$ for any $C' \in (C_1, C_4)_{\mathcal{F}(A, B)}$. To see this take any $C' \in (C_1, C_4)_{\mathcal{F}(A, B)}$ (the black circle in Fig. \ref{fig: case 4 proof1}). Assume that it intersects $\ell'$ at $P_3$. Now take any $r_1 \in {R}' \cap \mathcal{H}_2$. Suppose that the line passing through $r_1$ and perpendicular to $line(B, r_1)$, intersects $\ell'$ at $P_0$. We have $d(A, P_3) < d(A, P_1) < d(A, P_0)$, where the first inequality follows from definition of $C'$ and the second inequality follows from definition of $P_1$. So we have $\angle P_3r_1B > \angle P_1r_1B > \angle P_0r_1B = \pi/2$. Since $B$ and $P_3$ are diametrically opposite on $C'$ and $\angle P_3r_1B > \pi/2$, we conclude that that $r_1 \in encl(C')$. This proves that ${R}' \cap \mathcal{H}_2 \subset encl(C')$. 
 Also, ${R}' \cap \mathcal{H}_2^c \subset encl(C')$ because ${R}' \cap \mathcal{H}_2^c \subset \overline{encl(C_1)} \cap \mathcal{H}_2^c \subset \overline{encl(C')} \cap \mathcal{H}_2^c$, where the first containment is true because $C(R)=C_1$ and the second containment follows from Property \ref{fam1}. So we have $R' \cap \mathcal{H}_2 \subset encl(C)$.

 Now, as $C_2 \in (C_1, C_4)_{{F}(A, B)}$, we have  $\mathcal{R}' \subset encl(C_2)$. Now let $\mathcal{H}_3$ be the open half plane delimited by $\ell''$ that contains $\mathcal{R}'$. So $\mathcal{R}' \subset encl(C_2) \cap \mathcal{H}_3$. But it follows from Property \ref{fam1} that $encl(C_2) \cap \mathcal{H}_3 \subset encl(C_3)$. Hence $\mathcal{R}' \subset encl(C_3)$, as required.

 \begin{figure}[h!]
\centering
\fontsize{8pt}{8pt}\selectfont
\def\svgwidth{0.6\textwidth}
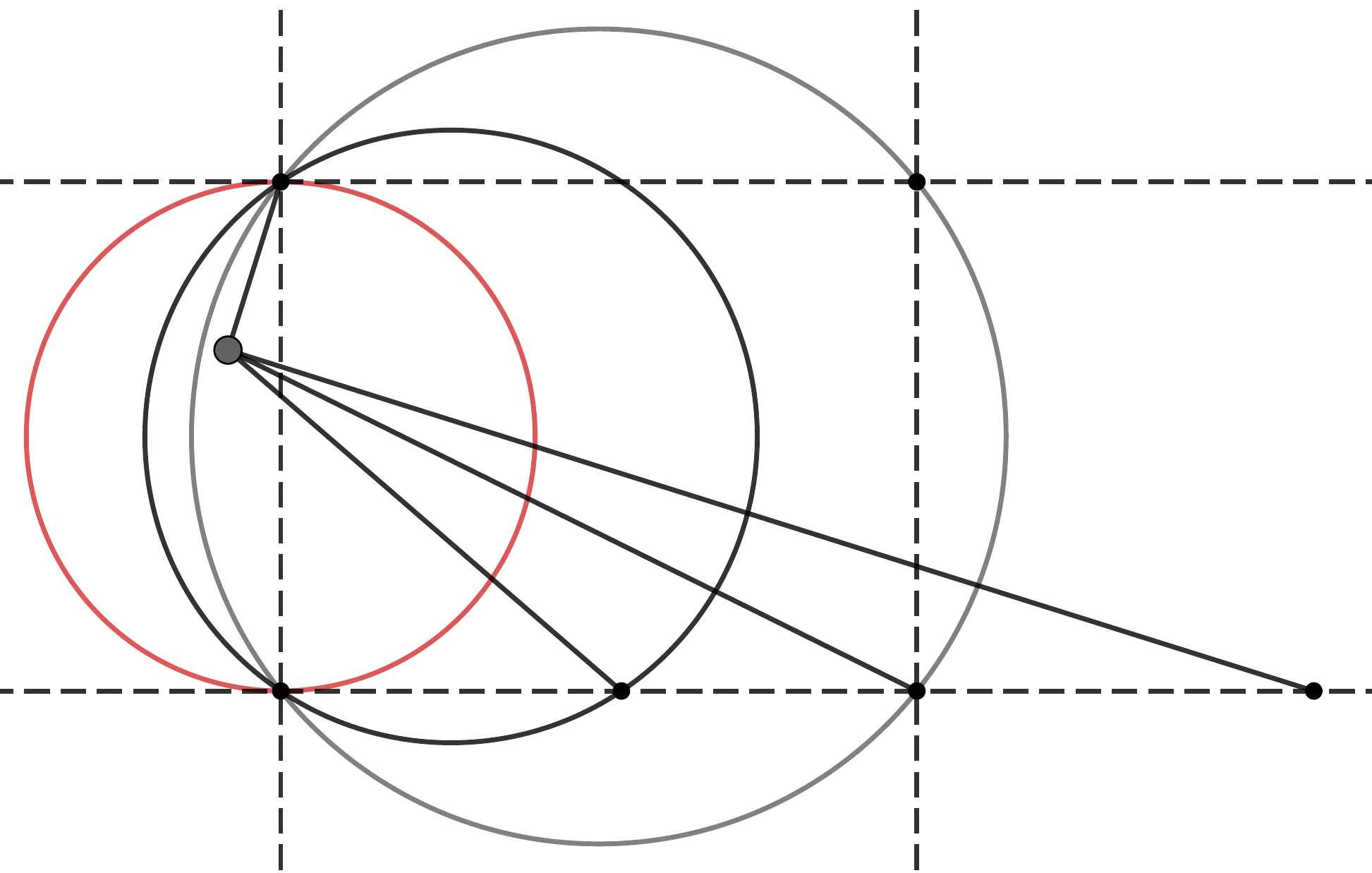
\caption{Illustrations supporting the proof of Lemma \ref{proof: l4}}
\label{fig: case 4 proof1}
\end{figure}

We have shown that $C_3$ is the minimum enclosing circle of the new configuration $R^{new}$. Furthermore, $C_3$ has exactly two robots on it, i.e., $r$ at $P$, and $r'$ at $B$. We have to show that the new configuration is asymmetric. Let $\mathcal{L}_1 = line(P, B)$ and $\mathcal{L}_2$ the perpendicular bisector of $\overline{seg}(B,P)$. If $R^{new}$ has rotational symmetry, then $R'$ also has rotational symmetry and the center of their minimum enclosing circles must coincide. But $c(R') \in seg(A, B)$ (because $R'$ has reflectional symmetry with respect to $\ell = line(A,B)$) and $c(R^{new}) \in seg(P, B)$. This is a contradiction as $seg(A, B) \cap seg(P, B) = \emptyset$. If the $R^{new}$ has reflectional symmetry, then the axis of symmetry is either $\mathcal{L}_1$ or $\mathcal{L}_2$. If  $R^{new}$ has reflectional symmetry with respect to $\mathcal{L}_1$, then $R'$ also has reflectional symmetry with respect to $\mathcal{L}_1$. This implies that $c(R') \in \mathcal{L}_1$, to be more precise, $c(R') \in seg(P, B)$. This is a contradiction as $c(R') \in seg(A, B)$ and $seg(A, B) \cap seg(P, B) = \emptyset$. So now assume that $R^{new}$ has reflectional symmetry with respect to $\mathcal{L}_2$. So $R'$ also has reflectional symmetry with respect to $\mathcal{L}_2$. Hence  $c(R') \in \mathcal{L}_2$. Since any point on $\mathcal{L}_2$ is equidistant from $B$ and $P$, we have $d(c(R'),B) = d(c(R'),P)$. This means that $P$ ( = the new position of $r$) is on the circle with center at $c(R')$ and passing through $B$ ( = the position of $r'$), i.e., $P \in C(c(R'), d(c(R'), r'))$. But this contradicts condition (4). Therefore, we conclude that $R^{new}$ is asymmetric.
\end{proof}

\subsection{Subphase 1.2}\label{1.2}
Let us suppose that, $\texttt{a} \wedge \neg \texttt{s} \wedge \neg \texttt{c}$ holds. Our goal  is to make the configuration symmetry safe. Since $\neg \texttt{c}$ holds, according to our definition,  the configuration will be symmetry safe if there is no robot at $c(R)$, there is a unique closest and a unique second closest robot to $c(R)$ and they are not collinear  with $c(R)$. Now, if there is a robot at $c(R)$ it will move in the same way as in Subphase 1.1, Case 1. Now consider the  case, where there is no robot at $c(R)$. If there is no  unique robot closest to the center, then the non-critical robot on $C^{1}_{\uparrow}(R)$ with the minimum view, say at point $x$, will decide to move to a point $y$ such that $y$ should satisfy the following conditions:

\begin{enumerate}[label=(\arabic*)]
 \item $\mathcal{Z}(x, y)  \subset encl(C^{1}_{\uparrow}(R)) \setminus \{c(R)\}$ 
 \item $\mathcal{Z}(x, y) \cap \ell = \emptyset$ for any reflection axis $\ell$ of $R \setminus \{r\}$.
\end{enumerate}

 When there is a unique robot $r'$ closest to the center, if there is only one robot  on $C^{2}_{\uparrow}(R)$ and not on $line(c(R), r')$, then we are done. Otherwise, among all the non-critical robots lying on $C^{2}_{\uparrow}(R)$, the one with the minimum view say at point $x$, moves to a point $y$ such that the following holds:
 
 \begin{enumerate}[label=(\arabic*)]
 \item $Cone(x,\mathcal{Z}(x, y))  \subset encl(C^{2}_{\uparrow}(R)) \setminus \overline{encl(C^{1}_{\uparrow}(R)}\}$ 
 \item $\mathcal{Z}(x, y) \cap line(c(R), r') = \emptyset$.
\end{enumerate}

\begin{lemma}
 
 If the algorithm is in Subphase 1.2, then after finitely many rounds we have $\texttt{s}$.

\end{lemma}

\subsection{Subphase 1.3}\label{1.3}

Let us suppose that, $\texttt{s} \wedge \neg \texttt{c}$ holds. Our objective   is to make the configuration $\texttt{a} \wedge  \texttt{c}$.  As the configuration  is asymmetric (since $\texttt{s} \implies \texttt{a}$), there is a robot $r_1$ with minimum view among all the non-critical robots lying on $C(R)$. Suppose that it is at $x$. Then $r_1$ will decide move to a point $y$ such that 
\begin{enumerate}
    \item $Cone(x, \mathcal{Z}(x, y)) \cap R = \emptyset$
    \item $Cone(x, \mathcal{Z}(x, y)) \subset ext(C^{2}_{\uparrow}(R)) \cap encl(C(R))$.
\end{enumerate} 
Clearly after each non-critical robot on $C(R)$ moves inside,  the initial property that there is a unique closest and unique second closest robot to the center and they are not collinear with the center, is retained. Hence the configuration remains asymmetric. Therefore all the non-critical robots on $C(R)$ will sequentially move inside and we will have  $\texttt{a} \wedge  \texttt{c}$. It should be noted here that after the final step when $\texttt{c}$ is achieved $\texttt{s}$ may not hold. This is because the definition of a symmetry safe configuration is different when $\texttt{c}$ is true. However in the intermediate steps, when $\texttt{c}$ was not true, $\texttt{s}$ was obviously true, i.e., in all intermediate steps we had $\texttt{s} \wedge \neg \texttt{c}$.


\begin{lemma}
 
 If the algorithm is in Subphase 1.3, then after finitely many rounds we have $\texttt{a} \wedge  \texttt{c}$.

\end{lemma}

\subsection{Phase 2}\label{sec: p2}

\subsubsection{Motive and Overview}

Phase 1 was a preprocessing step where a configuration was prepared in which there is no symmetry and all robots on the minimum enclosing circle are critical. Actual formation of the pattern will be done in two steps, in Phase 2 and Phase 3. In Phase 2, the robots on the minimum enclosing circle will reposition themselves according to the target pattern and then in Phase 3, the robots inside the minimum enclosing circle will move to complete the pattern. The standard approach to solve the \textsc{Arbitrary Pattern Formation} problem, however, is exactly the opposite. Usually, the part of the pattern inside the minimum enclosing circle is first formed and then the pattern points on the minimum enclosing circle are occupied by robots. In this approach, the minimum enclosing circle is kept invariant throughout the algorithm. Keeping the minimum enclosing circle fixed is important because it helps to fix the coordinate system with respect to which the pattern is formed. During the second step, a robot on the minimum enclosing circle may have to move to another point on the circle. In order to keep the minimum enclosing circle unchanged, it has to move exactly along the circumference.  However, it is not possible to execute such movement in our model. An error in movement in this step  will change the minimum enclosing circle and the progress made by the algorithm will be lost. Placing the robots at the correct positions on the minimum enclosing circle is a difficult issue in our model. In fact, it can be proved that it is impossible to deterministically obtain a configuration with $\geq 4$ robots on the minimum enclosing circle if the initial configuration does not have so. For this reason, we shall work with 2 or 3 (critical) robots on the minimum enclosing circle as obtained from Phase 1 (or may be from the beginning). So in Phase 2, we start with an asymmetric configuration where all robots on the minimum enclosing circle are critical. The objective of this phase is to move these critical robots so that their relative positions on the minimum enclosing circle is consistent with the target pattern. For this, we shall choose a set of two or three pattern points from the minimum enclosing circle of the target pattern. We shall call this set the bounding structure of the target pattern (defined formally in Section \ref{sec: bound}). Essentially, the objective of Phase 2 is to approximate this structure by the critical robots.

\subsubsection{The Bounding Structure}\label{sec: bound}

\begin{figure}[h!]
\centering
\subcaptionbox[Short Subcaption]{
        \label{fig: bs1}
}
[
    0.48\textwidth 
]
{
    \fontsize{8pt}{8pt}\selectfont
    \def\svgwidth{0.48\textwidth}
    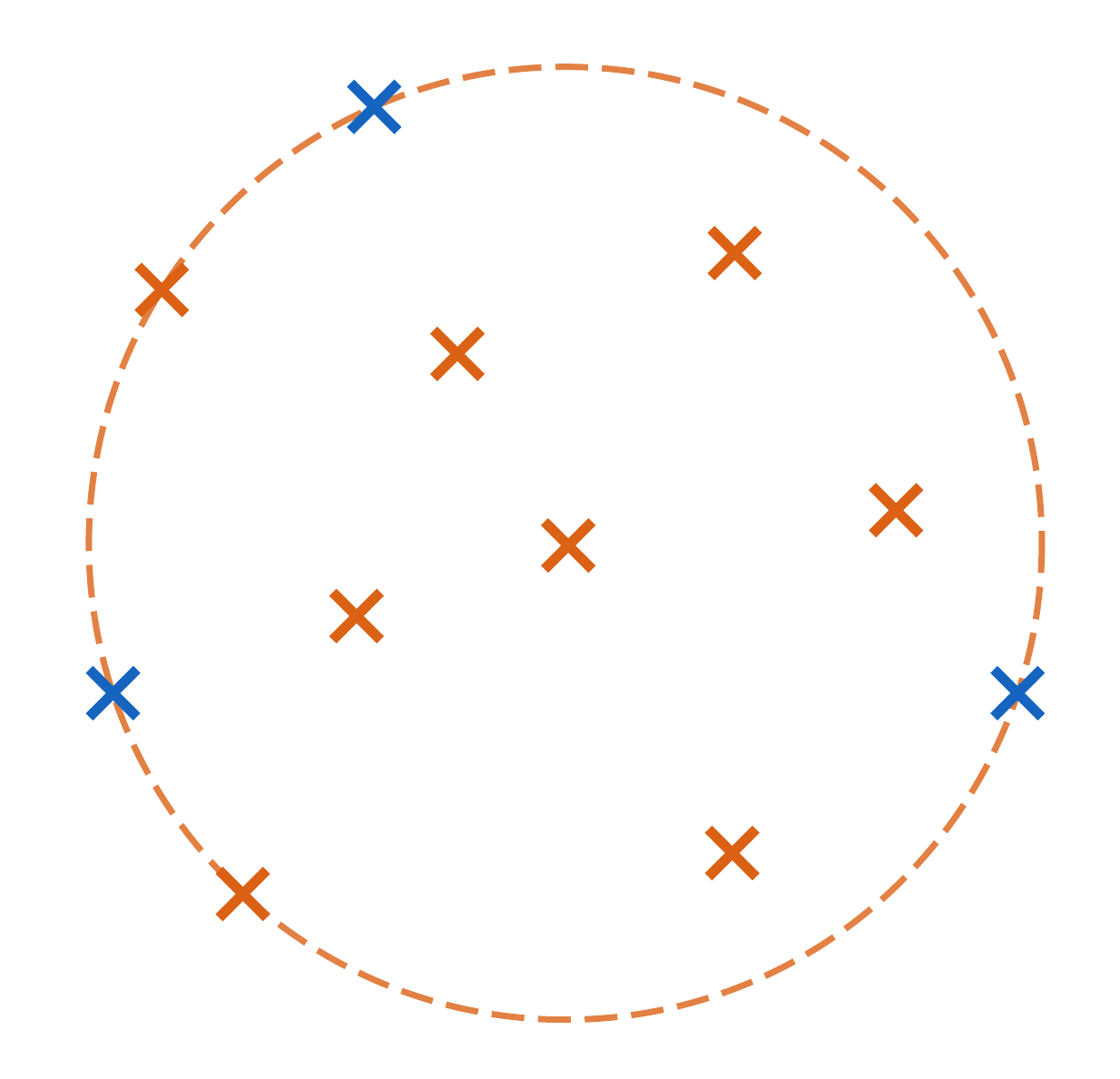
}
\hfill
\subcaptionbox[Short Subcaption]{
        \label{fig: bs2}
}
[
    0.48\textwidth 
]
{
    \fontsize{8pt}{8pt}\selectfont
    \def\svgwidth{0.46\textwidth}
    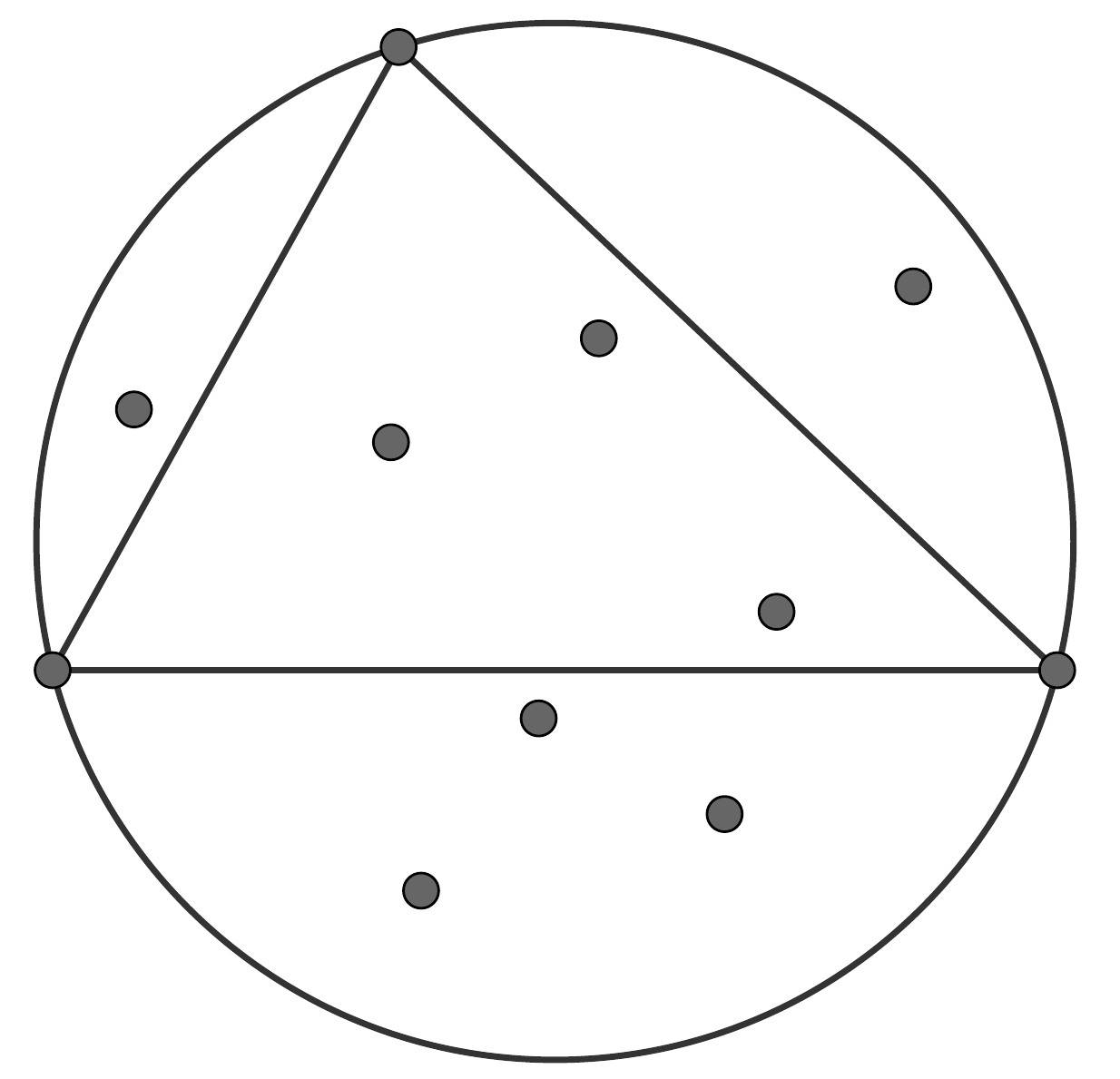
}
\\
\subcaptionbox[Short Subcaption]{
        \label{fig: bs3}
}
[
    0.48\textwidth 
]
{
    \fontsize{8pt}{8pt}\selectfont
    \def\svgwidth{0.48\textwidth}
    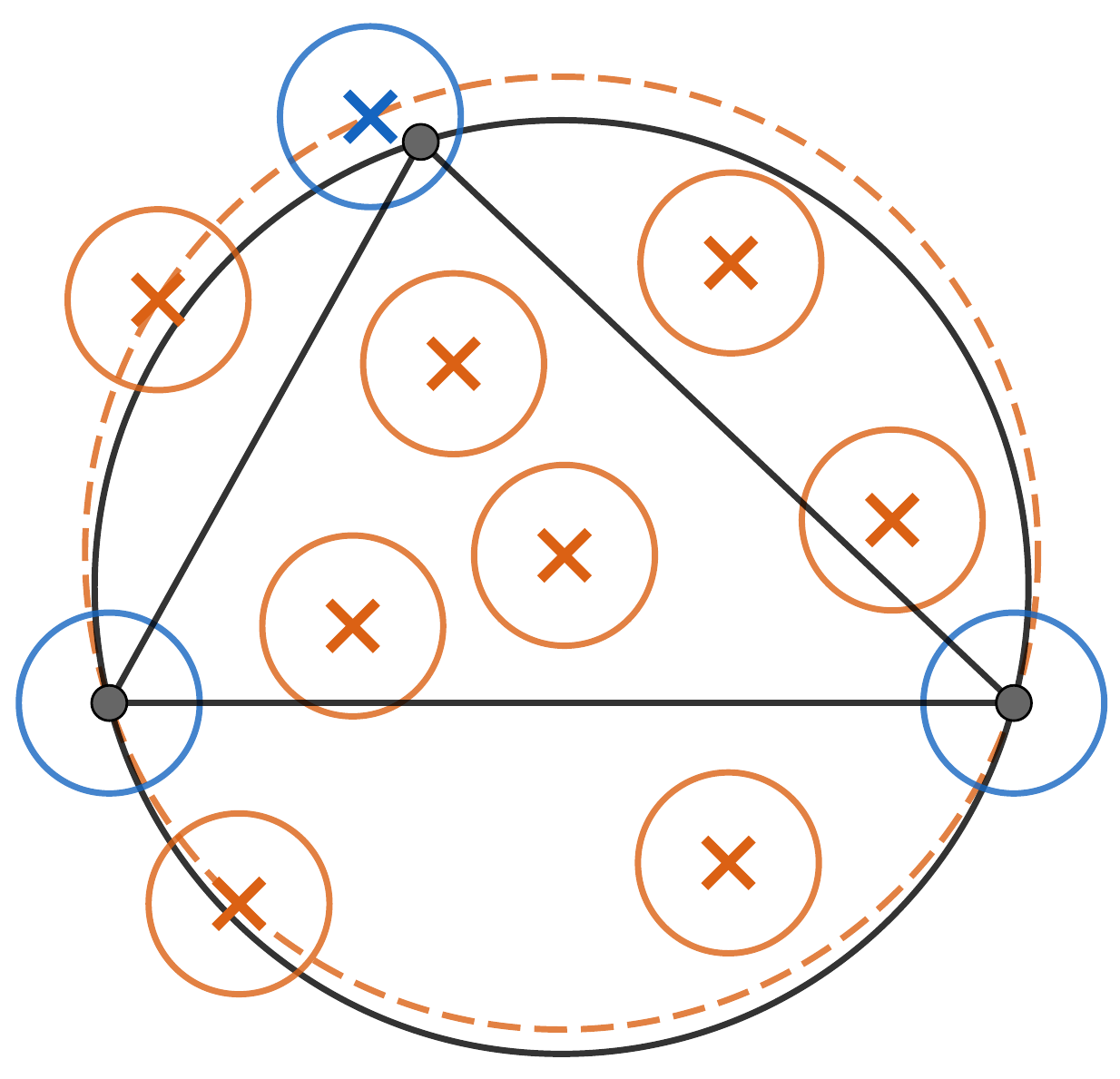
}
\hfill
\subcaptionbox[Short Subcaption]{
        \label{fig: bs4}
}
[
    0.48\textwidth 
]
{
    \fontsize{8pt}{8pt}\selectfont
    \def\svgwidth{0.48\textwidth}
    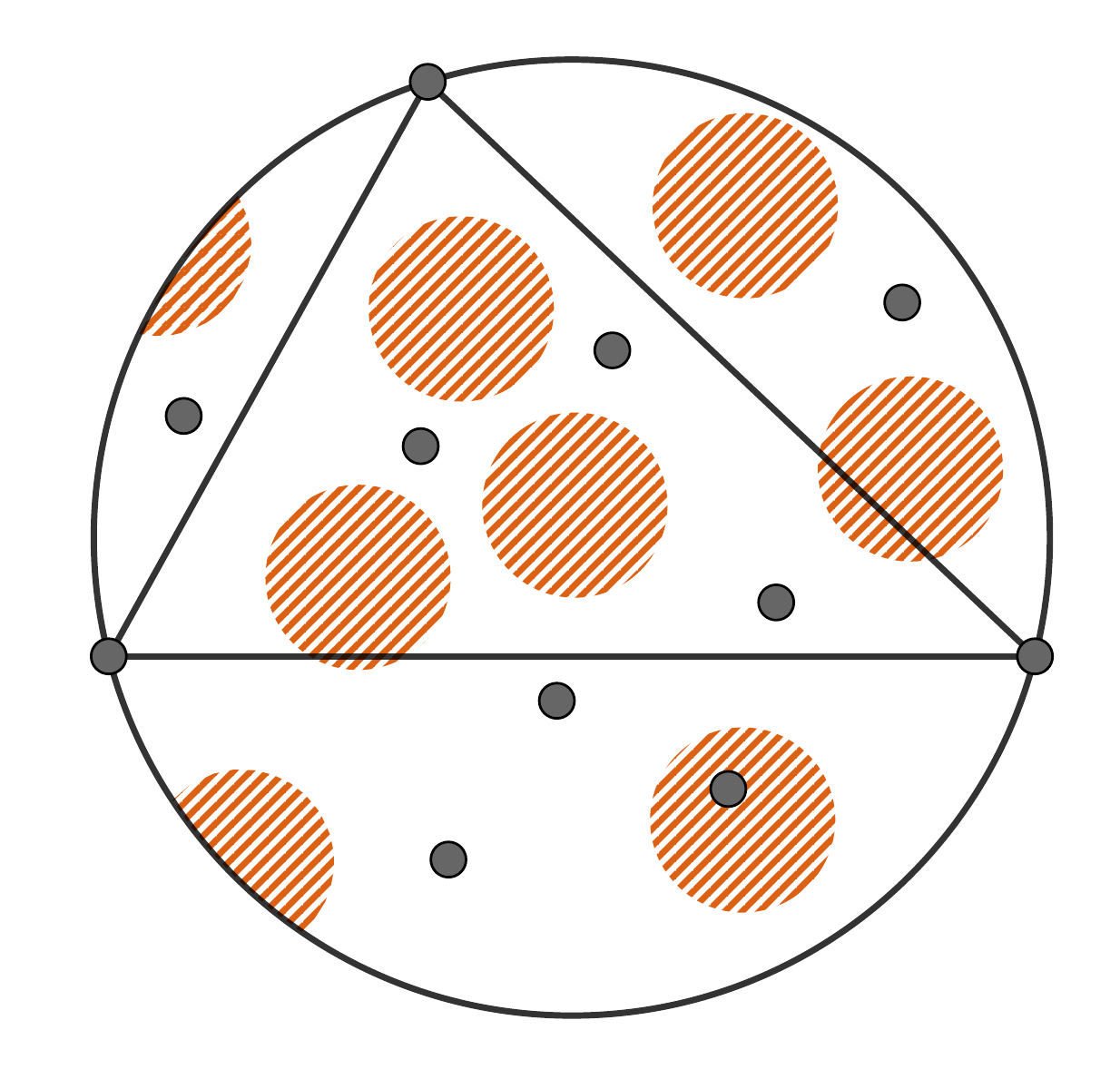
}
\caption[Short Caption]{a) The input pattern $F$. The bounding structure $b_F$ consists of the blue pattern points. b)-c) The bounding structure is formed by the robots. d) To obtain a final configuration, each shaded region must have a robot inside it.}
\label{fig: bs}
\end{figure}

If Algorithm \ref{algo: bf} is applied on the target pattern $F$, then we obtain a set $B_F \subseteq C(F) \cap F$ of pattern points such that $B_F$ is a minimal set of points of $C(F) \cap F$ such that $CC(B_F) = C(F)$. By minimal set we mean that no proper subset of $B_F$ has this property. Clearly by Property \ref{prop: cc}, $B_F$ either consists of two antipodal points or three points that form an acute angled triangle. We call $B_F$ the \emph{bounding structure} of $F$ (See Fig. \ref{fig: bs1}). We shall say that the bounding structure of $F$ is formed by the robots if one of the following holds.  

\begin{enumerate}
 \item $B_F$ has exactly two points, $C(R)$ also has exactly two robots on it and $R$ is symmetry safe.
 
 \item $B_F$ has exactly three points and $C(R)$ also has exactly three robots on it (See Fig. \ref{fig: bs}). Let $B_F = \{f_{i_1},f_{i_2},f_{i_3}\}$ and $C(R) \cap R = \{r_1, r_2, r_3\}$. $R$ is symmetry safe (i.e. $\Delta r_1r_2r_3$ is scalene) and furthermore, if $seg(r_1, r_2)$ is the largest side of the triangle formed by $r_1, r_2, r_3$ and $seg(f_{i_1},f_{i_2})$ is a largest side of the triangle formed by $f_{i_1},f_{i_2},f_{i_3}$, then there is an embedding $f_i \mapsto P_i$ of $F$ on the plane identifying $seg(f_{i_1},f_{i_2})$ with $seg(r_1, r_2)$ so that
 
 \begin{itemize}
  \item $r_3 \in B(P_{i_3}, \epsilon D)$, ($D =$ diameter of $C({P_1, \ldots, P_n}))$   
  \item $B(P_i, \epsilon D) \cap encl(CC(r_1, r_2, r_3)) \neq \emptyset$ for all $i \in \{1, \ldots, n\}$ 
 \end{itemize}
\end{enumerate}

\begin{algorithm}[H]
    \setstretch{.1}
    \SetKwInOut{Input}{Input}
    \SetKwInOut{Output}{Output}
    \SetKwProg{Fn}{Function}{}{}
    \SetKwProg{Pr}{Procedure}{}{}

    \Input{A pattern $F = \{f_1, \ldots, f_n\}$}
    
    
    Let $C(F) \cap F = \{f_{j_1}, \ldots, f_{j_k}\}$, where $j_1 < \ldots < j_k$
    
    $B_F \leftarrow \{f_{j_1}, \ldots, f_{j_k}\}$
    
    \For{$l \in 1, \dots, k$}{

      \If{$f_{j_l}$ is non-critical in $F$}{
      
	$F \leftarrow F \setminus \{f_{j_l}\}$\\
	$B_F \leftarrow B_F \setminus \{f_{j_l}\}$

      }
    
    }
    
    Return $B_F$
\caption{}
    \label{algo: bf} 
\end{algorithm}

\subsubsection{Description of the Algorithm}

The algorithm is in Phase 2 if $\texttt{a} \wedge \texttt{c} \wedge \neg\texttt{b}$ holds ($\texttt{b} =$ ``the bounding structure is formed''). The objective is to have $\texttt{b}$. We describe the algorithm for the following  cases: $C(R)$ has  three robots and the bounding structure also has three points (Case 1), $C(R)$ has three robots and the bounding structure has  two points (Case 2), $C(R)$ has two robots and the bounding structure has  three points (Case 3) and $C(R)$ has  two robots and the bounding structure also has two points (Case 4).


\paragraph{Case 1.}

\begin{figure}[]
\centering
\subcaptionbox[Short Subcaption]{
        \label{fig: p2c1 blue}
}
[
    0.6\textwidth 
]
{
    \fontsize{8pt}{8pt}\selectfont
    \def\svgwidth{0.55\textwidth}
    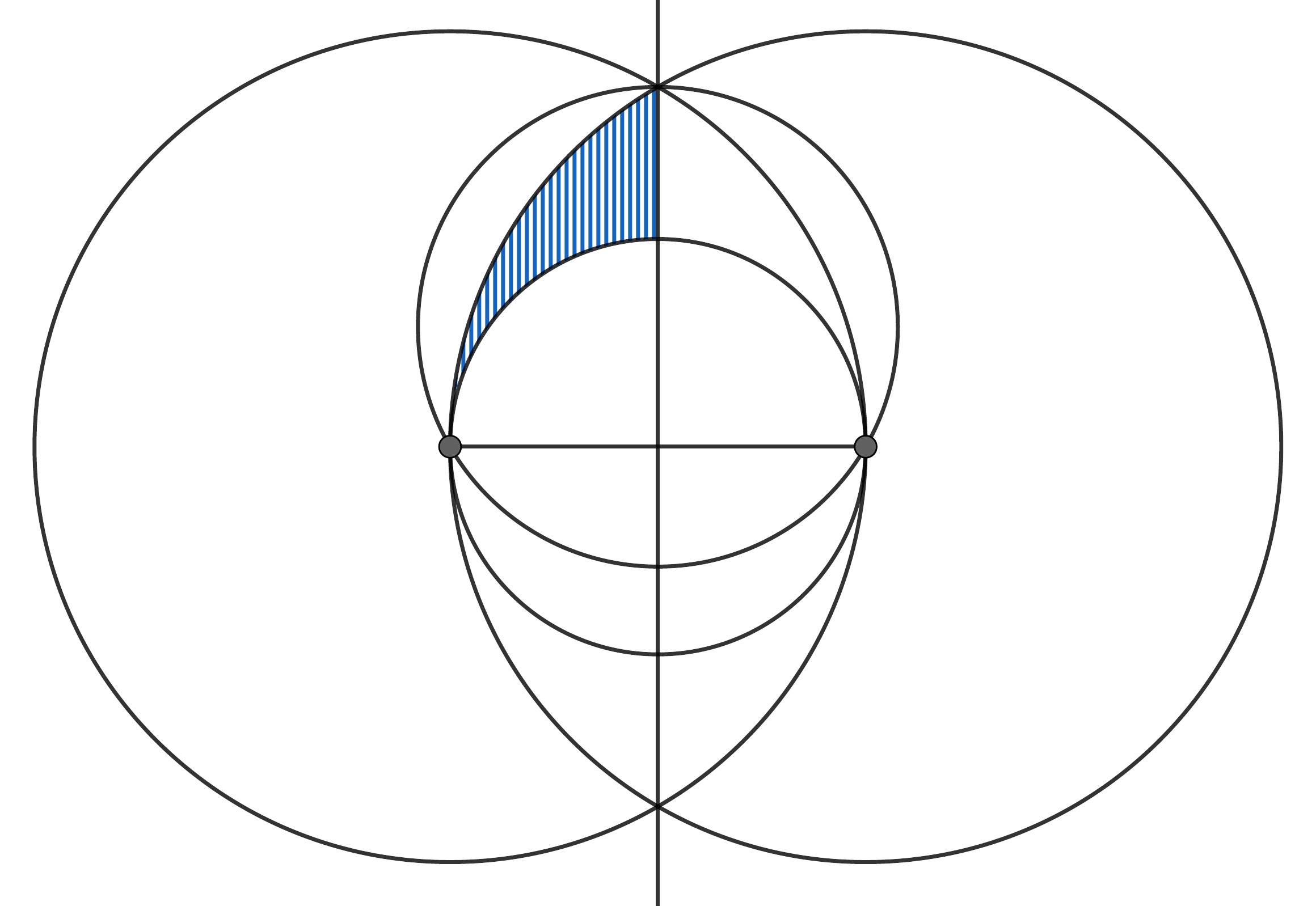
}
\\
\subcaptionbox[Short Subcaption]{
     \label{fig: p2c1 red}
}
[
    0.6\textwidth 
]
{
    \fontsize{8pt}{8pt}\selectfont
    \def\svgwidth{0.55\textwidth}
    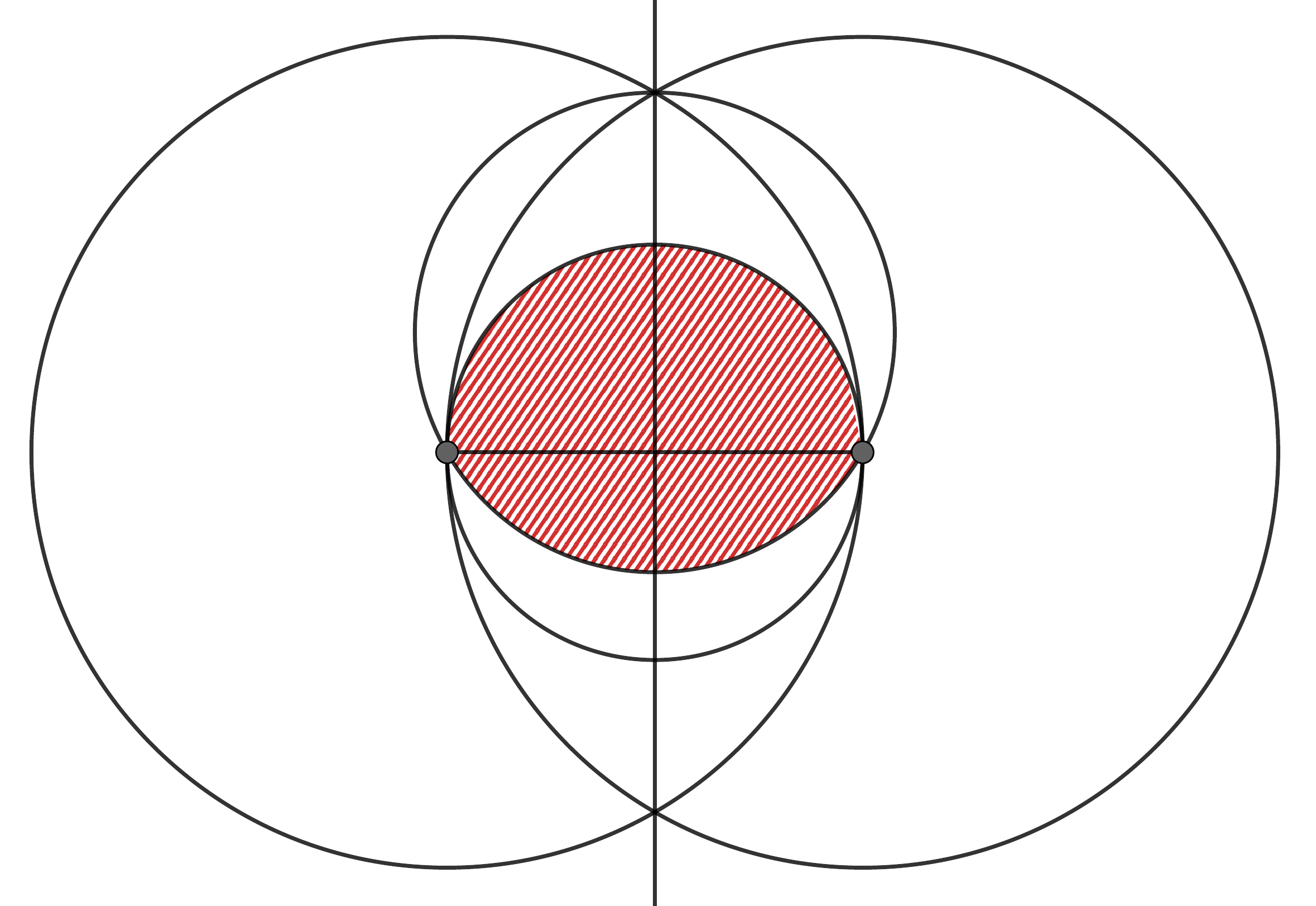
}
\caption[Short Caption]{Illustrations for Phase 2, Case 1.}
\label{fig p2case1}
\end{figure}

Assume that $C(R)$ has exactly three robots and the bounding structure consists of exactly three points. So the goal is to transform the triangle of the robots on $C(R)$ so that the bounding structure of $F$ is formed. Let $C(R) \cap R = \{r_1,r_2,r_3\}$.  If $\Delta r_1r_2r_3$ is not scalene, then we shall make it so by using similar techniques from Subphase 1.1, Case 3. So now assume that $\Delta r_1r_2r_3$ is scalene. Let $\overline{seg}(r_1,r_2)$ be the largest side of $\Delta r_1r_2r_3$. In that case, $r_3$ will be called the \emph{transformer robot}. This robot will move to form the bounding structure of $F$. Let $L$ be the perpendicular bisector of $\overline{seg}(r_1,r_2)$. Since no two sides of the triangle are of equal length, $r_3 \notin L$. Let $\mathcal{H}$ be the open half-plane delimited by $L$ that contains $r_3$. Without loss of generality, assume that $r_1 \in \mathcal{H}$. Let $L_1$ be the line parallel to $L$ and passing through $r_1$. Let $\mathcal{H}'$ be the open half-plane delimited by $L_1$ that contains $L$. Since $\Delta r_1r_2r_3$ is acute angled, $r_3 \in \mathcal{H}'$. Let $\mathcal{H}''$ be the open half-plane delimited by $line(r_1,r_2)$ that contains $r_3$. Let $C_1 = C(r_1, d(r_1,r_2))$ and $C_2 = C(r_2, d(r_2,r_1))$. Since $\overline{seg}(r_1,r_2)$ is (strictly) the largest side of $\Delta r_1r_2r_3$, $r_3 \in encl(C_1) \cap encl(C_2)$. If $C_3 = CC(r_1, r_2)$, then $r_3 \in ext(C_3)$ as $\Delta r_1r_2r_3$ is acute angled. Now take the largest side of the bounding structure $B_F$. In case of a tie, use the ordering of the points in the input $F$ to choose one of them. Embed the bounding structure $B_F$ on the plane identifying this side with $\overline{seg}(r_1, r_2)$ so that the third point of the bounding structure is mapped to a point $P \in \overline{\mathcal{H}} \cap \mathcal{H}''$. Since the bounding structure is acute angled, $P \in \mathcal{H}'$. Also, $P \in ext(C_3)$ for the same reason. Furthermore, since a largest side of the bounding structure is identified with $\overline{seg}(r_1, r_2)$, $P \in  \overline{encl(C_1)} \cap \overline{encl(C_2)}$. So we have $r_3 \in \mathcal{H} \cap \mathcal{H}' \cap \mathcal{H}'' \cap encl(C_1) \cap encl(C_2) \cap ext(C_3) = \mathcal{U}_{blue}$ (the blue open region in Fig. \ref{fig: p2c1 blue}) and $P \in \overline{\mathcal{H}} \cap \mathcal{H}' \cap \mathcal{H}'' \cap \overline{encl(C_1)} \cap \overline{encl(C_2)} \cap ext(C_3) = \mathcal{U'}_{blue}$. Notice that $\mathcal{U'}_{blue}$ consists of the open region $\mathcal{U}_{blue}$ and some parts of its boundary. Our objective is to move the robot $r_3$ to a point near $P$. The entire trajectory of the movement should lie inside the region $\mathcal{U}_{blue}$. However, before this movement, we have to make sure that the configuration satisfies some desirable properties described in the following.

Let $C_4$ be the circle passing through $r_1, r_2$ and the point in $\mathcal{H}''$ where $C_1$ and $C_2$ intersect each other. We shall say that the transformer robot is \emph{eligible to move} if $R \cap encl(C(R)) \subset encl(C_3) \cap encl(C_4) =  \mathcal{U}_{red}$ (the red region in Fig. \ref{fig: p2c1 red}). The transformer robot will not move until it becomes eligible. So the robots in $encl(C(R))$ that are not in $ \mathcal{U}_{red}$, should move inside this region first. As discussed in Section \ref{sec: safe zone}, the robots will fix some specific disk $B(p, l) \subset \mathcal{U}_{red}$ and move through the cone defined by it. The robots should move sequentially. So we ask the robot outside $\mathcal{U}_{red}$ that is closest to $p$ to move first. If there are multiple such robots, then the one with the minimum view is chosen. Recall that the configuration is asymmetric as $\Delta r_1r_2r_3$ remains scalene and also $r_3$ remains the transformer robot. So when we have $R \cap encl(C(R)) \subset \mathcal{U}_{red}$,   $r_3$ will become eligible to move and will move inside the region  $B(P, \epsilon D) \cap \mathcal{U}_{blue}$ via some trajectory that lies inside the blue region $\mathcal{U}_{blue}$. This can be done by the scheme described in Section \ref{sec: safe zone}.

\begin{lemma}\label{lemma p2c1}
If the algorithm is in Phase 2, Case 1 at some round, then after finitely many rounds we have  $\texttt{b}$.
\end{lemma}

\begin{proof}
Clearly all the robots of $R \setminus \{ r_1, r_2, r_3\}$ will move inside $\mathcal{U}_{red}$ in finitely many steps. Then $r_3$ becomes eligible to move. Recall that the movement of $r_3$ is restricted inside $\mathcal{U}_{blue}$. We claim that during this movement $r_3$ will remain the transformer robot. For this, we will show that during the movement 1) $\texttt{a} \wedge \texttt{c}$ is true, 2) the minimum enclosing circle of the configuration passes through only $r_1, r_2, r_3$ and 3) $\Delta r_1r_2r_3$ is scalene with $r_1 r_2$ being the strictly largest side. 

During the movement $r_3$ can be at any point in $\mathcal{U}_{blue}$.   Take any such point $Q \in \mathcal{U}_{blue}$.  It is easy to see that $\Delta Qr_1r_2$ is scalene with  $seg(r_1,r_2)$ being the largest side. Furthermore, $\Delta Qr_1r_2$ is acute-angled. Notice that $CC(Q,r_1,r_2) \in (C_3,C_4)_{\mathcal{F}(r_1, r_2)}$. Hence by Property \ref{fam}, $encl(C_3) \cap encl(C_4) = \mathcal{U}_{red} \subset encl(CC(Q,r_1,r_2))$. Recall that all robots except $r_1$, $r_2$, $r_3$ are inside $\mathcal{U}_{red}$. Hence if $r_3$ is at $Q$ during its movement, then the minimum enclosing circle of the configuration is $CC(r_1, r_2, r_3)$ with three of them critical robots. Also  $\texttt{a}$ holds as $\Delta r_1r_2r_3$ is scalene.  This implies that $r_3$ remains the transformer robot during the movement.

\begin{figure}[htb!]
\centering
\fontsize{8pt}{8pt}\selectfont
\def\svgwidth{0.35\textwidth}
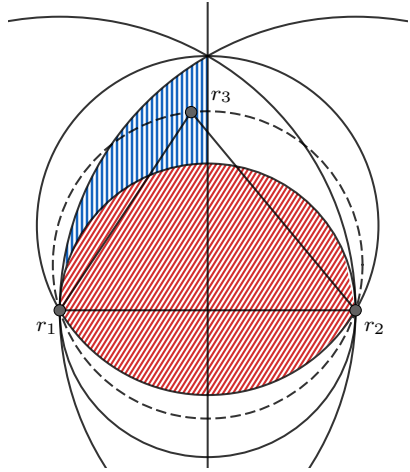
\caption{Illustration supporting the proof of Lemma \ref{lemma p2c1}.}
\label{fig: proof p2c1}
\end{figure}

When $r_3$  moves inside $B(P, \epsilon D) \cap \mathcal{U}_{blue}$, we shall have $\texttt{b}$ as required. However, there are details regarding its movement that need to be mentioned here. As discussed in Section \ref{sec: safe zone}, $r_3$ needs to set a disk $B(P', l) \subset B(P, \epsilon D) \cap \mathcal{U}_{blue}$ (depending only on the shape of the region) so that $cone(r_3, B(P', l)) \subset \mathcal{U}_{blue}$ and move through it. However, notice that the blue region $\mathcal{U}_{blue}$ is not a convex region and hence $seg(r_3, P')$ is not necessarily lying completely inside $\mathcal{U}_{blue}$. So first $r_3$ may have to move so that $seg(r_3, P')$ is inside the blue region $\mathcal{U}_{blue}$. This can be done easily. Then $r_3$ can move accordingly as discussed in Section \ref{sec: safe zone}. 
\end{proof}

\paragraph{Case 2.}

\begin{figure}[htb!]
\centering
\fontsize{8pt}{8pt}\selectfont
\def\svgwidth{0.5\textwidth}
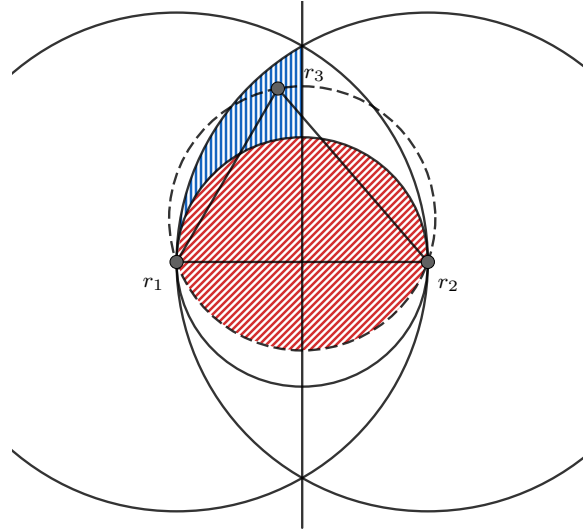
\caption{Illustrations for Phase 2, Case 2.}
\label{fig: p2c2}
\end{figure}

In Case 2, $C(R)$ has exactly three robots and the bounding structure consists of exactly two points. Let $C(R) \cap R = \{r_1,r_2,r_3\}$.  As before, $\Delta r_1r_2r_3$ will be made scalene. Let  $\overline{seg}(r_1,r_2)$ be the largest side. Then $r_3$ will be the transformer robot. The plan is to move the transformer robot inward so that it is no longer on the minimum enclosing circle of the configuration. Let $C_1, C_2, C_3, \mathcal{H}, \mathcal{H}', \mathcal{H}''$ denote the same as in Case 1. As before, we have $r_3 \in \mathcal{H} \cap \mathcal{H}' \cap \mathcal{H}'' \cap encl(C_1) \cap encl(C_2) \cap ext(C_3) = \mathcal{U}_{blue}$ (the blue region in Fig. \ref{fig: p2c2}). We shall say that the transformer robot is \emph{eligible to move} if 1) $R \cap encl(C(R)) \subset encl(C_3) \cap encl(C(R)) $ (the red region in Fig. \ref{fig: p2c2}) and 2) $R \setminus \{r_3\}$ is a symmetry safe configuration. The robots in $encl(C(R))$ that are not already in $encl(C_3)$ will first move inside $encl(C_3)$. After that we have $C(R \setminus \{r_3\}) = C_3$ and it passes through only two robots, i.e., $r_1$ and $r_2$. So $R \setminus \{r_3\}$ will be symmetry safe if there is a unique robot closest to $O$, the midpoint of $seg(r_1,r_2)$, and it is not on $seg(r_1,r_2)$ or its perpendicular bisector. This can be achieved easily. When $r_3$ becomes eligible to move, it will move inside the region $encl(C_3)$. During its movement, when it has not entered $encl(C_3)$, its trajectory should remain inside  $\mathcal{U}_{blue}$. Also, when it enters $encl(C_3)$, it should remain in $ext(C)$ where $C = C_{\uparrow}^{1}(R \setminus \{r_3\})$. So its entire trajectory should be inside the region  $\mathcal{H} \cap \mathcal{H}' \cap \mathcal{H}'' \cap encl(C_1) \cap encl(C_2) \cap ext(C)$ and it should not collide with any robot upon entering $encl(C_3)$. This can be done by the scheme described in \ref{sec: safe zone}.

\begin{lemma}
If the algorithm is in Phase 2, Case 2 at some round, then after finitely many rounds we have  $\texttt{b}$.
\end{lemma}

\begin{proof}
After finitely many rounds $r_3$ will become eligible to move. Then, as in Case 1, $r_3$ will move through the blue region  until it enters $encl(C_3)$. While in the blue region, $r_3$ will remain the transformer robot by similar arguments as in Case 1. After it enters $encl(C_3)$ we have $CC(r_1, r_2) = C(R)$. Recall that $R \setminus \{r_3\}$ was made symmetry safe before the movement of $r_3$ started. Since $r_3$ remains in  $ext(C_{\uparrow}^{1}(R \setminus \{r_3\}))$, the final configuration is symmetry safe. Hence the bounding structure is formed.



\end{proof}

\paragraph{Case 3.}

\begin{figure}[htb!]
\centering
\fontsize{8pt}{8pt}\selectfont
\def\svgwidth{0.5\textwidth}
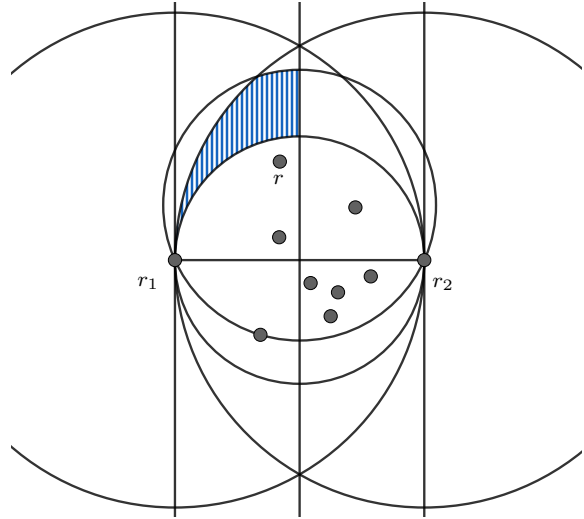
\caption{Illustrations for Phase 2, Case 3.}
\label{fig: p2c3}
\end{figure}

In Case 3, $C(R)$ has exactly two robots and the bounding structure consists of exactly three points. Let $C(R) \cap R = \{r_1, r_2\}$. Here the strategy is to move outward one of the robots from $encl(C(R))$, say $r$, so that the minimum enclosing circle of the configuration becomes the circumcircle of $r, r_1$ and $r_2$. We shall call $r$ the transformer robot. The robot farthest from $c(R)$ will be chosen as the transformer robot. In case of a tie, it is broken using the asymmetry of the configuration. Let $\mathcal{H}$ be the open half plane delimited by $line(r_1, r_2)$ that contains $r$. Let $L_1$ and $L_2$ be the lines perpendicular to $line(r_1, r_2)$ and passing through respectively $r_1$ and $r_2$. Let $L$ be the perpendicular bisector of $seg(r_1, r_2)$. Without loss of generality, assume that $r \in \mathcal{S}(L_1, L) \cup L$. Let $C_1 = C(r_1, d(r_1,r_2))$, $C_2 = C(r_2, d(r_2,r_1))$ and $C_3 = CC(r_1, r_2)$. Let $C_4$ be the largest circle from the family $\{C \in \mathcal{F}(r_1,r_2) \mid \text{center of $C$ lies in $\mathcal{H}$ and $R \subset \overline{encl}(C)$} \}$. The algorithm asks $r$ to move into the region $encl(C_1) \cap encl(C_2) \cap ext(C_3) \cap encl(C_4) \cap \mathcal{H} \cap \mathcal{S}(L_1,L)$ (the blue region in Fig. \ref{fig: p2c3}). Again, this can be done by the scheme described in \ref{sec: safe zone}.

\begin{lemma}
If the algorithm is in Phase 2, Case 3 at some round, then after finitely many rounds we have  $\texttt{b}$.
\end{lemma}

\begin{proof}

Here the transformer robot $r$, which is a farthest robot from $c(R)$ has to get inside the blue region shown in Fig.\ref{fig: p2c3}. Until $r$ reaches there, the minimum enclosing circle of the configuration remains unchanged and it remains the transformer robot since its movement is outward. When it moves to a point in the blue region, let $R^{new}$ denote the resulting configuration. We have to show that $CC(r,r_1,r_2) = C(R^{new})$. Notice that $CC(r,r_1,r_2) \in (C_3,C_4)_{\mathcal{F}(r_1,r_2)}$. Hence by Property \ref{fam}, $encl(C_3) \cap encl(C_4) \subset encl(CC(r,r_1,r_2))$. But recall that $R \setminus \{r_1,r_2 \} \subset encl(C_3) \cap \overline{encl(C_4)}$. Hence $R \setminus\{r_1, r_2, r \} \subset CC(r,r_1,r_2)$. By similar arguments as used in previous cases  $\Delta rr_1r_2 $ is acute-angled. Hence by Property \ref{prop: cc}, $CC(r, r_1, r_2) = C(R^{new})$. Furthermore, the configuration is asymmetric as  $\Delta rr_1r_2$ is scalene. Hence the situation reduces to Case 1. So the bounding structure will be formed after finitely many rounds by Lemma \ref{lemma p2c1}.
\end{proof}

\paragraph{Case 4.}

In this case, $C(R)$ has two exactly robots and the bounding structure also has exactly two points. So we have $\neg \texttt{b}$ because the configuration is not symmetry safe. The configuration can be easily made symmetry safe by previously discussed techniques.

\begin{lemma}
If the algorithm is in Phase 2, Case 4 at some round, then after finitely many rounds we have  $\texttt{b}$.
\end{lemma}

\subsection{Phase 3}

\subsubsection{Motive and Overview}

The algorithm is in Phase 3 if $\texttt{b}$ holds. The objective of this phase is to form the pattern approximately.
Notice that when $\texttt{b}$ holds, the configuration is symmetry safe and hence asymmetric. This will allow the robots to agree on a coordinate system in which the target will be formed (approximately). During this process, $\texttt{b}$ has to be preserved because otherwise the agreement in coordinate system will be lost.

The termination condition of the algorithm is that both $\texttt{b}$ holds (i.e., it is a Phase 3 configuration) and the configuration is $\epsilon$-close to $F$. Therefore, even if the initial configuration is $\epsilon$-close to $F$ (i.e., the pattern $F$ is already formed approximately), the algorithm will still go through the earlier phases to have $\texttt{b}$  and then approximately form the pattern while preserving $\texttt{b}$. The reason why we take this approach is because in general, even if the configuration is $\epsilon$-close to $F$, the robots may not be able to efficiently identify this. This is a basic difficultly of the problem. However, when $\texttt{b}$ holds there is a way to fix a particular embedding of $F$ in the plane and then the only thing to check is whether there are robots close to each point of the embedding. For Phase 3, there are two cases to consider: $B_F$ has exactly two points (Case 1) and $B_F$ has exactly three points (Case 2).

\subsubsection{Case 1}

For Case 1, let us first describe how we shall fix a common coordinate system. Let $\{r_1, r_2\} = C(R) \cap R$. Let $\ell = line(r_1,r_2)$ and $\ell'$ be the line passing through $c(R)$ and perpendicular to $\ell$. Let $r_l$ be the unique robot that is closest to $c(R)$. Also it is in $encl(C(R)) \setminus (\ell \cup \ell')$. Such a robot exists because $\texttt{b}$ holds. We set a \emph{global coordinate system} whose center is at $c(R)$, $X$ axis along $\ell$, $Y$ axis along $\ell'$. The positive directions of $X$ and $Y$ axis are such that $r_l$ lies in the positive quadrant. 

Now we choose an embedding of the pattern $F$ that will be approximated. Perform a coordinate transformation (rotation) on the target pattern $F$ so that the bounding structure is along the $X$ axis. Let $F'$ denote the input after this transformation. Consider the pattern points on $C^{1}_{\uparrow}(F')$ except the points of the bounding structure (notice that $C^{1}_{\uparrow}(F')$ may have points from the bounding structure when $C^{1}_{\uparrow}(F') = C^{1}_{\downarrow}(F')$). Reflect the pattern with respect to $X$ axis or $Y$ axis or both, if required, so that at least one of them is in the closed positive quadrant ($X \geq 0, Y \geq 0$). Let $F''$ denote the pattern thus obtained. Therefore, if $\{f_i, f_j\}$ be the bounding structure, then we have  1) $f_i, f_j$ on the $X$ axis and 2) at least one point from $C^{1}_{\uparrow}(F'') \cap (F'' \setminus \{f_i, f_j\})$ in the closed positive quadrant. Each robot applies coordinate transformations on $F$ and obtains the same pattern $F''$. Let $f_l$ denote the first pattern point from  $C^{1}_{\uparrow}(F'') \cap (F'' \setminus \{f_i, f_j\})$ that is in the closed positive quadrant. The pattern $F''$ is mapped in the plane in the global coordinate system and scaled so that the bounding structure is mapped onto $\overline{seg}(r_1, r_2)$. These points are called the \emph{target points}. The set of target points are denoted by $T$.

\begin{figure}
\centering
\subcaptionbox[Short Subcaption]{
        \label{}
}
[
    0.48\textwidth 
]
{
    \fontsize{8pt}{8pt}\selectfont
    \def\svgwidth{0.4\textwidth}
    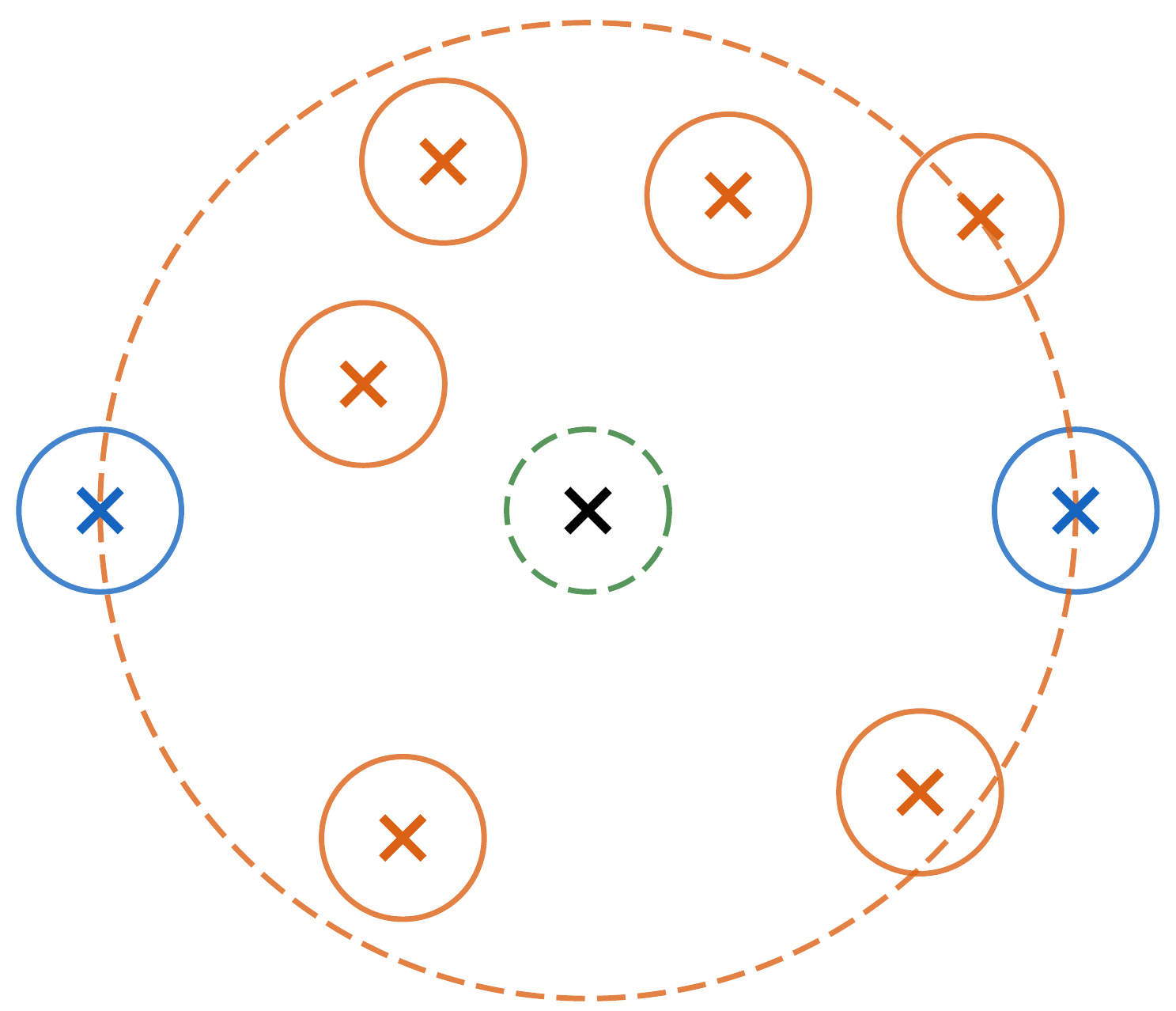
}
\hfill
\subcaptionbox[Short Subcaption]{
        \label{}
}
[
    0.48\textwidth 
]
{
    \fontsize{8pt}{8pt}\selectfont
    \def\svgwidth{0.4\textwidth}
    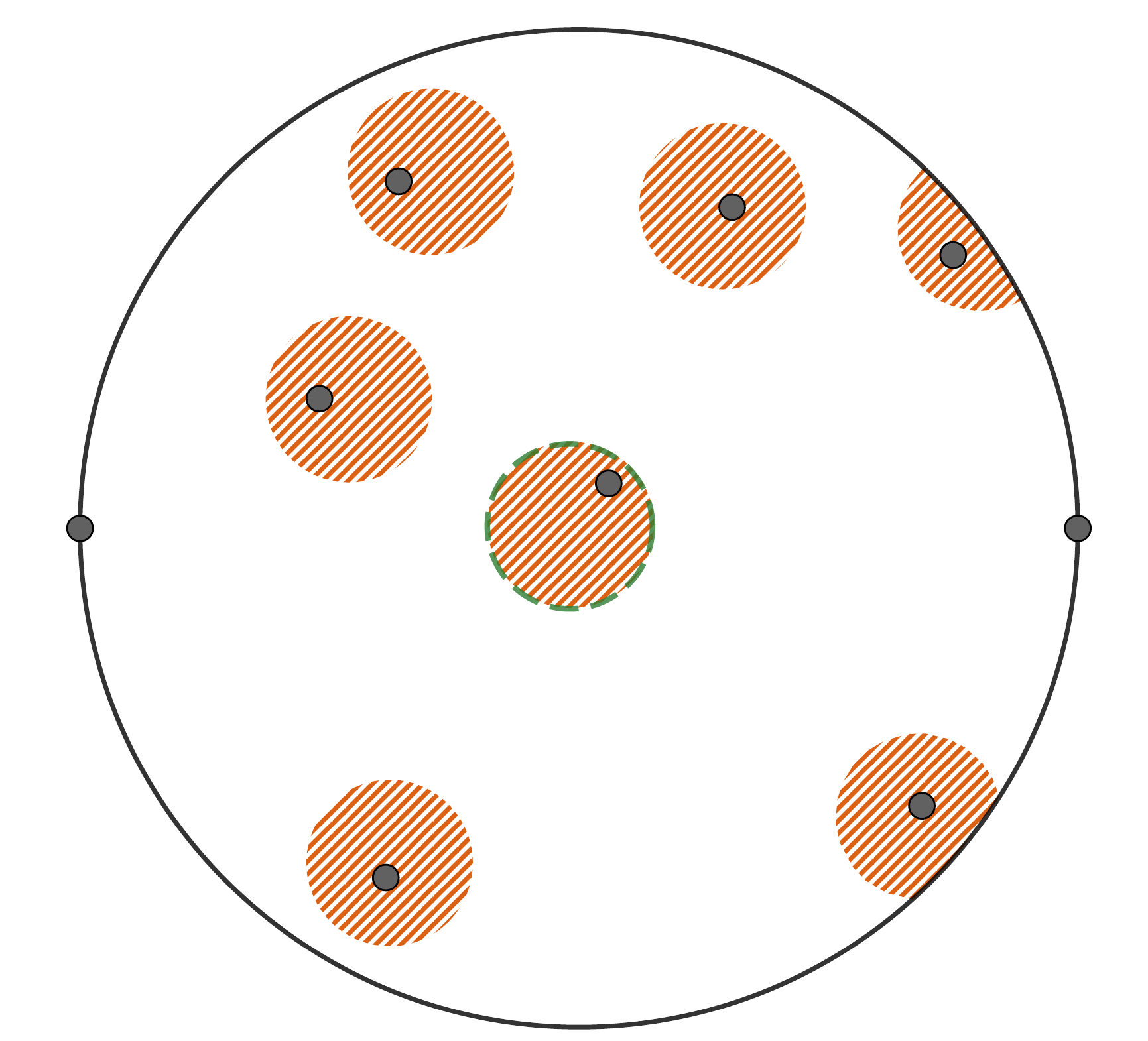
}
\\
\subcaptionbox[Short Subcaption]{
        \label{}
}
[
    0.48\textwidth 
]
{
    \fontsize{8pt}{8pt}\selectfont
    \def\svgwidth{0.4\textwidth}
    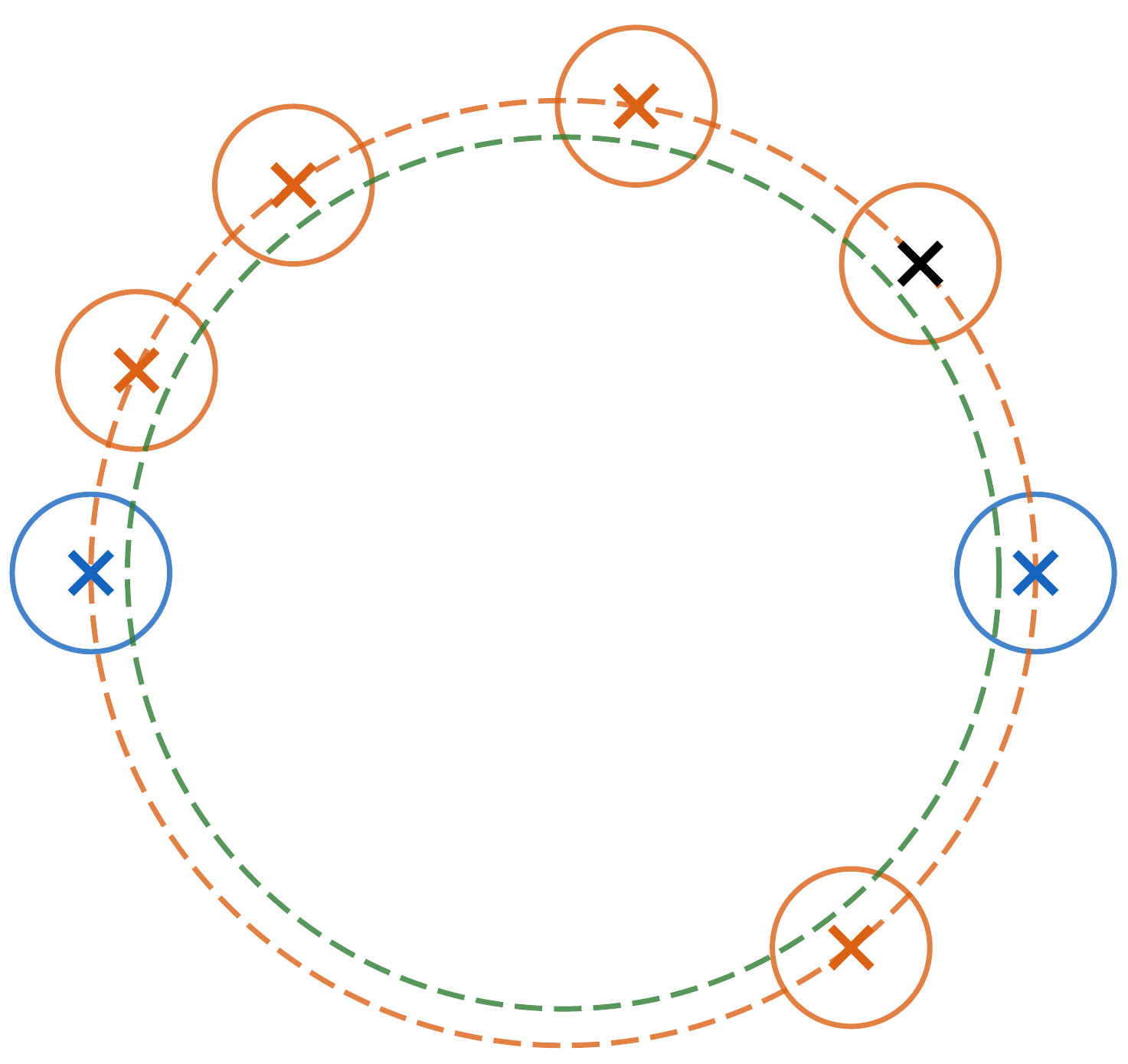
}
\hfill
\subcaptionbox[Short Subcaption]{
        \label{}
}
[
    0.48\textwidth 
]
{
    \fontsize{8pt}{8pt}\selectfont
    \def\svgwidth{0.4\textwidth}
    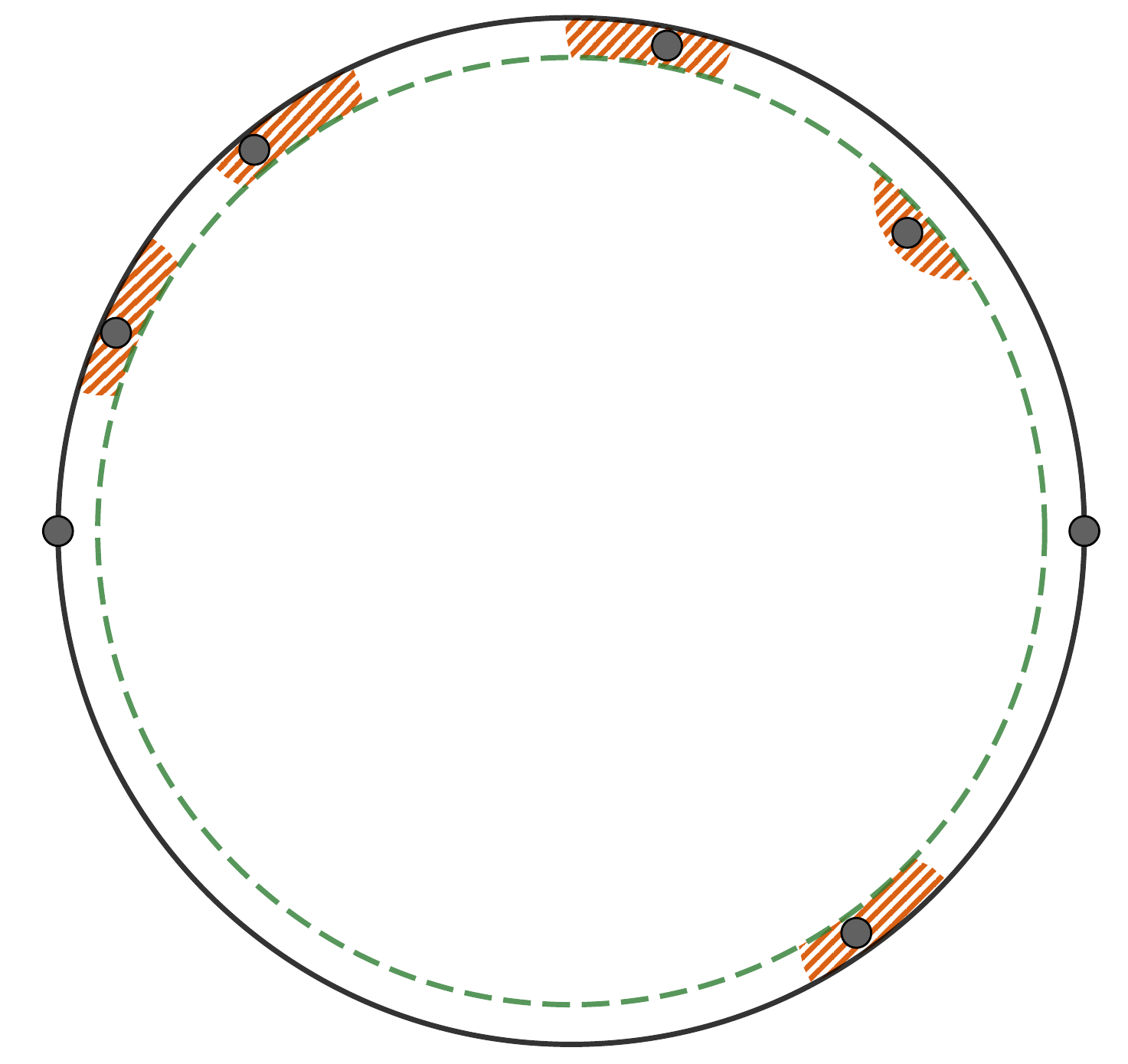
}
\\
\subcaptionbox[Short Subcaption]{
        \label{}
}
[
    0.48\textwidth 
]
{
    \fontsize{8pt}{8pt}\selectfont
    \def\svgwidth{0.4\textwidth}
    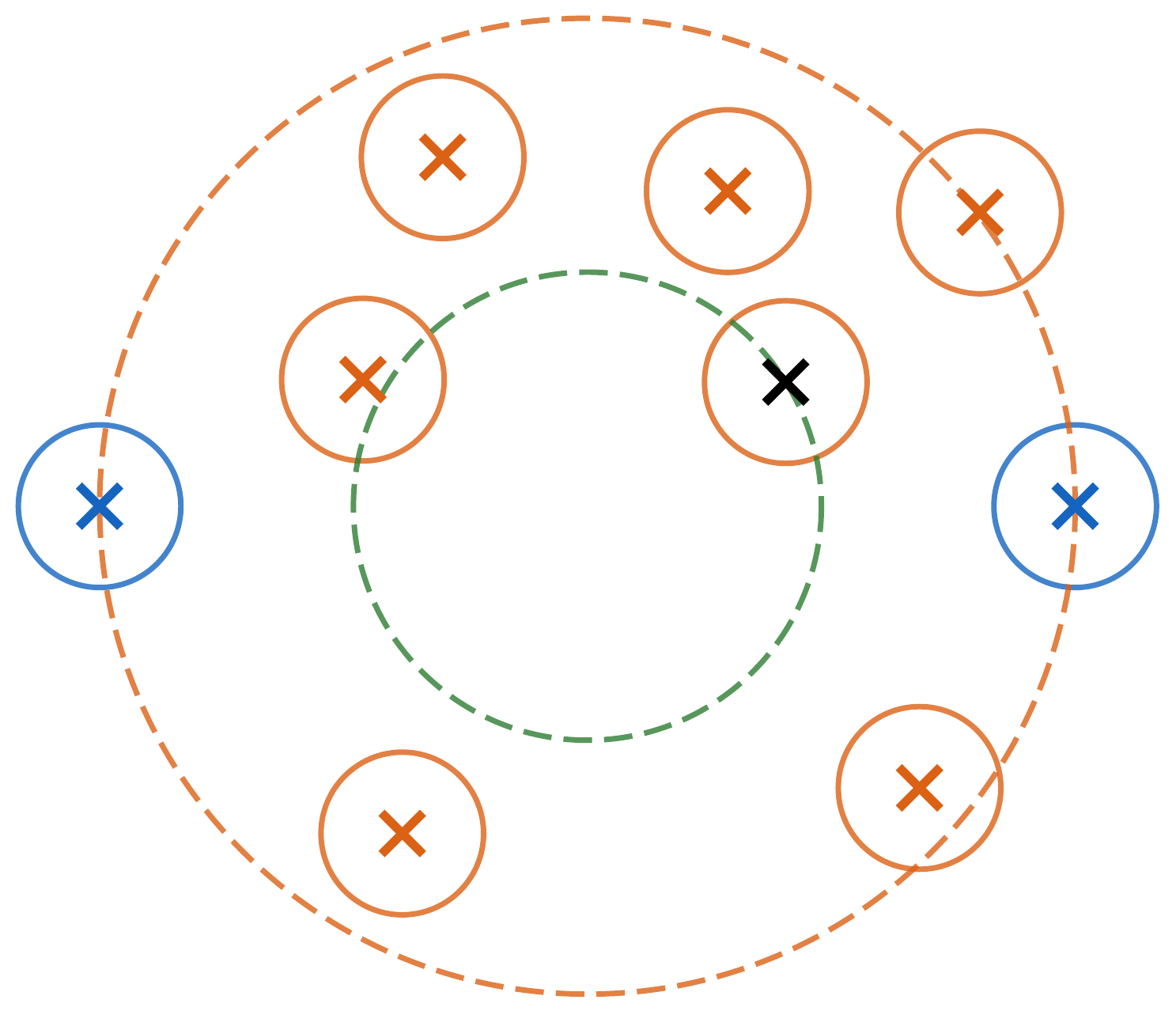
}
\hfill
\subcaptionbox[Short Subcaption]{
        \label{}
}
[
    0.48\textwidth 
]
{
    \fontsize{8pt}{8pt}\selectfont
    \def\svgwidth{0.4\textwidth}
    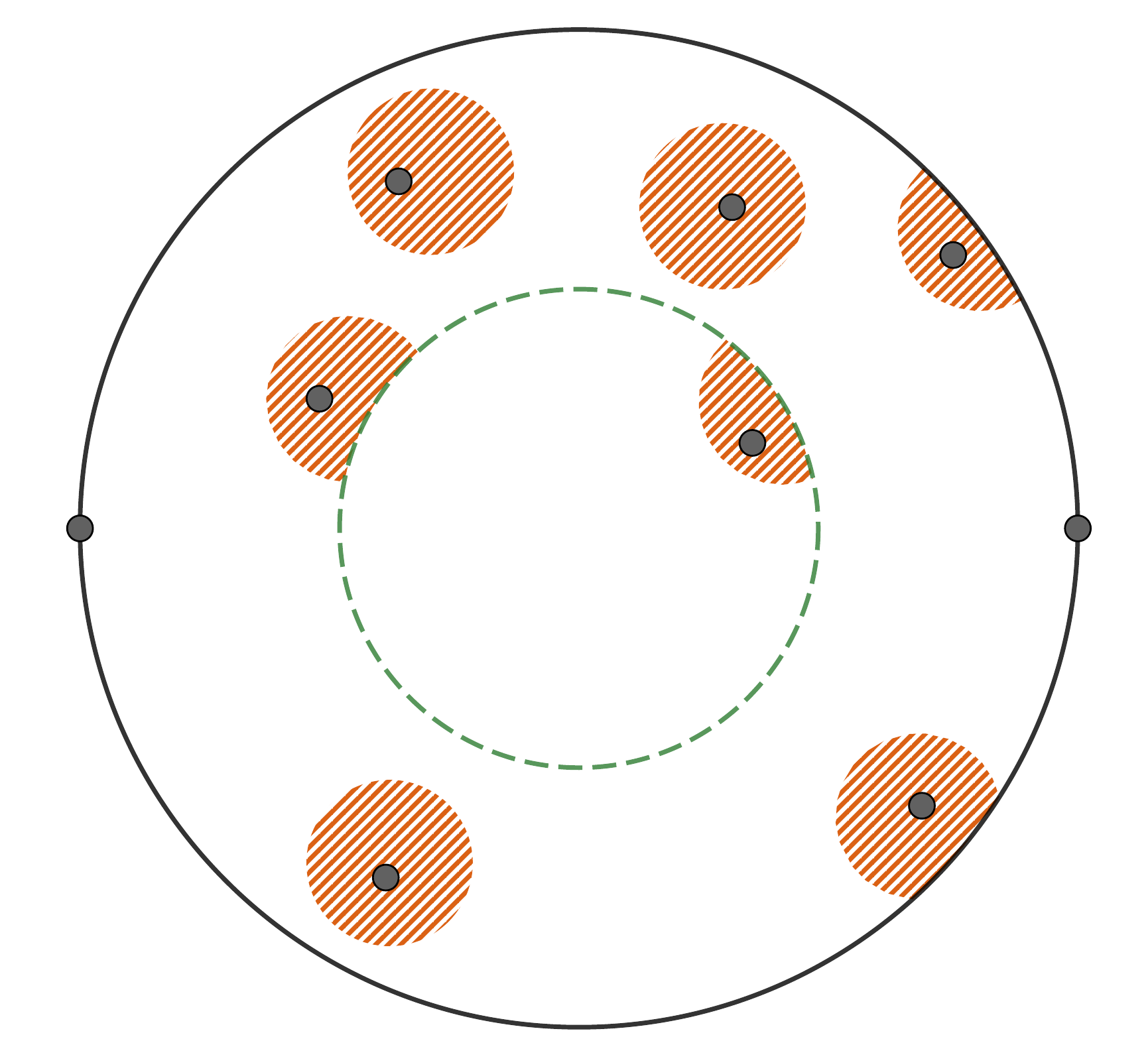
}
\caption[Short Caption]{Illustrations for Phase 3, Case 1. In each row, the input pattern $F$ is shown on the left and a final configuration approximating $F$ is shown on the right. In each case, points of the  bounding structure are shown in blue, $t_l$ is shown in black and the green circle represents $C_l$.}
\label{fig: cl}
\end{figure}

Notice that the robot $r_l$, being the unique robot on $C^{1}_{\uparrow}(R)$ and also being in an open quadrant (defined by $\ell \cup \ell'$),  plays crucial role in fixing the common coordinate system. This will be preserved throughout the algorithm. In particular, $r_l$ will remain in such a position even in the final configuration. The target point that $r_l$ will approximate in the final configuration will be the target point corresponding to $f_l$. Let us call it $t_l$. Notice that $t_l$ is on $C^{1}_{\uparrow}(T)$ and in the closed positive quadrant. As $r_l$ is in the open quadrant, it does not need to move out of it to approximate $t_l$. Now since $r_l$ needs to remain the closest robot from the center, we will define a circle $C_l$, that depends only on the position of $t_l$,  and require that in the final configuration we have $r_l$ inside this circle and all robots is outside the circle. If $D$ is the diameter of $C(T)$, i.e., $D = d(r_1,r_2)$, then we define the circle $C_l$ as (See also Fig. \ref{fig: cl}) 
 
 \begin{itemize}
  \item if $t_l \in C^{1}_{\uparrow}(T) = c(T)$, then $C_l = C(c(T), \epsilon D)$,
  
  \item if $t_l \in C^{1}_{\uparrow}(T) = C(T)$, then $C_l = C(c(T), (1 - \epsilon) \frac{D}{2})$,
  
  \item otherwise, $C_l = C^{1}_{\uparrow}(T)$.
 \end{itemize}

  We shall say that a target point $t \neq t_l$ is \emph{realized} by a robot $r$, if $r$ is the unique closest robot to $t$ and $r \in B(t, \epsilon D) \cap \overline{ext}(C_l) \cap \overline{encl}(C(R))$. We shall say that $t_l$ is \emph{realized} by a robot $r$ if all target points $t \neq t_l$ are realized, $r$ is the robot closest to $t_l$ and $r \in B(t, \epsilon D) \cap encl(C_l)$. Hence, if $t_l$ is realized then it implies that all target points are realized, i.e., the given pattern is formed. We call this the final configuration (See also Fig. \ref{fig: cl}).



 Now the objective is to realize all the target points. This will be done in the following way.  
 
 \begin{enumerate}[label=(\arabic*)]

  \item First the robot $r_l$ moves inside $encl(C_l)$, if not already there. The movement should be such that \texttt{s} remains true.
  
  \item  When $r_l$ is inside $encl(C_l)$, The robots from $R \setminus \{r_l\}$ sequentially realize all the target points of $T \setminus \{t_l\}$. During this, \texttt{s} should remain true, in particular, $r_l$ should remain as the unique robot closest to $c(R)$. 
  
  \item When the target points of $T \setminus \{t_l\}$ are realized, the robot $r_l$ will then realize $t_l$. Again, \texttt{s} should remain true and $r_l$ should remain as the unique robot closest to $c(R)$.

 \end{enumerate}

  The movement strategy for (1) and (3) are straightforward. So we now discuss (2) in detail. For any non-final configuration in this phase, the target points can be partitioned as $T = T_1 \cup T_2 \cup \{t_l\}$, where $T_1$ is the set of realized target points and $T_2 \cup \{t_l\}$ are the unrealized target points. Also, the robots can be partitioned as $R = R_1 \cup R_2 \cup \{r_l\}$, where $R_1$ is the set of robots realizing target points and $R_2 \cup \{r_l\}$ are the rest. Notice that $r_1$ and $r_2$ are at the two target points corresponding to the bounding structure and hence $r_1, r_2 \in R_1$. So our goal here is to make $T_2 = \emptyset,  R_2 = \emptyset$. When $T_2 \neq \emptyset, R_2 \neq \emptyset$, we choose the closest pair from the set $R_2 \times T_2$. In case of tie between say $(r, t)$ and $(r',t')$ with $t \neq t'$, the target with lexicographically smaller coordinates (with respect to the global coordinate system) is chosen. In case of a tie between $(r, t)$ and $(r',t)$ the robot with lexicographically smaller coordinates is chosen. Let $(r,t)$ be the chosen pair. We call the robot $r$ the \emph{traveler} and $t$ its \emph{destination}. The goal is for $r$ to realize $t$.

  Let us denote the region $B(t, \epsilon D) \cap \overline{ext}(C_l) \cap \overline{encl}(C(R))$ as $\mathcal{R}(t)$. Recall that the objective here is to have $r$ inside the region $\mathcal{R}(t)$ and to be strictly closest to $t$.  Now $r$ is either in $\mathcal{R}(t)$ or not in $\mathcal{R}(t)$. In the first case, although $r \in \mathcal{R}(t)$, there may be at least another robot $r' \in \mathcal{R}(t)$ with $d(r,t) = d(r',t)$. So it will move closer to $t$ so that it realizes $t$. So now we consider the later case where $r \notin \mathcal{R}(t)$ and $r$ has to get inside $\mathcal{R}(t)$. $\mathcal{R}(t)$ is either $B(t, \epsilon D)$ or some other region which can be of two types (a region bounded by two circular arcs or four circular arcs) as shown in Fig. \ref{fig: cl}. As discussed in Section \ref{sec: safe zone}, in this case the robot will take some fixed disk $B(\tilde{t}, l) \subseteq \mathcal{R}(t)$ and try to move inside it. We set a fixed rule regarding how the ball $B(\tilde{t}, l)$ will be chosen and it depends only upon the shape of the region $\mathcal{R}(t)$. We then call $\tilde{t}$ \emph{the modified destination of} $r$. We use the terminology and the notation $B(\tilde{t}, l)$ in general, i.e., even when $\mathcal{R}(t) = B(t, \epsilon D)$, in which case $B(\tilde{t}, l) = B(t, \epsilon D)$. So the robot $r$ has to get inside $B(\tilde{t}, l)$. We now describe the algorithm.

   Let $S = \{\overline{B}(t', d(t', r')) \mid t' \in T_1 \text{ and } r' \text{ is the robot } \text{realizing } t'\} \cup \{\overline{B}(c(T), d(c(T), r_l))\}$ be the set of disks around the points of $T_1 \cup \{c(T)\}$ with their radii being equal to their distances from their robots. While moving, the robot $r$ should not move inside any of these disks. This is because if $r$ enters $\overline{B}(t', d(t', r'))$, then $r'$ is no longer realizing $t'$. Similarly if $r$ enters $\{\overline{B}(c(T), d(c(T), r_l))\}$, then $r_l$ is no longer strictly closest to the center.  Therefore the disks in $S$ will be treated as obstacles that the robot $r$ needs to avoid while moving. If none of the disks from $S$ intersect ${seg}(r, \tilde{t})$, then the robot can follow the movement strategy described in Section \ref{sec: safe zone}. Now assume that some disks from $S$ are intersecting $seg(r,\tilde{t})$. If $\overline{B}(c, d(c,r'))$ be such a disk (where $c$ is either a target point from $T_1$ or $c(T)$), then we say that $r'$ is obstructing the traveler. First consider the case where all disks from $S$ that intersect $seg(r,\tilde{t})$ have their centers not lying on $seg(r,\tilde{t})$. If $\overline{B}(c, d(c,r'))$ be such a disk, then $r'$ will move closer to $c$ so that the disk gets smaller and it does not intersect $seg(r,\tilde{t})$. If there are multiple such robots, then movements are sequentialized based on the view of their corresponding target points. Hence, after some rounds, there will not be any robot obstructing $r$ and it can find a safe zone to move through. Now consider the case where the center of some disk from $S$ is collinear with $r$ and $\tilde{t}$. In this case, $r$ will move to a point inside $Cone(r, B(\tilde{t}, l))$ so that there are no such collinearities as described in Section \ref{sec: safe zone}.

   We have to show that  $r$ remains the traveler robot and $t$ remains its target until it gets inside $\mathcal{R}(t)$. For this, we first claim that after an intermediate move by $r$, its distance from $t$ reduces. This is not difficult to see, but still not completely obvious as $r$ intends to move towards $\tilde{t}$ which may be different from $t$. Suppose that the  position of $r$ before the movement was $x$. Now $r$ have moved to some point inside $Cone(x, B(\tilde{t}, l))$. But since $B(\tilde{t}, l) \subseteq B(t, \epsilon D)$, we have $Cone(x, B(\tilde{t}, l)) \subseteq Cone(x, B(t, \epsilon D))$. Hence, $r$ have moved to some point inside $Cone(x, B(t, \epsilon D))$. Hence, its distance from $t$ has reduced by Property \ref{prop: cone reduce}. From this it implies that $r$ remains the traveler. This is because 1) $(r,t)$ was a closest pair from $R_2 \times T_2$ before the move 2) all distances $d(r', t')$ with $r' \in R_2 \setminus \{r\}$ and $t' \in T_2$ have remained unchanged. Now we have to show that $t$ remains its destination. To see this, assume for the sake of contradiction that $r$ moves from $x$ to $y$ and we have $d(y, t') < d(y, t)$ for some $t' \in T_2 \setminus \{t\}$. We have $y \in Cone(x, B(\tilde{t}, l)) \subseteq Cone(x, B(t, \epsilon D))$. Let $L$ be the perpendicular bisector of $\overline{seg}(t, t')$ and $\mathcal{H}$ is the open half-plane delimited by $L$ that contains $t$. We have $x \in \overline{\mathcal{H}}$ as $(r,t)$ was a closest pair from $R_2 \times T_2$ before the move. Since $d(t, t') > 2 \epsilon D$, ${B}(t, \epsilon D) \subset \mathcal{H}$. Hence $Cone(x, B(t, \epsilon D)) \subset \mathcal{H}$. This implies that $y \in \mathcal{H}$ which contradicts our assumption that $d(y, t') < d(y, t)$.



  \begin{lemma}
 
 If the algorithm is in Phase 3, Case 1 then we have a final configuration after finitely many rounds.

\end{lemma}

   \subsubsection{Case 2}

  In Case 1, the robot $r_l$ played crucial role in keeping $\texttt{b}$ true throughout the execution and as a result having  an agreement regarding the embedding of the pattern $F$. In Case 2, the situation is simpler as the scalene triangle formed  by the robots on $C(R)$ determines a particular embedding of $F$ as described in Section \ref{sec: p2}. So as long as the remaining robots stay inside $encl(R)$, we have $\texttt{b}$ and consequently  an agreement regarding the embedding of $F$. So these robots will sequentially move to get inside the regions $B(t, \epsilon D) \cap {encl}(C(R))$, where $t$ are the target point, while preserving $\texttt{b}$. This will be achieved using the similar strategies as discussed in the previous case.

    \begin{lemma}
 
 If the algorithm is in Phase 3, Case 2 then we have a final configuration after finitely many rounds.

\end{lemma}

%
%
%

%
%
%
%
%
%
%
%

\subsection{The Main Result}\label{sec: correct}

Recall that a configuration with $\neg\texttt{u} \wedge (\neg\texttt{a} \vee \neg\texttt{c})$ is in Phase 1, a configuration with $\texttt{a} \wedge \texttt{c} \wedge \neg\texttt{b}$ is in Phase 2, and a configuration with $\texttt{b}$ is in Phase 3. It is easy to see that any configuration with $\neg \texttt{u} $ belongs to one of the three phases.  Phase 1 terminates with $\texttt{a} \wedge \texttt{c}$ which is either a Phase 2 or Phase 3 configuration.  Phase 2 terminates with $\texttt{b}$ which is a Phase 3 configuration. A final configuration is formed in Phase 3. Hence the algorithm solves the problem in $\mathcal{OBLOT} + \mathcal{SSYNC}$ from any configuration which is $\neg \texttt{u}$.

\section{The Algorithm for Asynchronous Robots}\label{sec: main async}

 Let us denote the algorithm presented in Section \ref{sec: main ssync} as $\mathbf{A}$. It works in $\mathcal{OBLOT} + \mathcal{SSYNC}$. Notice that a feature of this algorithm is that it is \emph{sequential} in the following sense. At any round during the execution of the algorithm, at most one robot decides to move. This immediately gives an algorithm that works in $\mathcal{FCOM} + \mathcal{ASYNC}$ using two colors $\{\texttt{busy}, \texttt{idle}\}$. The algorithm $\mathbf{A}$ can be seen as a function that maps the snapshot taken by a robot to a movement instruction.  We now construct an algorithm $\mathbf{A}'$ from $\mathbf{A}$ with two colors $\{busy, idle\}$ in the following way. Initially the colors of all robots are set to $\texttt{idle}$. If any robot finds some robot with light set to $\texttt{busy}$, then it does nothing. Otherwise, it applies $\mathbf{A}$ on its snapshot (ignoring colors). If $\mathbf{A}$ returns a non-null move, it sets its light to $\texttt{busy}$ and moves accordingly. If $\mathbf{A}$ returns a null move, it sets its light to $\texttt{idle}$ (recall that it does not know what its present color is) and does not make any move. It is easy to see that  $\mathbf{A}'$ solves the problem in $\mathcal{FCOM} + \mathcal{ASYNC}$.

   \section{Concluding Remarks}
   
   We have introduced a  model for robots with inaccurate movements. We have presented algorithms for pattern formation in $\mathcal{OBLOT + SSYNC}$ and $\mathcal{FCOM + ASYNC}$. Devising an algorithm for $\mathcal{OBLOT + ASYNC}$ is an interesting direction for future research. Another direction would be to consider robots with physical extent.

\bibliographystyle{plainurl}
\bibliography{epsilon_pattern}

 \end{document}